\documentclass[%
 aip,
 amsmath,amssymb,
reprint,%
author-year,%
]{revtex4-1}

\usepackage{graphicx}
\usepackage{dcolumn}
\usepackage{bm}

\usepackage[utf8]{inputenc}
\usepackage[T1]{fontenc}
\usepackage{mathptmx}
\usepackage{etoolbox}

\usepackage{natbib}

\usepackage{multirow}   
\usepackage{enumitem}   
\usepackage{booktabs}   
\usepackage{xcolor}		
\usepackage{caption}
\usepackage{subcaption}
\usepackage{anyfontsize}
\usepackage{siunitx}  
\usepackage{microtype} 

\usepackage{hyperref}
\usepackage{CJKutf8} 
\setcitestyle{aysep={,}} 
\makeatletter
\def\@email#1#2{%
 \endgroup
 \patchcmd{\titleblock@produce}
  {\frontmatter@RRAPformat}
  {\frontmatter@RRAPformat{\produce@RRAP{*#1\href{mailto:#2}{#2}}}\frontmatter@RRAPformat}
  {}{}
}%
\makeatother
\begin{document}
\begin{CJK*}{UTF8}{gbsn} 
\preprint{AIP/123-QED}

\title[]{Lattice Boltzmann simulation reveals supercritical bifurcation in flow mode transitions of power-law fluids in the four-roll mill}

\author{Yuan Yu~({\CJKfamily{gbsn}余愿})}
\affiliation{School of Mathematics and Computational Science, Xiangtan University, Xiangtan 411105, China}
\affiliation{National Center for Applied Mathematics in Hunan, Xiangtan 411105, China}
\affiliation{Hunan Key Laboratory for Computation and Simulation in Science and Engineering, Xiangtan University, Xiangtan 411105, China}

\author{Xiao Jiang~({\CJKfamily{gbsn}江校})}\altaffiliation[Electronic mail: ]{xjiang@smail.xtu.edu.cn}
\affiliation{School of Mathematics and Computational Science, Xiangtan University, Xiangtan 411105, China}

\author{Qingqing Gu~({\CJKfamily{gbsn}顾青青})}\altaffiliation[Author to whom correspondence should be addressed: ]{guqq@usst.edu.cn}
\affiliation{School of Energy and Power Engineering, University of Shanghai for Science and Technology, Shanghai 200093, China}

\author{Chuandong Lin~({\CJKfamily{gbsn}林传栋})}
\affiliation{Sino-French Institute of Nuclear Engineering and Technology, Sun Yat-sen University, Zhuhai 519082, China}

\author{Qingyong Zhu~({\CJKfamily{gbsn}朱庆勇})}\altaffiliation[Electronic mail: ]{mcszqy@mail.sysu.edu.cn}
\affiliation{School of Aeronautics and Astronautics, Sun Yat-sen University, Shenzhen 518107, China}

\author{Hai-zhuan Yuan~({\CJKfamily{gbsn}袁海专})}\altaffiliation[Electronic mail: ]{yhz@xtu.edu.cn}
\affiliation{School of Mathematics and Computational Science, Xiangtan University, Xiangtan 411105, China}

\date{\today}

\begin{abstract}
The four-roll mill has been traditionally viewed as a device generating simple extensional flow with a central stagnation point. Our systematic investigation using a two-relaxation-time regularized lattice Boltzmann (TRT-RLB) model reveals unexpected richness in the flow physics, identifying two previously unreported supercritical bifurcation modes: a quadrifoliate vortex mode featuring four symmetrical counter-rotating vortices, and a dumbbell-shaped quad-vortex mode where vortices detach from but remain symmetric about the stagnation point. The numerical framework, representing the first successful extension of TRT-RLB method to power-law fluid dynamics, enables comprehensive mapping of flow characteristics across Reynolds numbers ($1 \leq Re \leq 50$), power-law indices ($0.7 \leq n \leq 1.3$), and geometric configurations. The transition from quadrifoliate vortex mode exhibits distinct pathways depending on the power-law index: at relatively small $n$, the flow undergoes a direct supercritical bifurcation to simple extensional flow, while at relatively large $n$, it evolves through an intermediate dumbbell-shaped state. Among geometric parameters, the roller radius $r$ emerges as the dominant factor controlling bifurcation points and vortex dimensions, whereas the roller-container gap $\delta$ exerts minimal influence on flow regimes. The transitions between flow modes can be precisely characterized through the evolution of vortex dimensions and velocity gradients at the stagnation point, providing quantitative criteria for flow regime identification. These findings enrich our fundamental understanding of bifurcation phenomena in extensional devices and provide quantitative guidelines for achieving desired flow patterns in four-roll mill applications.
\end{abstract}

\maketitle
\end{CJK*} 

\section{Introduction}\label{section1}

\begin{figure*}
  \includegraphics[width=0.45\textwidth]{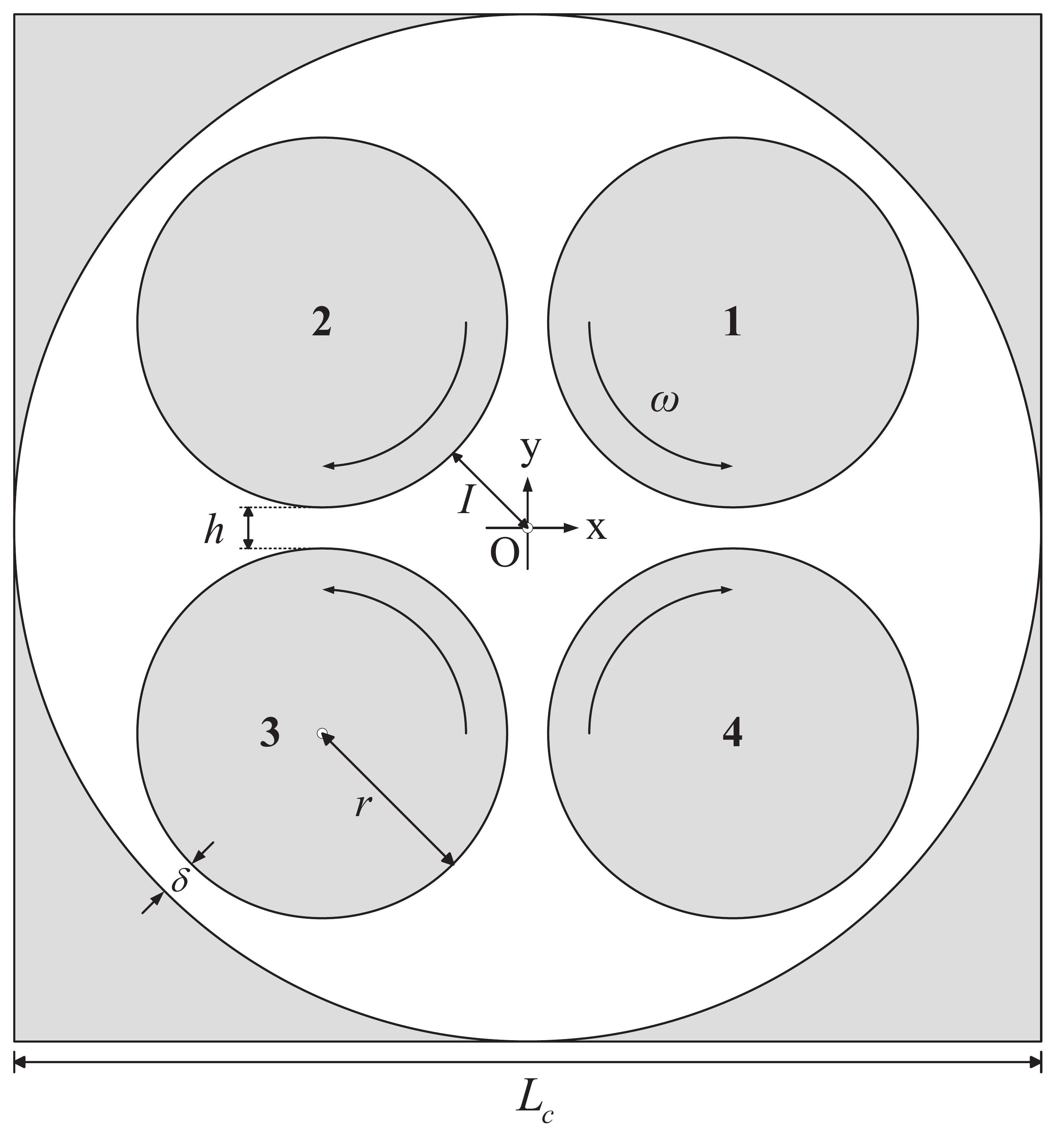}
  \caption{Cross-sectional schematic of the four-roll mill configuration. The rollers of length $L_{z}$ rotate at angular velocity $\omega$. The geometric parameters include: roller radius $r$, gap between adjacent rollers $h$, gap between rollers and container wall $\delta$, and container diameter $L_c$. Each roller rotates with equal speed but with alternating directions as indicated by the arrows. The flow is assumed two-dimensional due to the sufficiently long rollers ($L_{z} \gg L_c$).}
  \label{four_roll_mill}
\end{figure*}

While extensive research has explored four-roll mill dynamics using various numerical approaches, the simulation of power-law fluids in such complex geometries presents unique challenges. Traditional numerical methods often struggle with the nonlinear constitutive relationships and complex boundary conditions. The two-relaxation-time regularized lattice Boltzmann (TRT-RLB) method offers distinct advantages for such systems, including superior numerical stability and efficient handling of curved boundaries. This work represents the first successful extension of the TRT-RLB methodology to power-law fluid dynamics in four-roll mill configurations, enabling comprehensive investigation of flow mode transitions and bifurcation phenomena.

To achieve steady-state deformation of droplets or cells at a prescribed strain rate, a continuous extensional flow (also known as hyperbolic flow) must be generated in the fluid. This flow configuration is characterized by a stagnation point, where the local velocity vanishes while maintaining a non-zero strain rate. The four-roll mill, as shown in Fig.~\ref{four_roll_mill}, was invented by \citet{taylor_formation_1934} to generate simple extensional flow with a stagnation point. The device comprises four cylindrical rollers with their axes arranged in a square configuration, positioned symmetrically and parallel to one another. The simple extensional flows with stagnation points, generated by four-roll mills or similar devices, have been instrumental in various research areas: investigating drop and cell deformation and breakup \citep{PhysRevE.100.063107,BRYNGELSON2019171,PhysRevFluids.5.113603,HYMEL2020493,10.1122/8.0000600,CRMECA_2023__351_S2_163_0}, conducting extensional rheometry of complex fluids \citep{Haward2012ExtensionalRA,Haward2012OptimizedCF}, and examining macromolecular dynamics \citep{10.1126/science.1086070,10.1122/1.4972236,PhysRevFluids.1.053301,10.1122/1.4972237} and elastic instabilities \citep{PhysRevLett.96.144502,PhysRevLett.99.164503,hawardElasticInstabilitiesPlanar2016a}.

Research on non-Newtonian fluids using the four-roll mill has focused on several key areas. In polymer studies, \citet{millikenDeformationBreakupViscoelastic1991} employed a computer-controlled four-roll mill to investigate polymer drop deformation and breakup in immiscible Newtonian fluids. The dynamics of polymer systems was further explored by \citet{leeFlowLightScattering1999}, who examined isolated polymer chains in dilute solutions under steady simple extensional flow using flow light scattering, and by \citet{schroederObservationPolymerConformation2003}, who observed highly extended E. coli DNA molecules through fluorescence microscopy. Investigations into viscoelastic effects include studies by \citet{trethewayDeformationRelaxationNewtonian2001} and \citet{haExperimentalStudyDrop2001} on the deformation and relaxation of Newtonian droplets in Boger fluids, while \citet{davinoNumericalSimulationsDispersive2017} numerically analyzed the dispersive mixing of particles in viscoelastic fluids. The optical properties of non-Newtonian fluids have also been extensively studied, with several researchers investigating flow birefringence phenomena of polymers in simple extensional flow \citep{crowley_localized_1976,fuller_flow_1980,fuller_flow_1981,dunlap_dilute_1987}. These diverse investigations of polymer dynamics, viscoelastic effects, and flow birefringence highlight the four-roll mill as an indispensable apparatus for probing the fundamental physics of non-Newtonian fluids.
	
Various control techniques of the four-roll mill have been developed to achieve stable simple extensional flow and keep particles or droplets at the stagnation point. Initially, Taylor's manually controlled four-roll mill \citep{taylor_formation_1934} often resulted in unstable simple extensional flow. \citet{bentleyComputercontrolledFourrollMill1986} designed a computer-controlled four-roll mill, which stabilized the flow field and maintained particles at its center through automated roller speed adjustment. Subsequent studies focused on different control strategies: \citet{balasuriyaControllingUnsteadyAnalogue2013} proposed a method based on velocity perturbation control to maintain particles on hyperbolic trajectories, while \citet{vonaStabilizingViscousExtensional2021} developed machine learning algorithms for droplet stabilization. While these control techniques have achieved considerable success, their emphasis on technological solutions rather than fundamental flow physics suggests an incomplete understanding of the underlying flow mechanisms. A comprehensive investigation of flow modes and their transitions in the four-roll mill is therefore essential for both scientific understanding and practical applications.

Beyond control techniques, extensive research has explored how the geometric parameters of four-roll mills influence the generated simple extensional flow. \citet{Bentley1985DropDA} emphasized the importance of moderate mill dimensions: sufficiently large to accommodate highly deformed droplets before breakup, yet small enough to maintain stable simple extensional flow while minimizing construction and filling costs. Through experimental studies, \citet{lagnadoVisualizationThreedimensionalFlow1990} observed that while homogeneous simple extensional flow develops in the central region at specific Reynolds numbers $ Re $, stable vortices emerge along the top and bottom walls when $ Re $ exceeds a critical value. This transition was primarily attributed to the ratio of roller length to the gap between adjacent rollers. In a numerical investigation comparing various mill configurations, \citet{higdonKinematicsFourrollMill1993} determined that optimal simple extensional flow occurs when the roller radius $ r $, the gap $ h $ between adjacent rollers, and the square container size $ L $ satisfy $ r/(h/2+r)=0.625 $ and $ L/(h/2+r)>3.0 $. \citet{andreottiExperimentTwoAspects2001} proposed that simple extensional flow requires the gap $ h $, roller radius $ r $, and distance $ I $ from rollers to the mill center to satisfy:
\begin{equation}\label{1eq1}
  (r+I)^2=2(r+h/2)^2,
\end{equation}
where $ (r+I) $ represents the distance from roller center to mill center, and $ (r+h/2) $ denotes the distance from roller center to axis. They further outlined specific geometric conditions to minimize the impact of roller-container spacing on central simple extensional flow. A circular container was recommended over a square one, as corner regions in square containers generate complex recirculating flows that may influence central flow modes. Furthermore, excessive gaps $ \delta $ between rollers and container walls can induce Couette-Taylor-like flows around rollers, characterized by superposed toroidal sectors with alternating rotation directions. Similarly, insufficient gaps between adjacent rollers may produce complex recirculating flows in the central region. \citet{akbaridoustAssessmentMiniatureFourroll2018} experimentally demonstrated that reducing inter-roller gaps and increasing roller radius enhances the strain rate at the stagnation point, attributed to increased linear roller velocity and consequent flow velocity gradients. Notably, both \citet{andreottiExperimentTwoAspects2001} and \citet{akbaridoustAssessmentMiniatureFourroll2018} found that the stagnation point strain rate is directly proportional to roller angular speed, and that lower fluid viscosity leads to more unstable simple extensional flow.

Previous research on four-roll mills has primarily focused on optimizing control techniques and geometric configurations to achieve stable simple extensional flow in the central region, thereby maintaining particles or droplets at the stagnation point. However, a systematic investigation of the underlying flow mechanisms remains unexplored. While extensive research has explored four-roll mill dynamics using various numerical approaches \citep{wang2012numerical}, the simulation of power-law fluids in such complex geometries presents unique challenges. Traditional numerical methods often struggle with the nonlinear constitutive relationships and complex boundary conditions. The two-relaxation-time regularized lattice Boltzmann (TRT-RLB) method offers distinct advantages for such systems, including superior numerical stability and efficient handling of curved boundaries \citep{yu_two-relaxation-time_2024,YU2025129135}.
This study examines the flow mechanisms of Newtonian and power-law fluids in the four-roll mill, revealing previously unreported flow modes. A distinctive mode is observed where the flow deviates from simple extensional behavior, forming four symmetrical vortices around the central point with alternating rotational directions between adjacent vortices. While the central point remains stagnational, it maintains simple extensional flow in its immediate vicinity. The correlation between central point shear rate, vortex formation, and governing parameters such as Reynolds number $Re$, power-law index $n$, roller-container gap $\delta$, and roller radius $r$ is investigated, with special attention to the critical conditions for vortex suppression.
For systematic numerical investigation of various flow modes, the TRT-RLB model \citep{yu_two-relaxation-time_2024} is employed, ensuring high accuracy and stability in power-law fluid simulations. High-order accuracy near boundary regions is achieved through the implementation of the one-point second-order curved boundary scheme \citep{tao_one-point_2018}. The remainder of this paper is structured as follows: Section \ref{section1} reviews the research progress on four-roll mills; Section \ref{section2} presents the problem description and governing equations; Section \ref{section3} details the TRT-RLB model, boundary condition implementation, and numerical validation; Section \ref{section4} discusses the simulation results; and Section \ref{section5} summarizes the main findings.

\section{Problem description and governing equation}\label{section2}
	
\subsection{Structure of the four-roll mill}\label{section2.1}

Fig.~\ref{four_roll_mill} shows the cross-sectional view of the four-roll mill used in this study, where the rollers are assumed to be sufficiently long such that the flow can be approximated as two-dimensional (2D) in the central region. This idealization is justified when the roller length is much larger than other geometric dimensions, allowing us to focus on the essential flow physics in the transverse plane. The origin of the coordinate system, denoted as point $O(0,0)$, coincides with the center of the mill. The circular container has a diameter of $L_c$, and the gap between adjacent rollers is $h$. Four identical rollers of radius $r$ are symmetrically arranged around point $O$, labeled as rollers 1 through 4, with their centers located at $(r+h/2,r+h/2)$, $(-r-h/2,r+h/2)$, $(-r-h/2,-r-h/2)$, and $(r+h/2,-r-h/2)$, respectively. The distance from each roller edge to point $O$ is denoted as $I$, yielding a center-to-origin distance of $(I+r)$, which satisfies the geometric relationship proposed by \citet{andreottiExperimentTwoAspects2001} in Eq.~(\ref{1eq1}).

Given the assumption of infinite roller length $L_{z}$ along the $z$-axis, the flow field exhibits no variation in the longitudinal direction ($\partial u/\partial z=0$) and no axial velocity component ($u_z=0$). Each roller rotates at a constant angular velocity $\omega$, producing a uniform tangential velocity of $U_c=\omega r$ at the roller surface. To establish simple extensional flow near point $O$, adjacent rollers rotate with equal angular speeds but in opposite directions: rollers 2 and 4 rotate clockwise, while rollers 1 and 3 rotate counterclockwise.

\subsection{Governing equations}\label{section2.2}

The 2D continuity equation and momentum equations for the generalized Newtonian fluid used in this study are as follows:

\begin{subequations}
  \begin{align}
    \frac{\partial \rho}{\partial t} + \frac{\partial (\rho u_\alpha)}{\partial x_\alpha} &= 0, \\
    \frac{\partial (\rho u_\alpha)}{\partial t} + \frac{\partial (\rho u_\alpha u_\beta)}{\partial x_\beta} &= -\frac{\partial (p \delta_{\alpha\beta})}{\partial x_\alpha} + \frac{\partial \tau_{\alpha\beta}}{\partial x_\beta}.
  \end{align}
\end{subequations}
where $\rho$ denotes fluid density, $u_{\alpha}$ represents velocity component in the $\alpha$ direction, $x_{\alpha}$ is the spatial coordinate, $t$ represents time, $p$ is pressure, $\delta_{\alpha\beta}$ is the Kronecker delta, and $\tau_{\alpha\beta}$ represents the stress tensor, given by
\begin{equation}
  \tau_{\alpha\beta}=\mu\dot{\gamma}_{\alpha\beta},
\end{equation}
where $\dot{\gamma}_{\alpha\beta}$ is shear-rate tensor and $\mu$ is dynamic viscosity. For power-law fluids, the effective viscosity is
\begin{equation}
  \mu=\mu_0|\dot{\gamma}|^{n-1},
\end{equation}
where $\mu_0$ and $n$ represent viscosity coefficient and power-law index, respectively. The shear rate $|\dot{\gamma}|$ is defined as $|\dot{\gamma}|=\sqrt{2D_{\Pi}}$, with $D_{\Pi}$ being the second invariant of the strain rate tensor $S_{\alpha\beta}$ and expressed as
\begin{equation}
  D_{\Pi}=\sum_{\alpha,\beta=1}^2S_{\alpha\beta}S_{\alpha\beta},
\end{equation}
where $S_{\alpha\beta}$ is defined as 
\begin{equation}
  S_{\alpha\beta}=\frac{1}{2}\Big(\partial_{\beta}u_{\alpha}+\partial_{\alpha}u_{\beta}\Big),
\end{equation}
leading to the explicit expression for $|\dot{\gamma}|$: 
\begin{equation}
  |\dot{\gamma}|=\sqrt{2{{\left( {{\partial }_{x}}{{u}_{x}} \right)}^{2}}+2{{\left( {{\partial }_{y}}{{u}_{y}} \right)}^{2}}+{{\left( {{\partial }_{y}}{{u}_{x}}+{{\partial }_{x}}{{u}_{y}} \right)}^{2}}}.
\end{equation}
For power-law fluids, the effective Reynolds number $Re$ is defined using the characteristic length $h$ and velocity $U_c$:
\begin{equation}
  Re=\frac{\rho U_c^{2-n}h^n}{\mu_0}.
\end{equation}

\section{Numerical methods and validation}\label{section3}
	
In the following sections, the TRT-RLB model is introduced in \ref{section3.1}, the one-point second-order curved boundary scheme is described in \ref{section3.2}, the model is validated using power-law fluids in channel flow and Taylor-Couette flow problems in \ref{section3.3}, and the grid independence analysis is conducted in \ref{section3.4}.

\subsection{Two-relaxation-time regularized lattice Boltzmann model}\label{section3.1}

The TRT-RLB model proposed by \citet{yu_two-relaxation-time_2024} is employed in this study to ensure superior numerical stability and accuracy compared to other collision models. Its evolution equation is expressed as
\begin{align}	
  \label{2eq3}
  f_i\left(x_\alpha+e_{i \alpha} \Delta t, t +\Delta t\right) &=
  f_i^{eq}\left(x_\alpha,t\right)+\left(1-\frac{1}{\tau_{1}}\right) w_i \frac{\mathcal{H}_{i, \alpha}}{c_s^2} \mathcal{A}_\alpha^{neq} \notag \\
  & +\left(1-\frac{1}{\tau_{1}}\right) w_i \frac{\mathcal{H}_{i, \alpha \beta}}{2 c_s^4} \mathcal{A}_{\alpha \beta}^{neq} \notag \\
  & +\left(1-\frac{1}{\tau_{2}}\right) w_i \frac{\mathcal{H}_{i, \alpha \beta \gamma}}{6 c_s^6} \mathcal{A}_{\alpha \beta \gamma}^{neq} \notag \\
  & + \left(1-\frac{1}{2\tau_{1}}\right) F_i\Delta t+G_i \Delta t,
\end{align}
where $f_{i}$ represents the particle distribution function in the $i$-th direction, and discrete velocity space D$d$Q$q$. The term $e_{i\alpha}$ denotes the discrete velocity in the $i$-th direction of the D$d$Q$q$. $\tau_{1}$ and $\tau_{2}$ are dimensionless relaxation times, while $w_i$ and $c_s$ represent the weight coefficients and lattice sound speed, respectively. The equilibrium distribution function $f_i^{eq}$ is represented by a third-order Hermite expansion, given by
\begin{equation}\label{2eq4}
  f_i^{e q}=w_i \rho \left\{\mathcal{H}_i+\frac{\mathcal{H}_{i, \alpha}}{c_s^2} u_\alpha+\frac{\mathcal{H}_{i, \alpha \beta}}{2 c_s^4} u_\alpha u_\beta+\frac{\mathcal{H}_{i, \alpha \beta \gamma}}{6 c_s^6} u_\alpha u_\beta u_\gamma\right\},
\end{equation}
where $\rho$ is the density, satisfying the equation $p=\rho c_s^2$. The Hermite polynomials are defined as follows:
\begin{gather*}
  \mathcal{H}_i=1, \quad \\
  \mathcal{H}_{i, \alpha}=e_{i \alpha}, \quad \\
  \mathcal{H}_{i, \alpha \beta}=e_{i \alpha} e_{i \beta}-c_s^2 \delta_{\alpha \beta}, \quad \\
  \mathcal{H}_{i, \alpha \beta \gamma}=e_{i \alpha} e_{i \beta} e_{i \gamma}-c_s^2\left(e_{i \alpha} \delta_{\beta \gamma}+e_{i \beta} \delta_{\gamma \alpha}+e_{i \gamma} \delta_{\alpha \beta}\right),
\end{gather*}
and their non-equilibrium moments are
\begin{subequations}\label{2eq5}
  \begin{gather}
    \label{2eq5a}
    \begin{align}
      \mathcal{A}_\alpha^{neq}=\sum_i \mathcal{H}_{i, \alpha}\left(f_i-f_i^{eq}\right),
    \end{align}\\
    \label{2eq5b}
    \begin{align}
      \mathcal{A}_{\alpha \beta}^{neq}=\sum_i \mathcal{H}_{i, \alpha \beta}\left(f_i-f_i^{e q}\right),
    \end{align}\\
    \label{2eq5c}
    \begin{align}
      \mathcal{A}_{\alpha \beta \gamma}^{\mathrm {neq }}=\sum_i \mathcal{H}_{i, \alpha \beta \gamma}\left(f_i-f_i^{e q}\right).
    \end{align}
  \end{gather}
\end{subequations}
$G_i$ in Eq.~(\ref{2eq3}) is the compensatory source term added to eliminate error terms when recovering the NS equations. The expression for $G_i$ is given by
\begin{equation}
  \label{2eq6}
  G_i=-w_i \frac{\mathcal{H}_{i, \alpha \beta}}{6 c_s^6}\left(1-\frac{1}{2\tau_{1}}\right)\partial_\gamma \Phi_{\alpha \beta \gamma},
\end{equation}
where
\begin{gather}
  \Phi_{\alpha \beta \gamma}=\left\{
  \begin{aligned}
    \rho &u_x^3,\quad &&\alpha=\beta=\gamma=x, \\
    \rho &u_y^3,\quad &&\alpha=\beta=\gamma=y, \\
    &0, &&\quad otherwise,
  \end{aligned}
  \right. \\
  \partial_\gamma \Phi_{\alpha\beta\gamma}=\partial_x\left(\rho u_x^3\right)\delta_{\alpha x}\delta_{\beta x}+\partial_y\left(\rho u_y^3\right)\delta_{\alpha y}\delta_{\beta y}.
\end{gather}
Forcing term $F_i$ in Eq.~(\ref{2eq3}) is
\begin{equation}
  \label{2eq9} 
  F_i=w_i\left\{\frac{\mathcal{H}_{i, \alpha}}{c_s^2}F_\alpha+\frac{\mathcal{H}_{i, \alpha \beta}}{2 c_s^4}\left(F_\alpha u_\beta+u_\alpha F_\beta\right)\right\}.
\end{equation}
The D2Q9 discrete velocity model is used in this paper. Its velocity sets are given by
\begin{gather}
  \left(\begin{array}{c}e_{ix}\\ e_{iy}\end{array}\right) 
  = c \left[\begin{array}{ccccccccc}
    0 & 1 & 0 & -1 & 0 & 1 & -1 & -1 & 1 \\
    0 & 0 & 1 & 0 & -1 & 1 & 1 & -1 & -1
  \end{array}\right],
\end{gather}
and the corresponding weights are $w_0=4/9$, $w_{1,...,4}=1/9$,$w_{5,...,8}=1/36$. The speed of sound is $c_s=c/\sqrt{3}$, where $c=\Delta x/\Delta t$ is lattice speed. $\Delta x$ and $\Delta t$ are the lattice spacing and time step, respectively. The local density $\rho$ and velocity $u_\alpha$ are calculated from the distribution function $f_i$:
\begin{gather}
  \sum_i f_i=\sum_i f_i^{e q}=\rho, \label{2eq10} \\
  \sum_i e_{i\alpha}f_i+\frac{\Delta t}{2}F_\alpha=\rho u_\alpha.\label{2eq11} 
\end{gather}
The viscosity of fluid is related to $\tau_{1}$:
\begin{equation}
  \mu=\rho c_s^2(\tau_{1}-0.5)\Delta t.
\end{equation}
Following \citet{yu_two-relaxation-time_2024}, the relaxation time $\tau_2$ is set to 1.6 to ensure robust numerical stability when simulating power-law fluids with wide-ranging viscosity variations.

\begin{figure*}
  \centering
  \includegraphics[width=0.5\textwidth]{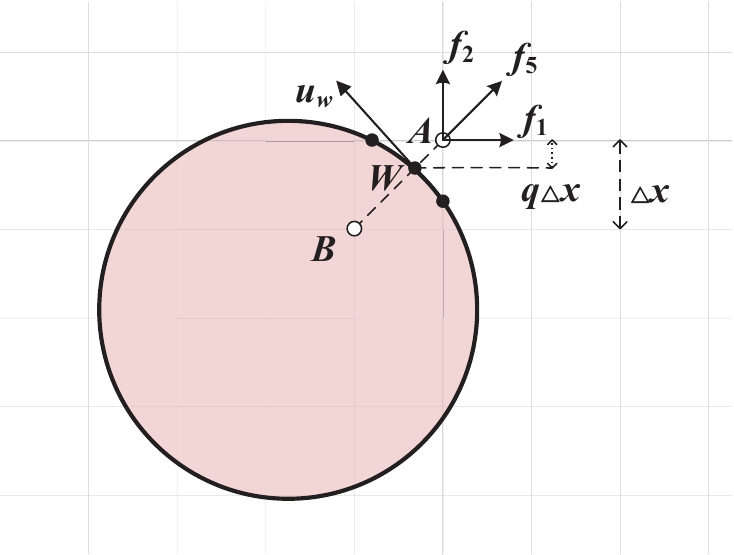}
  \caption{Schematic of the one-point second-order curved boundary scheme, where $x_{\alpha,A}$, $x_{\alpha,W}$, and $x_{\alpha,B}$ denote the fluid node, boundary point, and solid node, respectively.}
  \label{test_fig3.2.1}
\end{figure*}

\begin{figure*}
  \centering
  \includegraphics[width=0.6\textwidth]{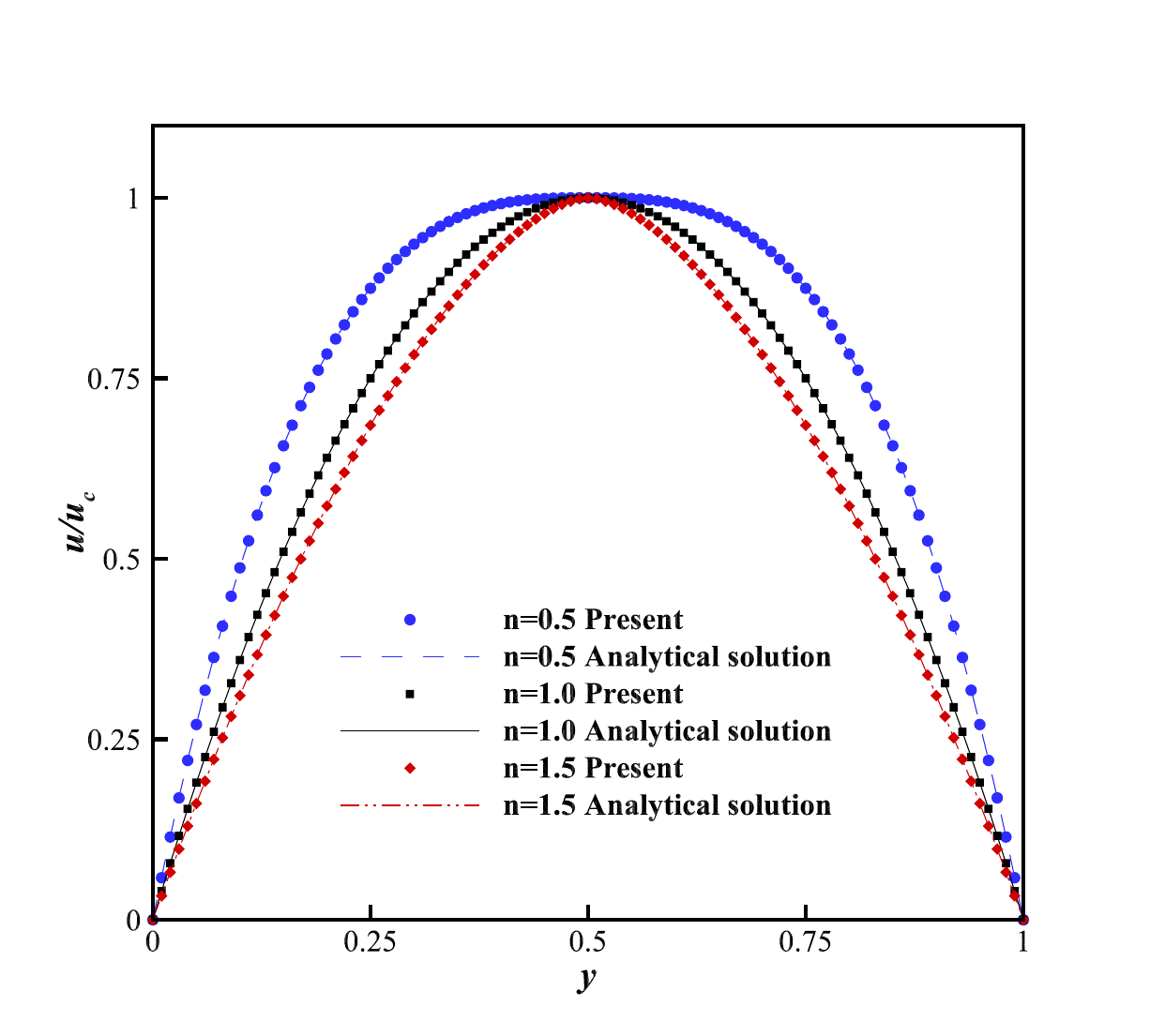}
  \caption{Comparison between numerical results (symbols) and analytical solutions (lines) for force-driven power-law channel flow at $Re=10$ with power-law indices $n=0.5$, 1.0, and 1.5.}
  \label{3test_fig1}
\end{figure*}

\begin{figure*}
  \centering
    \includegraphics[width=0.5\textwidth]{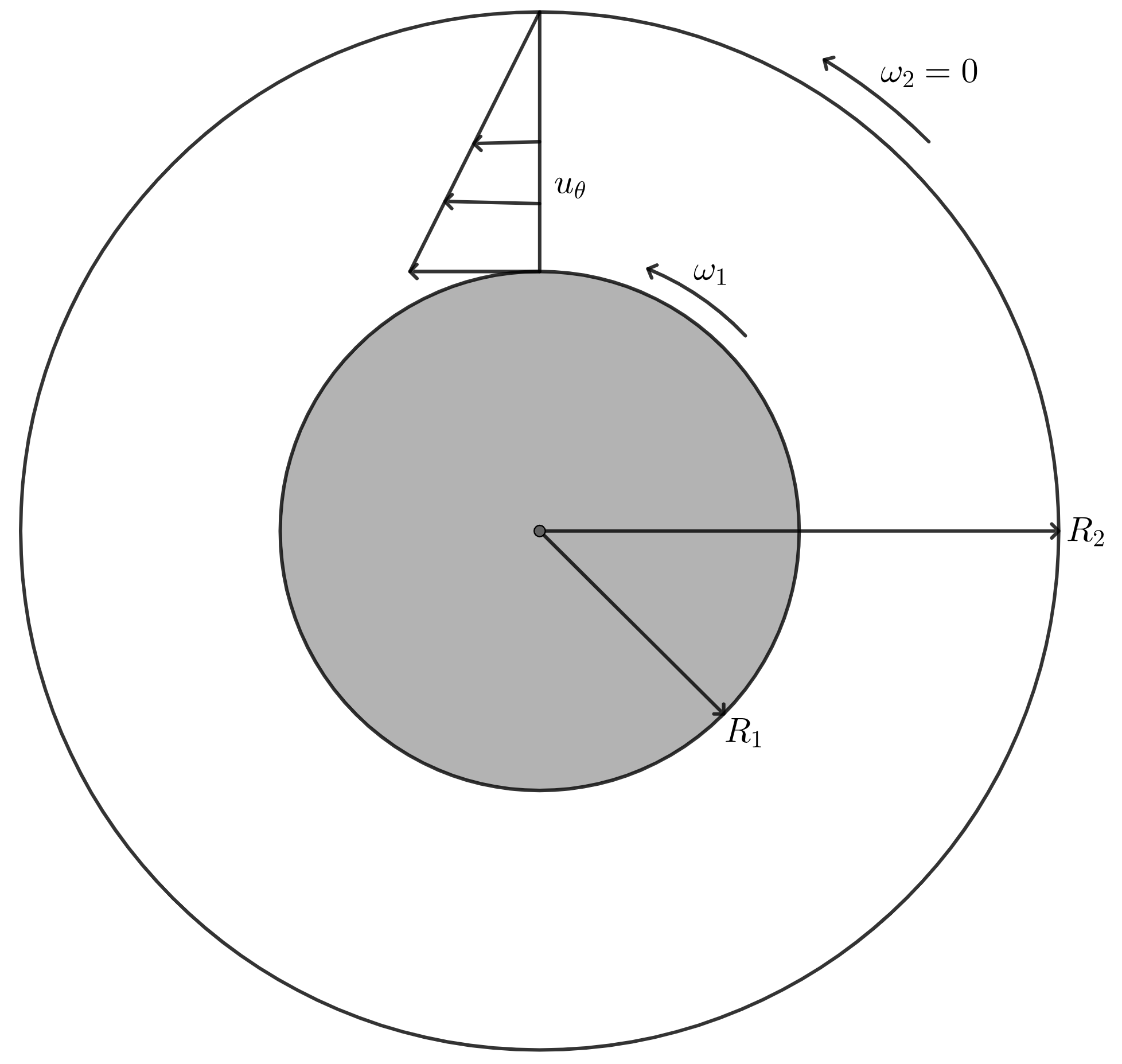}
    \caption{Schematic of Taylor-Couette flow: an inner cylinder of radius $R_1$ rotating at angular velocity $\omega_1$, and a stationary outer cylinder of radius $R_2$. This configuration serves as a benchmark case for validating the curved boundary treatment.}
    \label{3Cylindrical_Couette_flow}
\end{figure*}

\begin{figure*}
  \centering
  \includegraphics[width=0.6\textwidth]{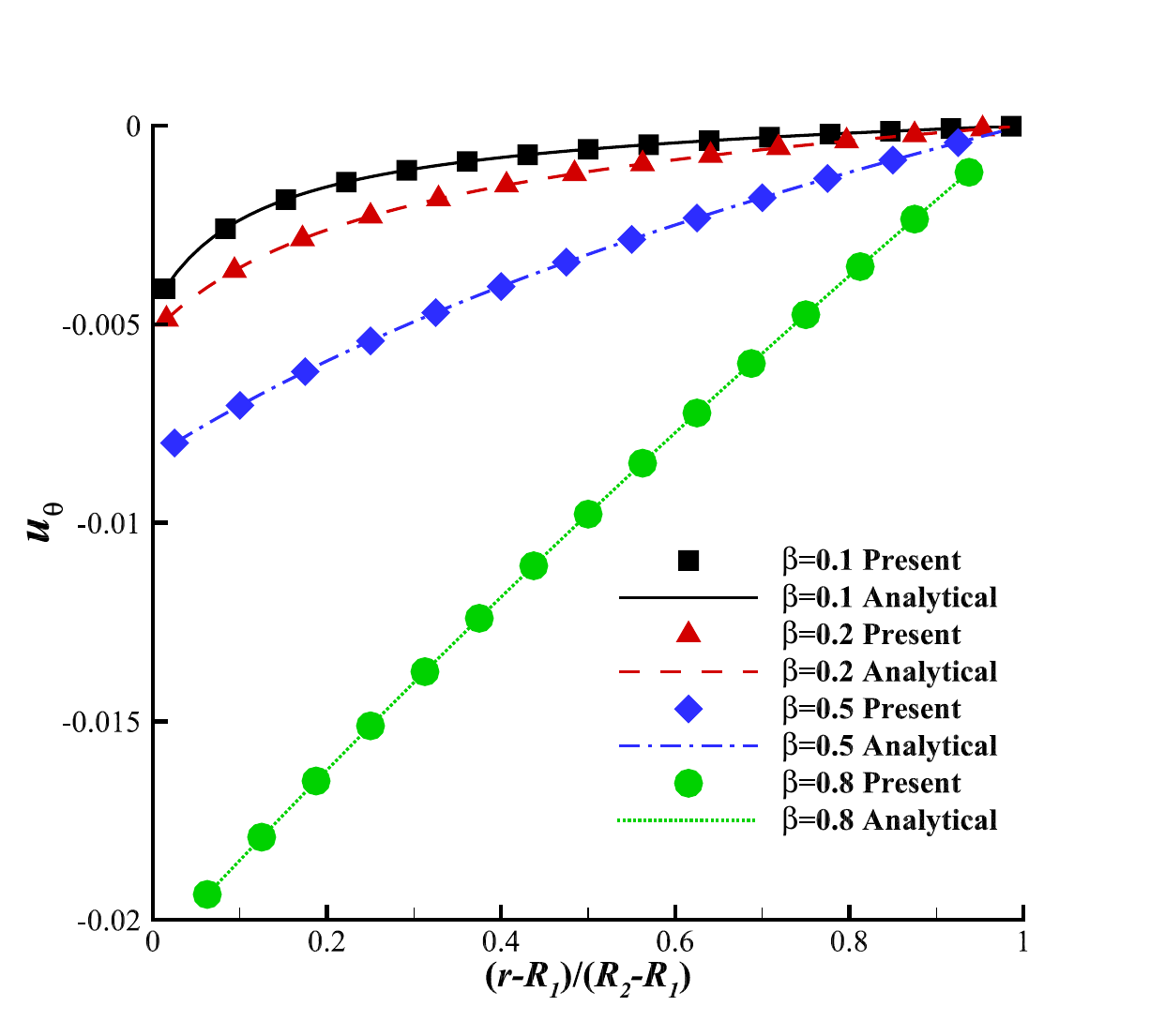}
  \caption{Velocity profiles of Taylor-Couette flow for different radius ratios ($\beta = 0.1\sim0.8$). Symbols represent numerical results and lines denote analytical solutions.}
  \label{3test_fig2}
\end{figure*}

\begin{table*}
  \captionsetup{justification=justified, singlelinecheck=false}
  \caption{Grid independence test of the four-roll mill problem for power-law fluid: Dimensionless velocity $u_x / u_c$ and shear rate $|\dot{\gamma}|$ at steady state for different grid sizes $N^2$ and power-law index $n$.}
  \centering
  \begin{tabular}{ccccccc}
    \toprule
    \multirow{2}{*}{$N \times N$} & \multicolumn{3}{c}{$u_x/u_c$} & \multicolumn{3}{c}{$|\dot{\gamma}|$} \\
    \cmidrule(lr){2-4} \cmidrule(lr){5-7}
    & $n=0.7$ & $n=1.0$ & $n=1.3$ & $n=0.7$ & $n=1.0$ & $n=1.3$ \\

    \addlinespace[5pt] 
    $500 \times 500$  & 1.5158  & 1.5352 	& 1.5457 	& 3.530E-03   & 2.959E-03		& 7.797E-04\\	
    $600 \times 600$  & 1.5136  & 1.5347 	& 1.5463 	& 3.525E-03   & 2.960E-03		& 7.806E-04\\
    $700 \times 700$  & 1.5131  & 1.5350 	& 1.5470 	& 3.521E-03   & 2.961E-03		& 7.813E-04\\
    $800 \times 800$  & 1.5121  & 1.5349 	& 1.5473 	& 3.518E-03	  & 2.961E-03		& 7.816E-04\\
    $900 \times 900$  & 1.5117  & 1.5349 	& 1.5477 	& 3.518E-03	  & 2.962E-03		& 7.820E-04\\
    $1000 \times 1000$ & 1.5113  & 1.5349 	& 1.5479 	& 3.517E-03		& 2.963E-03		& 7.821E-04\\
    \bottomrule
  \end{tabular}
  \label{Grid_tab}
\end{table*}

\subsection{Implementation of boundary condition}\label{section3.2}

The four-roll mill geometry features curved boundaries at both the rollers and container walls. To accurately resolve these curved boundaries, we implement the one-point second-order interpolation scheme developed by \citet{tao_one-point_2018}. This method not only achieves second-order accuracy but also maintains high computational efficiency through its straightforward parallelization. As illustrated in Fig.~\ref{test_fig3.2.1}, the scheme considers three key points along a straight line intersecting the boundary: $x_{\alpha ,A}$ denotes the nearest fluid node to the curved boundary, $x_{\alpha ,W}$ represents the intersection point on the curved boundary itself, and $x_{\alpha ,B}$ indicates the first solid node inside the boundary. The distribution function at point $A$ can be expressed as \citep{tao_one-point_2018}
\begin{equation}
  \begin{aligned}
    f_{i}\left(x_{\alpha ,A},t+\Delta t\right)=\frac{1}{1+q}\Big(f_{i}^{eq}\big(u_{\beta}\left(x_{\alpha ,W},t+\Delta t\right),\rho(x_{\alpha ,A},t)\big)\\+f_{\bar{i}}^{neq}\big(x_{\alpha ,A},t\big)+qf_{i}^{\prime}\big(x_{\alpha ,A},t\big)\Big),
  \end{aligned}
  \label{eq:3-10}
\end{equation}
where $\overline{i}$ satisfies $e_{\overline{i}\alpha}=-e_{i\alpha}$. $f_{i}^{eq}\left(u_{\beta}\left(x_{\alpha ,W},t+\Delta t\right),\rho_{A}(t)\right)$ represents the equilibrium distribution function with the velocity at point $W$ and density at point $A$. $f_{\bar{i}}^{neq}\left(x_{\alpha ,A},t\right)$ represents the non-equilibrium distribution function at $A$. $f'_i\left(x_{\alpha ,A},t\right)$ denotes the post-collision but pre-streaming distribution function at $A$. The parameter $q$ represents the relative distance from the boundary point $A$ to the point $W$ of solid surface, defined as $q={\left|x_{\alpha ,W}-x_{\alpha ,A}\right|}/{\left|x_{\alpha ,B}-x_{\alpha ,A}\right|}$.

\begin{figure*}
  \centering
  \begin{subfigure}{0.45\textwidth}
    \includegraphics[width=\linewidth]{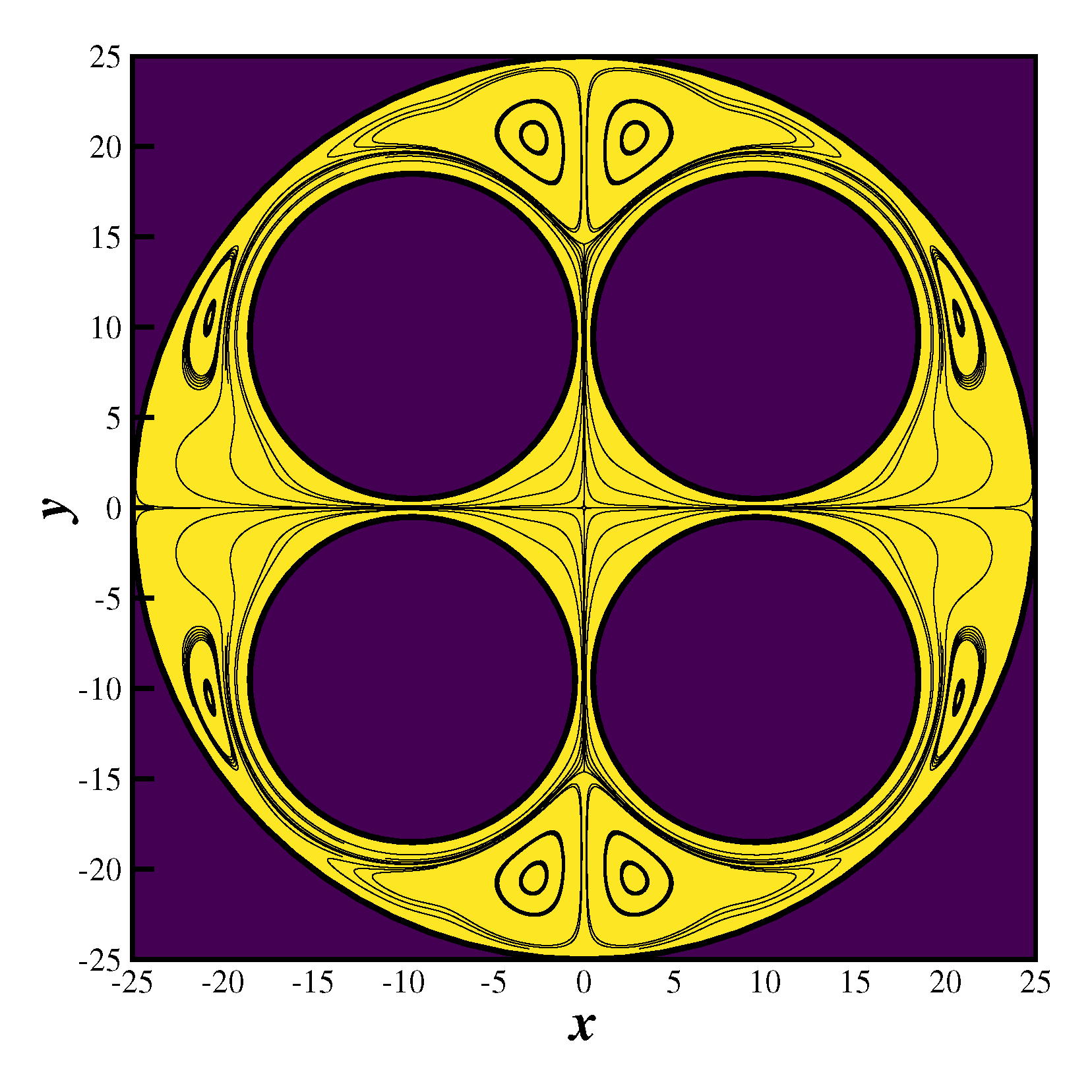}
    \caption{}
    \label{fig.4.1.1._b_h=1_n=1_Re=50_entire_streamline}
  \end{subfigure}
  \hfill
  \begin{subfigure}{0.45\textwidth}
    \includegraphics[width=\linewidth]{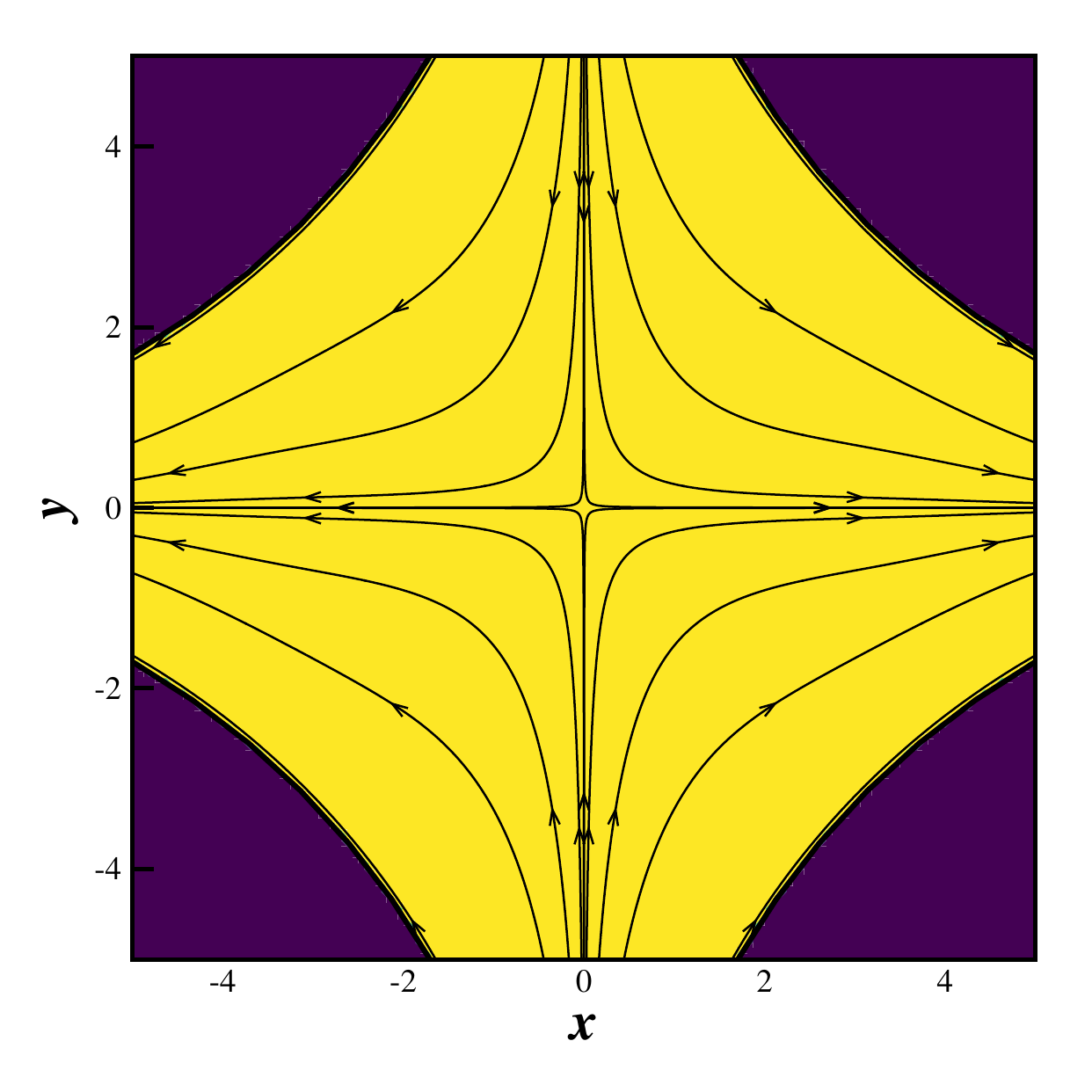}
    \caption{}
    \label{fig.4.1.1._b_h=1_n=1_Re=50_central_streamline}
  \end{subfigure}
  \caption{Streamline patterns in the four-roll mill for Newtonian fluid ($n=1$) at $Re=50$ with $h=1$, $r=9$, and $\delta \approx 2.56$: (a) entire flow domain and (b) central region. Yellow and purple regions represent fluid and solid domains, respectively.}
  \label{fig.4.1.1.h=1_n=1_Re=50_streamline}
\end{figure*}

\begin{figure*}
  \centering
  \begin{subfigure}{0.45\textwidth}
    \includegraphics[width=\linewidth]{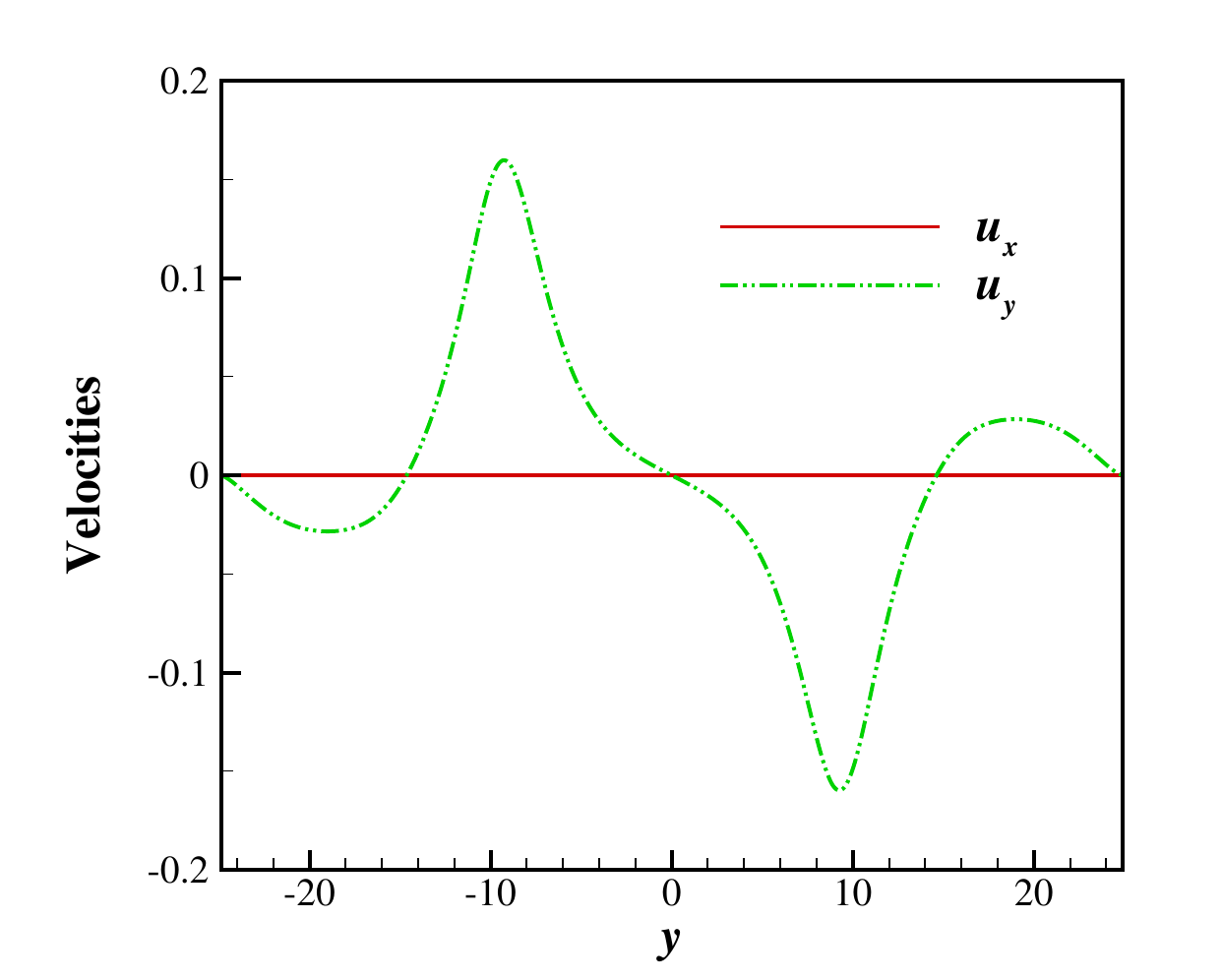}
    \caption{}
    \label{figures/4.1/fig.4.1.2._b_n=1_Re=50_ux_uy_y-axis-eps-converted-to.pdf}
  \end{subfigure}
  \hfill
  \begin{subfigure}{0.45\textwidth}
    \includegraphics[width=\linewidth]{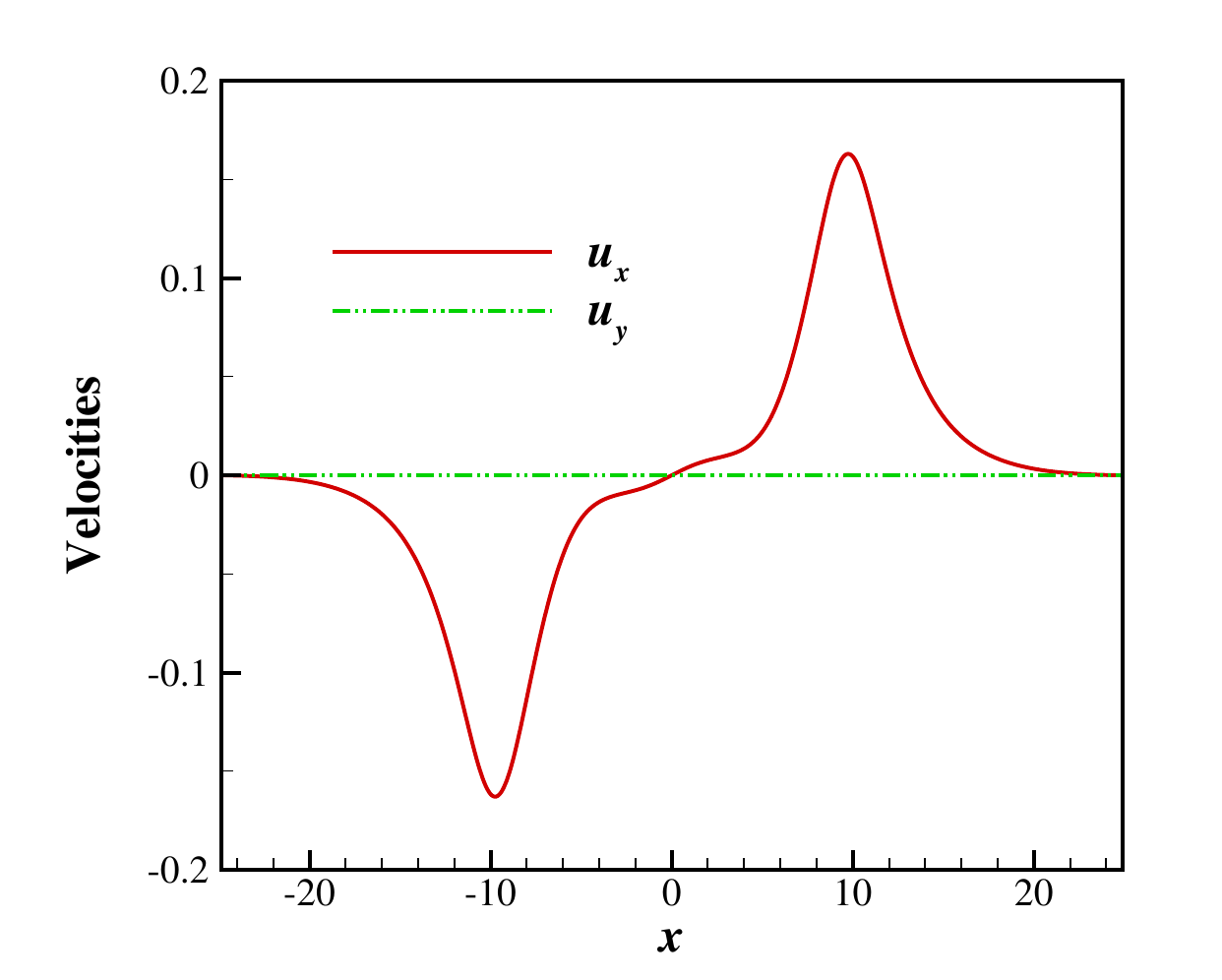}
    \caption{}
    \label{fig.4.1.2._b_n=1_Re=50_ux_uy_x-axis}
  \end{subfigure}
  \caption{Velocity distributions characterizing simple extensional flow in the four-roll mill at $Re=50$ for Newtonian fluid ($n=1$): (\textit{a}) $u_x$ and $u_y$ along the $y$-axis and (\textit{b}) $u_x$ and $u_y$ along the $x$-axis. The geometric parameters are $h = 1$, $\delta \approx 2.56$, and $r = 9$. The profiles demonstrate central symmetry about the stagnation point $O$ and the absence of cross-axial fluid transport ($u_x = 0$ along $y$-axis and $u_y = 0$ along $x$-axis).}
  \label{fig.4.1.2.uxuy_along(a)y-axis(b)x-axis_Re=50}
\end{figure*} 

\begin{figure*}
  \centering  
  \includegraphics[width=0.5\textwidth]{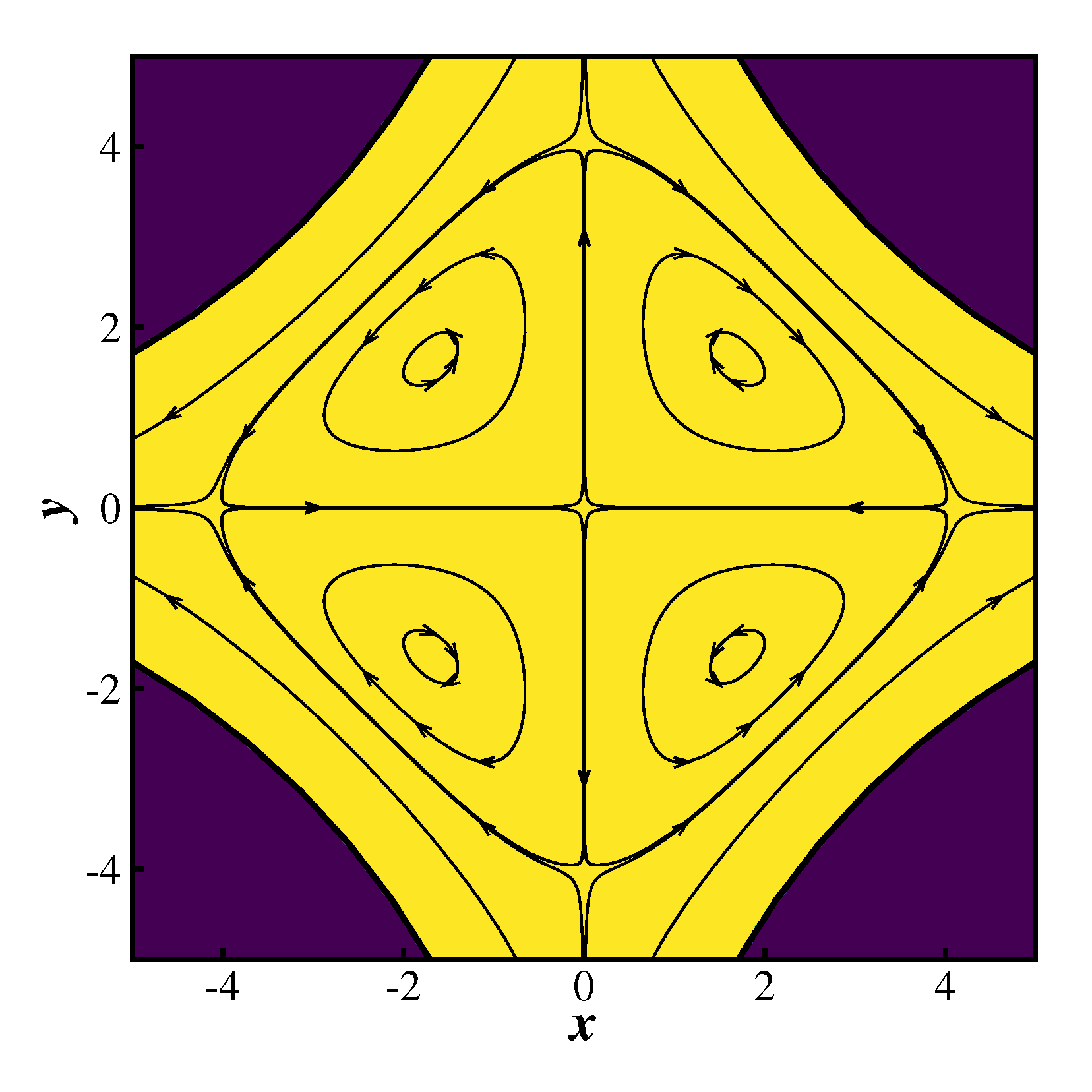}
    \caption{Discovery of the quadrifoliate vortex mode: streamline patterns in the central region of the four-roll mill for Newtonian fluid ($n=1$) at $Re=10$. Four symmetrical vortices of equal size form around the stagnation point, with adjacent vortices rotating in opposite directions. The geometric parameters are set to $h = 1$, $r = 9$, and $\delta \approx 2.56$.}
    
  \label{fig.4.1.3.Streamline_Re=10}
\end{figure*} 

\begin{figure*}
  \centering
  \begin{subfigure}{0.45\textwidth}
    \includegraphics[width=\linewidth]{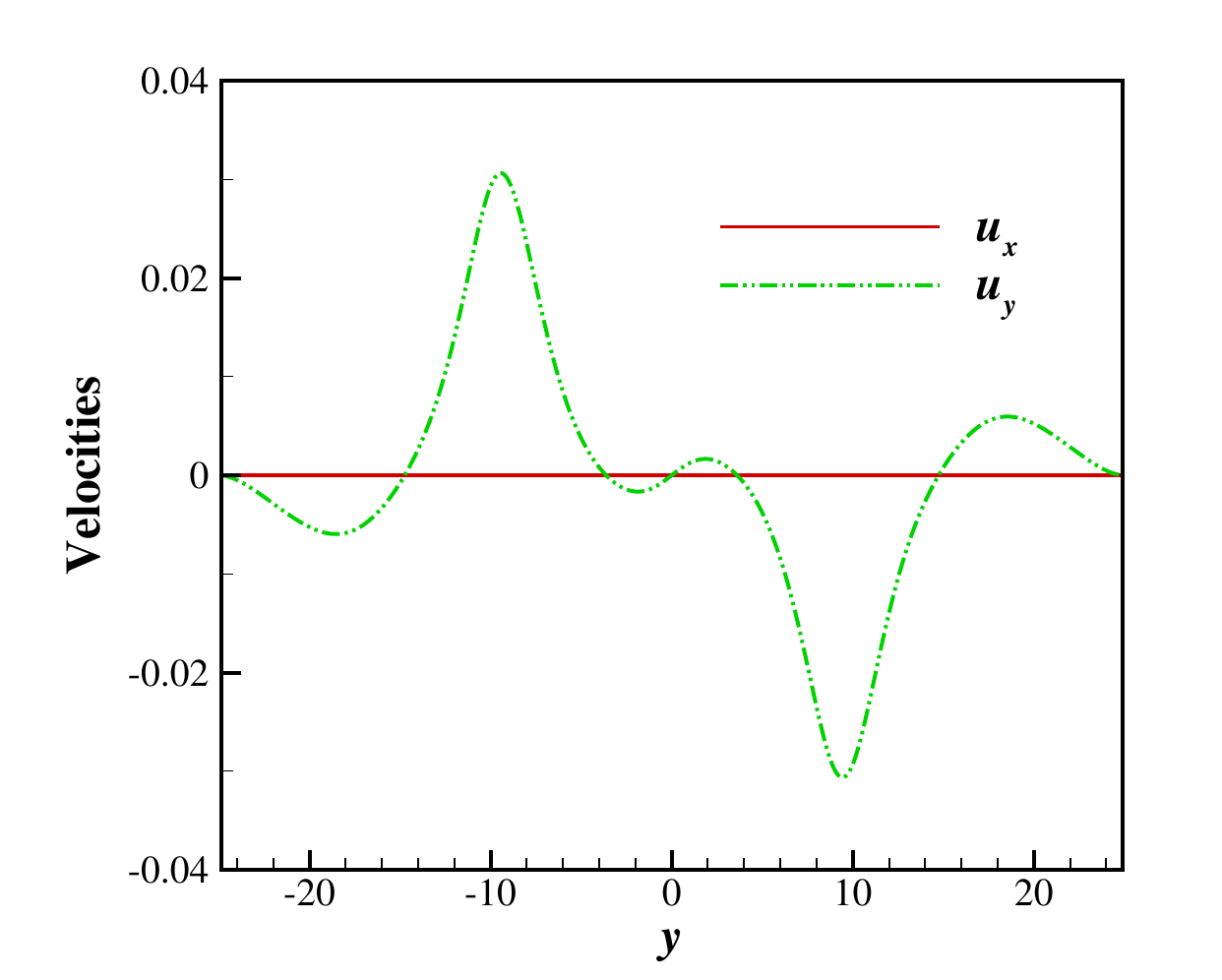}
    \caption{}
    \label{fig.4.1.4._b_n=1_Re=10_ux_uy_y_axis}
  \end{subfigure}
  \hfill
  \begin{subfigure}{0.45\textwidth}
    \includegraphics[width=\linewidth]{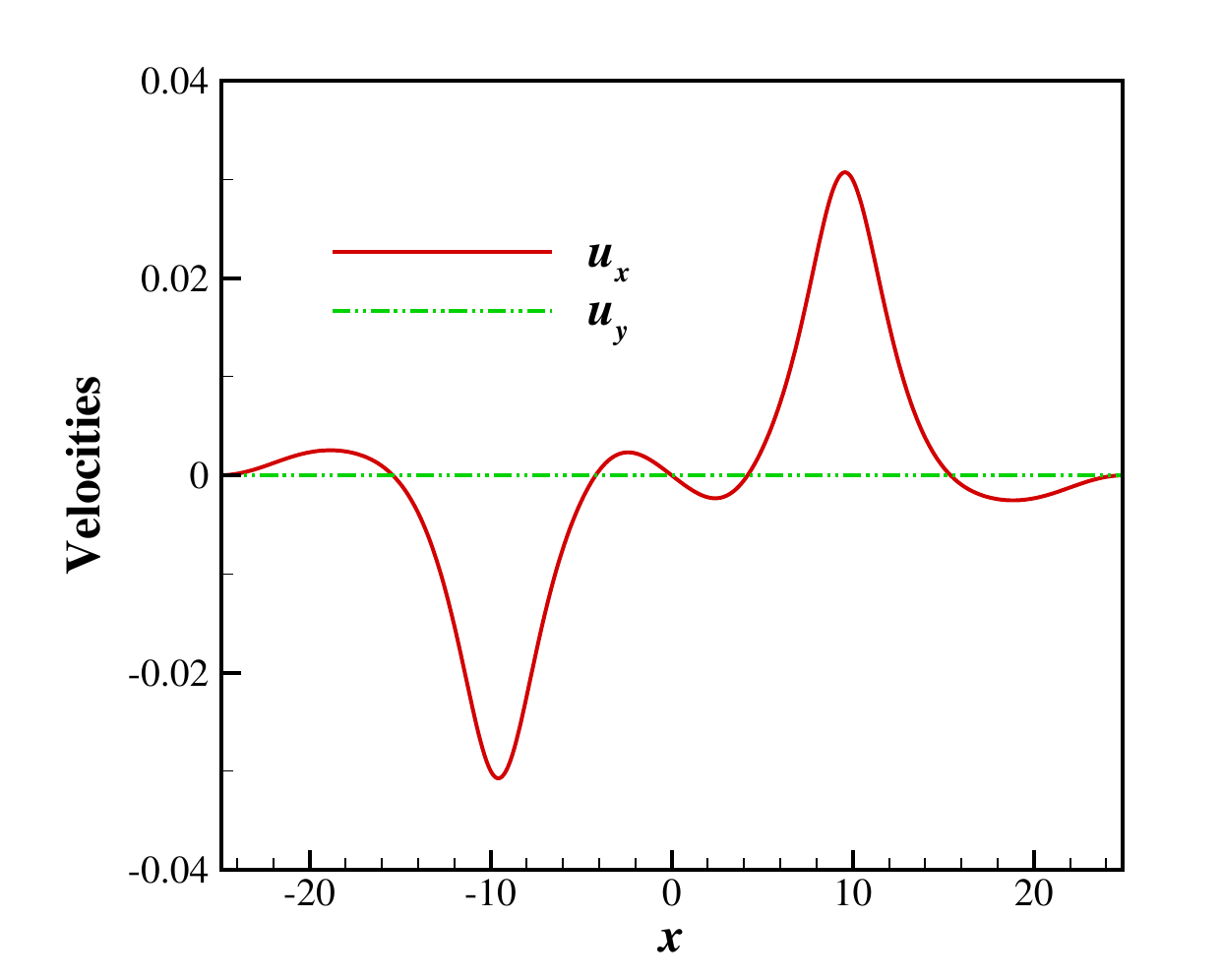}
    \caption{}
    \label{fig.4.1.4._b_n=1_Re=10_ux_uy_x_axis}
  \end{subfigure}
  \caption{Velocity distributions in the quadrifoliate vortex mode: profiles of $u_x$ and $u_y$ along (a) $y$-axis and (b) $x$-axis for Newtonian fluid ($n=1$) at $Re=10$. Geometric parameters: $h = 1$, $\delta \approx 2.56$, and $r = 9$.}
  \label{fig.4.1.4.uxuy_along(a)y-axis(b)x-axis_Re=10}
\end{figure*} 

\begin{figure*}
  \centering
  \includegraphics[width=0.6\textwidth]{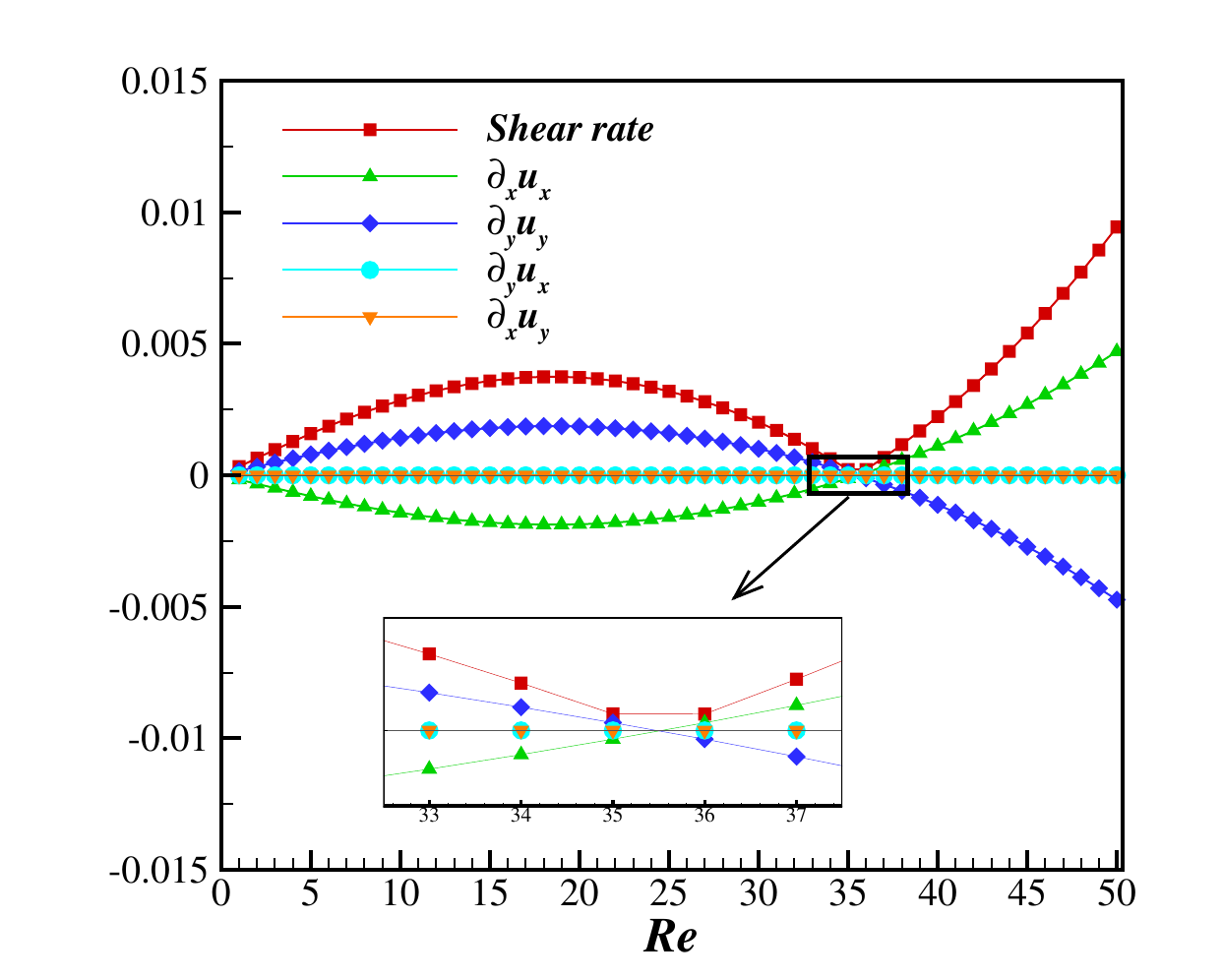}
  \caption{Variation of shear rate and velocity gradients at point $O$ with Reynolds number for a Newtonian fluid ($n=1$). The geometric parameters are fixed at $h = 1$, $\delta \approx 2.56$, and $r = 9$.}
  \label{fig.4.1.5.h=1_n=1_Re=1-50_shear_rate_and_velo_gradient}
\end{figure*} 

\begin{figure*}
  \centering
    \includegraphics[width=0.6\textwidth]{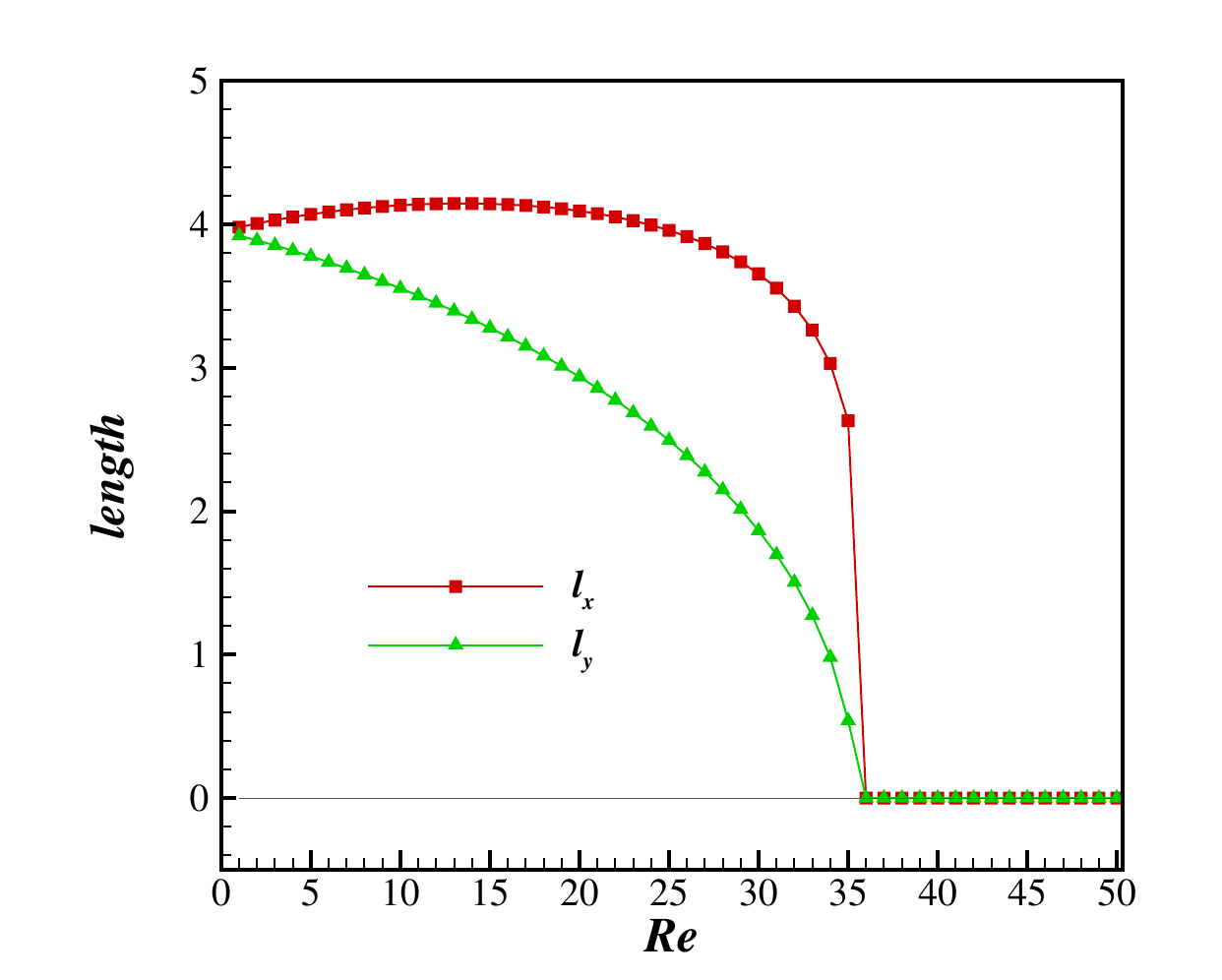}
    \caption{Vortex dimensions ($l_x$ and $l_y$) as functions of Reynolds number for Newtonian fluid ($n=1$) with geometric parameters $h = 1$, $\delta \approx 2.56$, and $r = 9$, where $l_x$ and $l_y$ represent the vortex extent in the horizontal and vertical directions, respectively.}
    \label{fig.4.1.6.h=1_n=1_Re=1-50_vortex_length}
\end{figure*} 
\subsection{Numerical validation}\label{section3.3}

Two benchmark cases are employed to validate our numerical implementation: force-driven channel flow of power-law fluids for validating the power-law model, and Taylor-Couette flow for verifying the accuracy of the one-point second-order curved boundary scheme. Since our focus is on steady-state solutions, the convergence criterion is defined as:
\begin{gather}\label{convergence_criterion}
  E_u=\max(\left|\frac{u_\alpha \left(x_\alpha ,t+10000 \Delta t\right)-u_\alpha\left(x_\alpha ,t\right)}{U_c}\right|)<10^{-7},
\end{gather}
where $E_u$ represents the maximum relative velocity difference over a period of 10000 time steps.

\subsubsection{Power-law fluid channel flow}
	
For a force-driven channel filled with power-law fluid, the analytical velocity profile is given by \citep{pontrelli_unstructured_2009}:
\begin{equation}
  u(y) = u_c\left[1-\left(\frac{|y-\frac{H}{2}|}{\frac{H}{2}}\right)^{(n+1)/n}\right],
\end{equation}
where $H$ denotes the channel height (characteristic length) and $u_c$ represents the maximum velocity (characteristic velocity). The latter is expressed as \citep{pontrelli_unstructured_2009}:
\begin{equation}
  u_c=\left(\frac{F_x}{\mu_0}\right)^{1/n}\left(\frac{n}{n+1}\right)\left(\frac{H}{2}\right)^{\left(n+1\right)/n},
\end{equation}
where $F_x$ represents the body force in the $x$-direction.

For the numerical validation, we set the following parameters: channel height $H=1$, maximum velocity $u_c=0.01$, and lattice speed $c=1$. The computational domain is discretized with a grid size of $N_x \times N_y = 10 \times 100$, yielding a spatial resolution of $\Delta x = H/N_y = 0.1$. The Reynolds number is fixed at $Re=10$, from which the viscosity coefficient is determined as $\mu_0 = \rho u_c^{2-n} H^n / Re$. 

It is worth noting that while traditional TRT implementations achieve no-slip conditions through the magic parameter $\Lambda=3/16$ \citep{Ginzburg2008TworelaxationtimeLB}, this constraint cannot be simultaneously satisfied with our choice of $\tau_2=1.6$, which is essential for maintaining numerical accuracy and stability across a wide range of viscosities. Therefore, the moment-based boundary scheme \citep{reis_lattice_2020} is implemented at the upper and lower walls to ensure no-slip conditions without relying on the magic parameter constraint. The numerical simulations are performed for three different power-law indices ($n = 0.5, 1.0, 1.5$), and the resulting normalized velocity profiles are compared with the analytical solutions. As demonstrated in Fig.~\ref{3test_fig1}, the excellent agreement between numerical and analytical results validates the implementation of our power-law model.

\subsubsection{Taylor-Couette flow}

To validate the accuracy of the one-point second-order scheme in handling curved moving boundaries, we employ the Taylor-Couette flow \citep{tao_one-point_2018, zhao_single-node_2017} as a benchmark case. As illustrated in Fig.~\ref{3Cylindrical_Couette_flow}, this flow configuration features fluid confined between two concentric cylinders: an inner cylinder of radius $R_1$ rotating at angular velocity $\omega_1$, and a stationary outer cylinder of radius $R_2$, providing an ideal test case for curved moving boundary treatment. The steady-state analytical solution for the velocity profile is given by \citep{tao_one-point_2018}:
\begin{equation}
  u_\theta=\frac{\left(R_1^2-\beta^2r^2\right)\omega_1}{\left(1-\beta^2\right)r},\quad u_r=0,
\end{equation}
where $u_\theta$ and $u_r$ are the angular and radial velocity components, respectively, $r$ is the radial distance, and the parameter $\beta$ represents the ratio of radius $R_1$ and $R_2$. In simulations, the detailed setup are given in \citet{tao_one-point_2018}. The velocity profiles for $\beta = 0.1\sim0.8$ are shown in Fig.~\ref{3test_fig2}, demonstrating perfect agreement with the analytical solutions. Thus the one-point second-order curved boundary scheme is validated in this study.

\begin{figure*}
  \centering
  \begin{subfigure}{0.45\textwidth}
    \includegraphics[width=\linewidth]{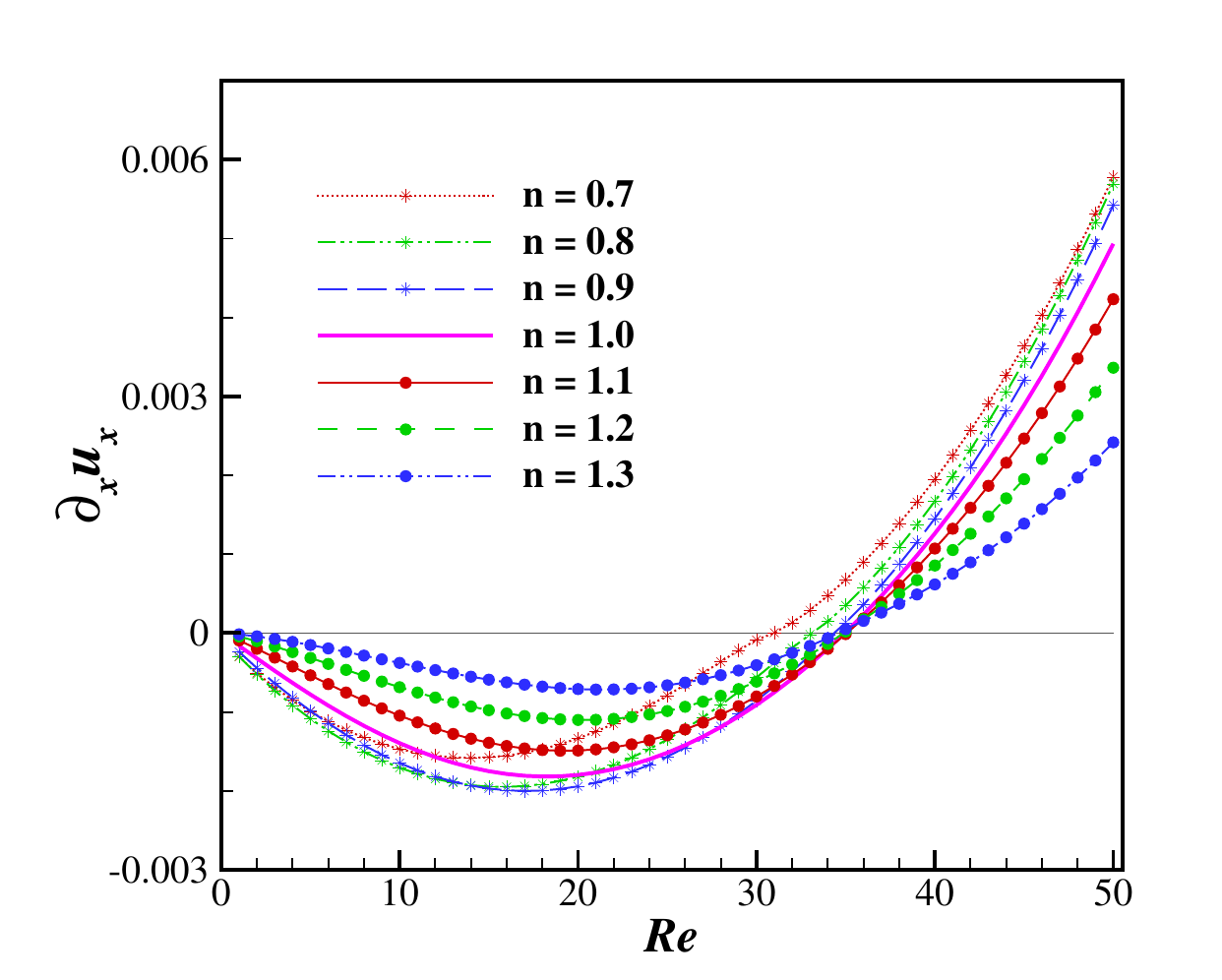}
    \caption{}
    \label{fig.4.2.1._b_n=0.7-1.3_duxdx}
  \end{subfigure}
  \hfill
  \begin{subfigure}{0.45\textwidth}
    \includegraphics[width=\linewidth]{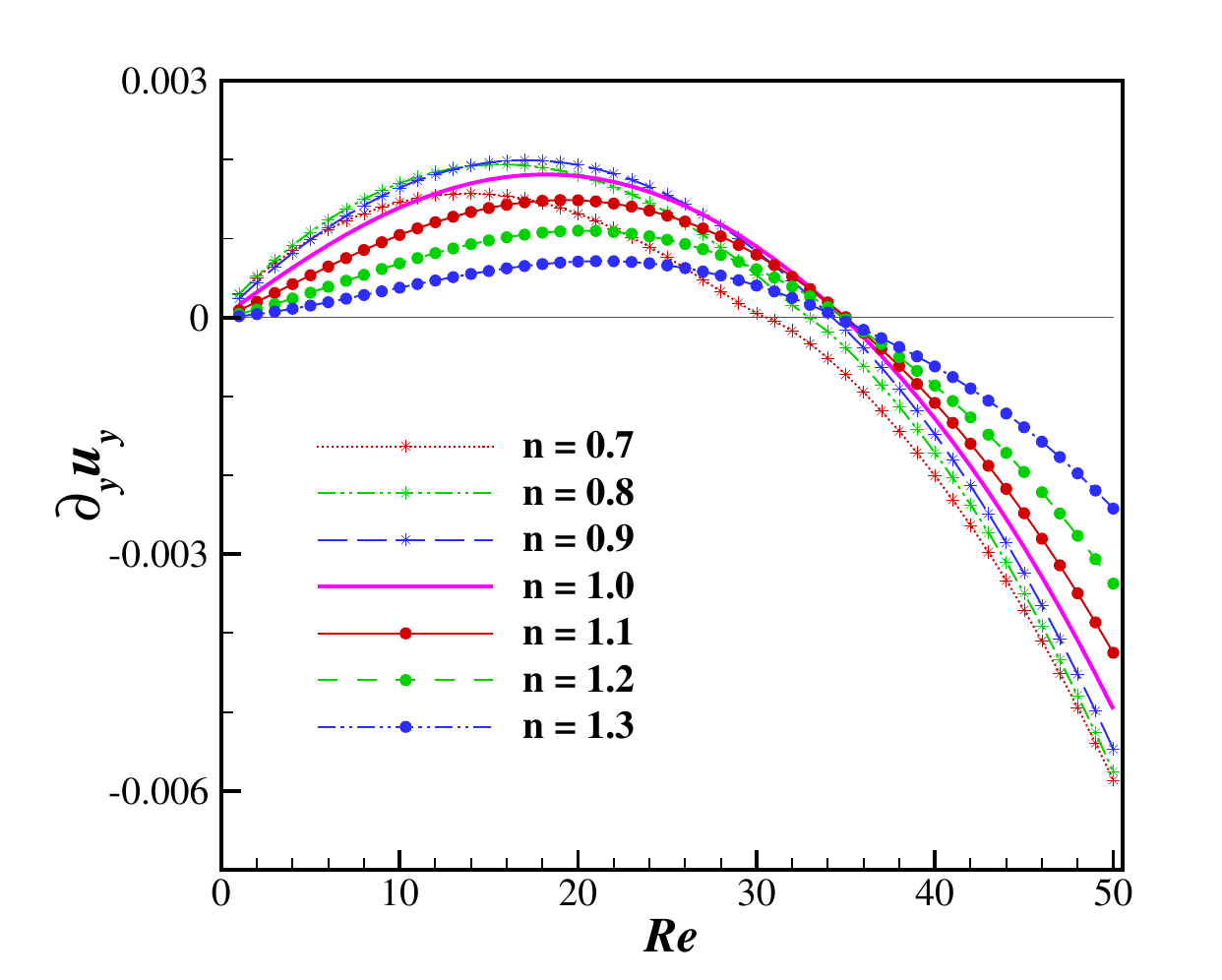}
    \caption{}
    \label{fig.4.2.1._b_n=0.7-1.3_duydy-eps-converted-to.pdf}
  \end{subfigure}
  \hfill
  \caption{Velocity gradients (a) $\partial_x u_x$ and (b) $\partial_y u_y$ at the stagnation point $O$ plotted against Reynolds number for different power-law indices $n$. The geometric parameters are fixed at $h = 1$, $\delta \approx 2.56$, and $r = 9$.}
  \label{fig.4.2.1.velocity_gradient_n=0.7-1.3}
\end{figure*} 

\begin{figure*}
  \centering
  \begin{subfigure}{0.45\textwidth}
    \includegraphics[width=\linewidth]{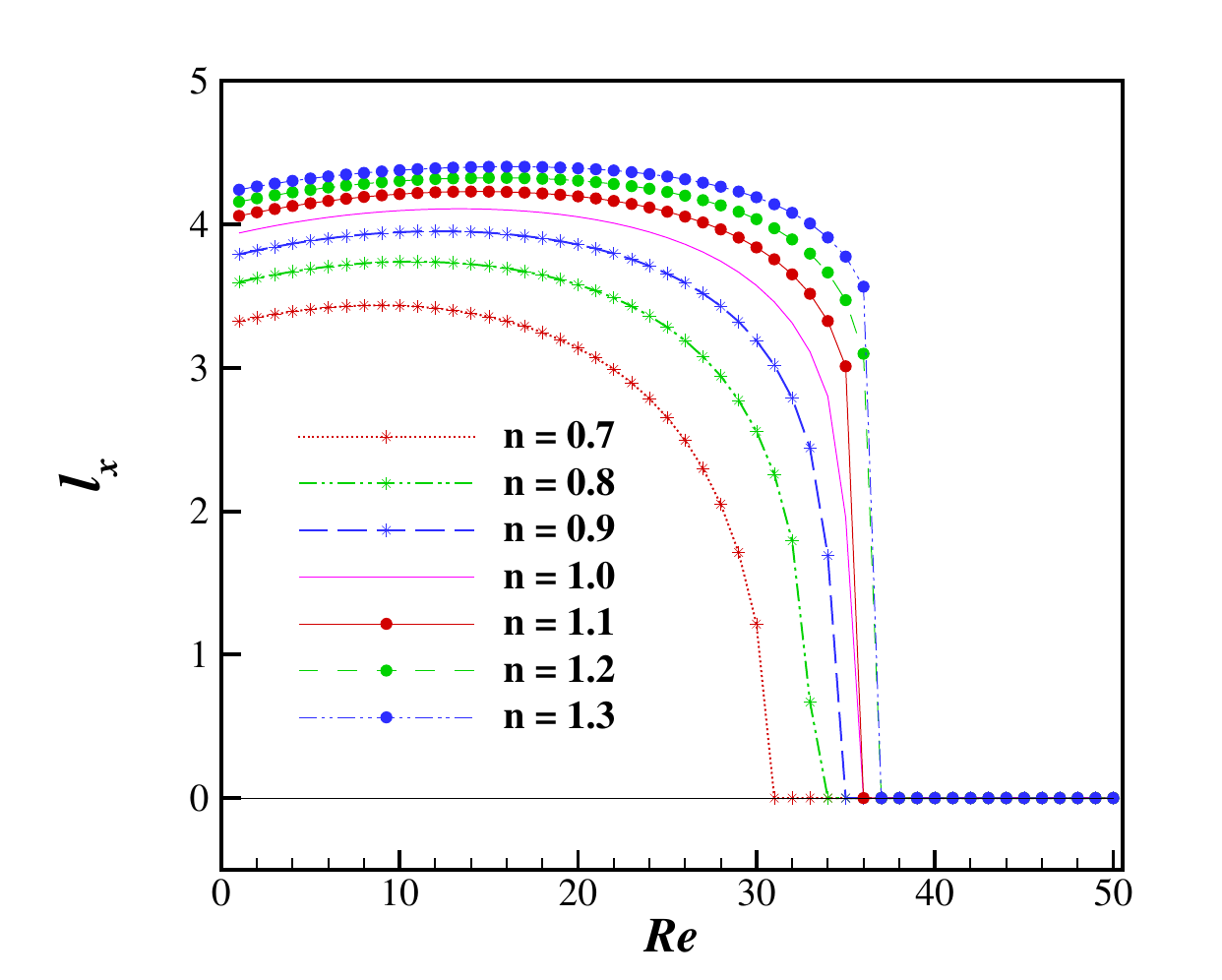}
    \caption{}
    \label{fig.4.2.2._b_n=0.7-1.3_lx}
  \end{subfigure}
  \hfill
  \begin{subfigure}{0.45\textwidth}
    \includegraphics[width=\linewidth]{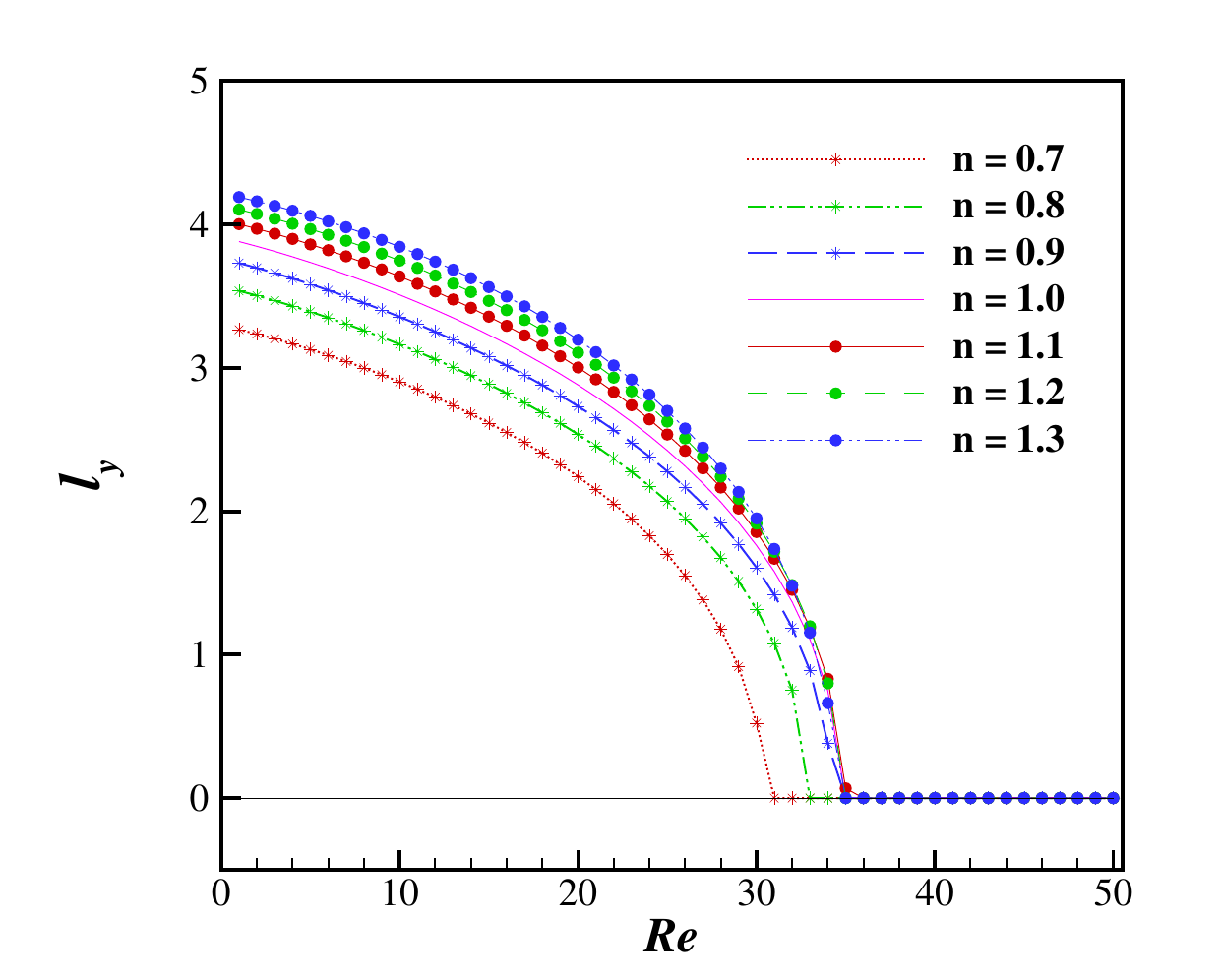}
    \caption{}
    \label{fig.4.2.2._b_n=0.7-1.3_ly}
  \end{subfigure}
  
  \caption{Variations of vortex lengths (\textit{a}) $l_x$ and (\textit{b}) $l_y$ with $Re$ for various values of power-law index, $h = 1$, $\delta \approx 2.56$, and $r = 9$.}
  \label{fig.4.2.2.vortex_length_n=0.7-1.3}
\end{figure*} 

\begin{figure*}
  \centering
    \includegraphics[width=0.6\textwidth]{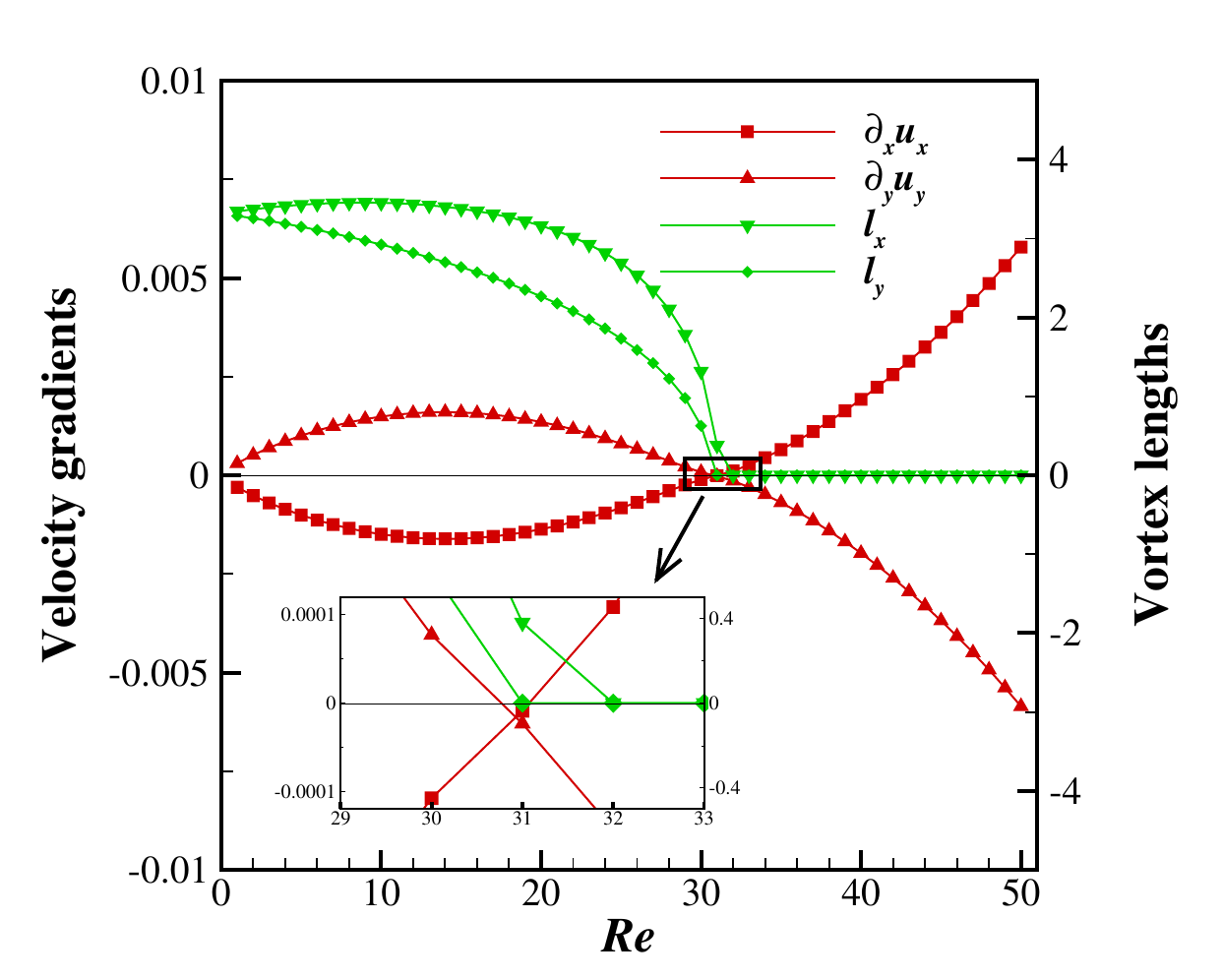}
    \caption{Variations of velocity gradients and vortex lengths with $Re$ for $n=0.7$.}
    \label{fig.4.2.3.h=1_n=0.7_velocity_gradients_vortex_lengths}

\end{figure*}

\begin{figure*}[htbp!]
  \centering
  \begin{subfigure}{0.3\textwidth}
    \includegraphics[width=\linewidth]{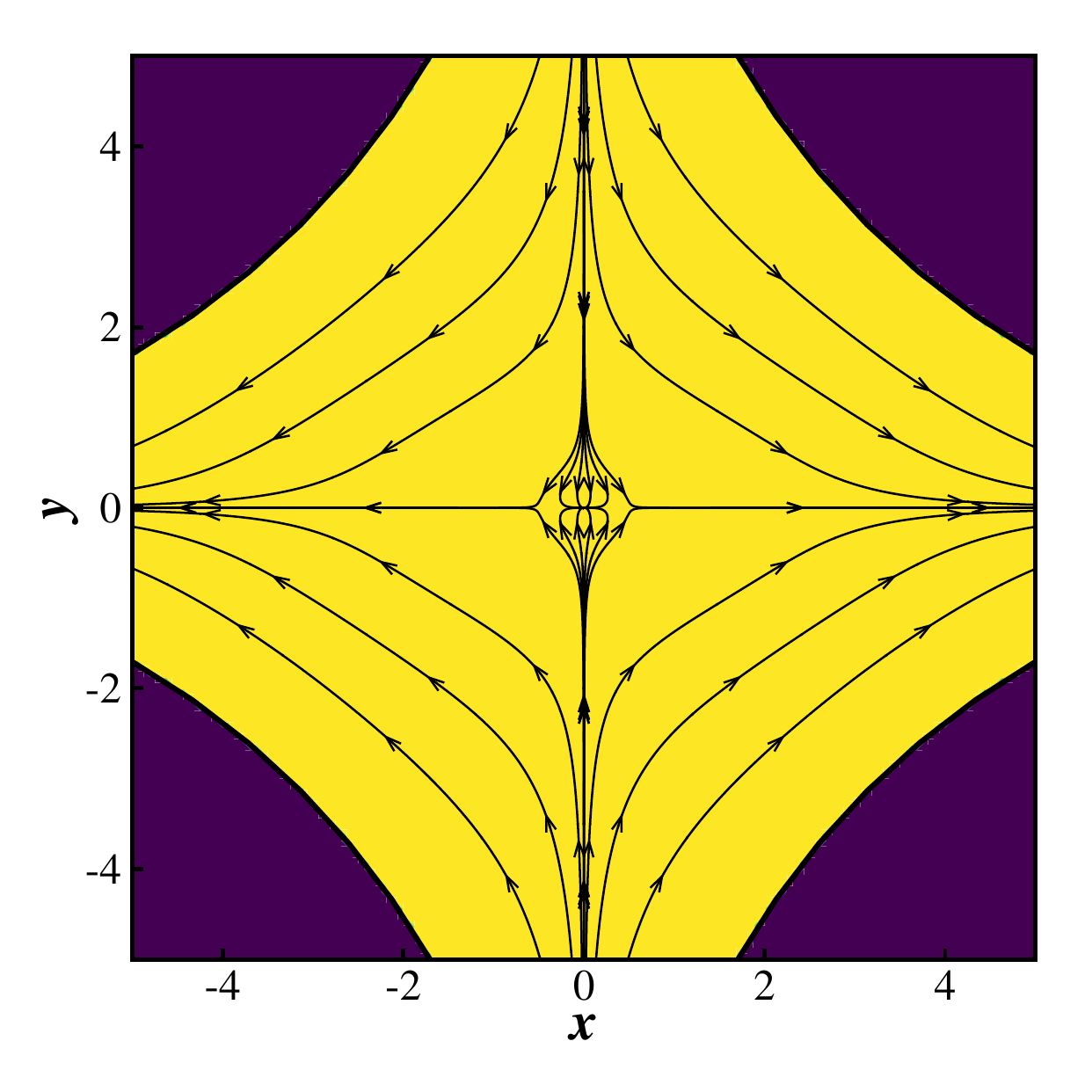}
    \caption{}
    \label{fig.4.2.4._b_h=1_n=0.7_Re=31_central_streamline_800}
  \end{subfigure}
  \hfill
  \begin{subfigure}{0.3\textwidth}
    \includegraphics[width=\linewidth]{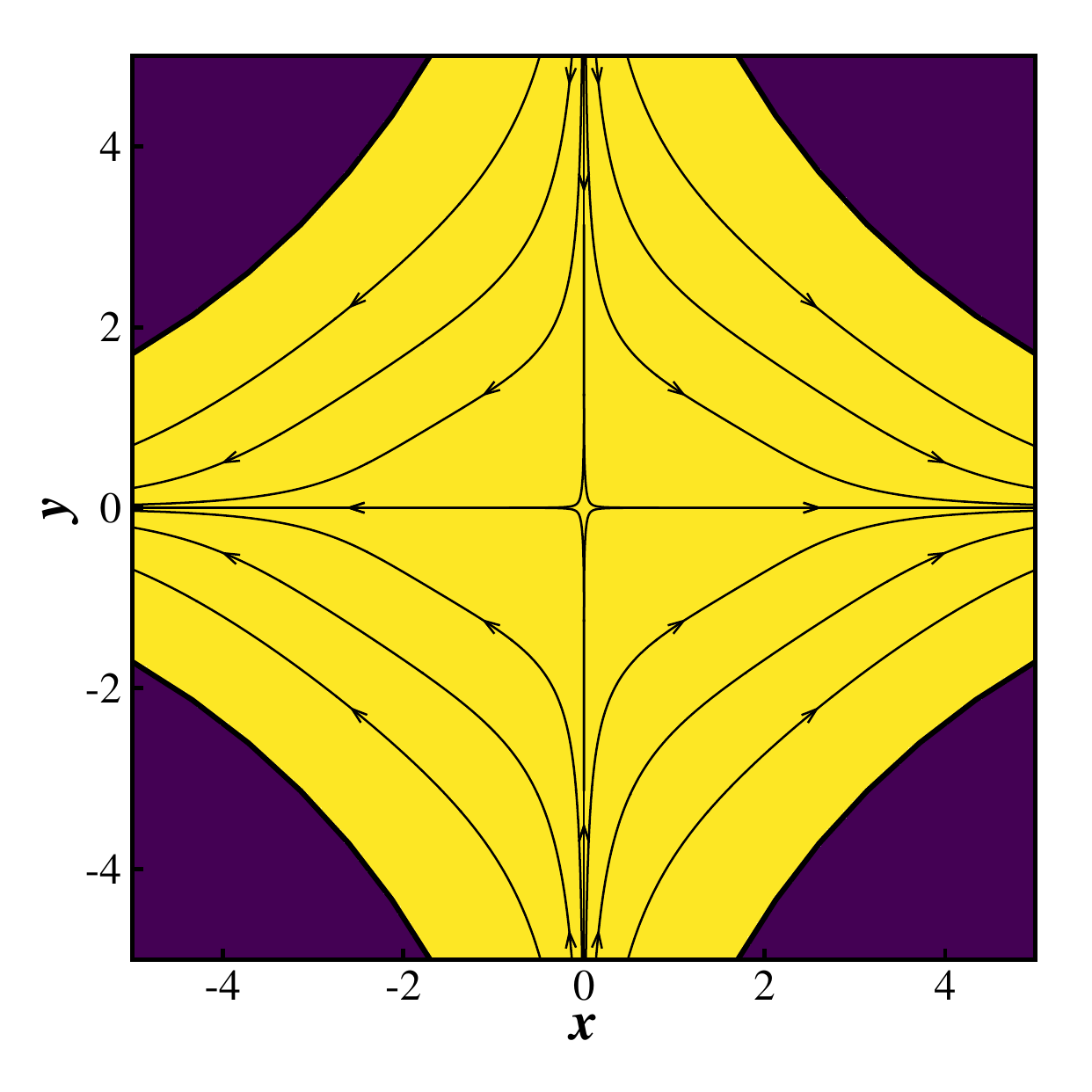}
    \caption{}
    \label{fig.4.2.4._b_h=1_n=0.7_Re=31_central_streamline_1600}
  \end{subfigure}
  \hfill
  \begin{subfigure}{0.3\textwidth}
    \includegraphics[width=\linewidth]{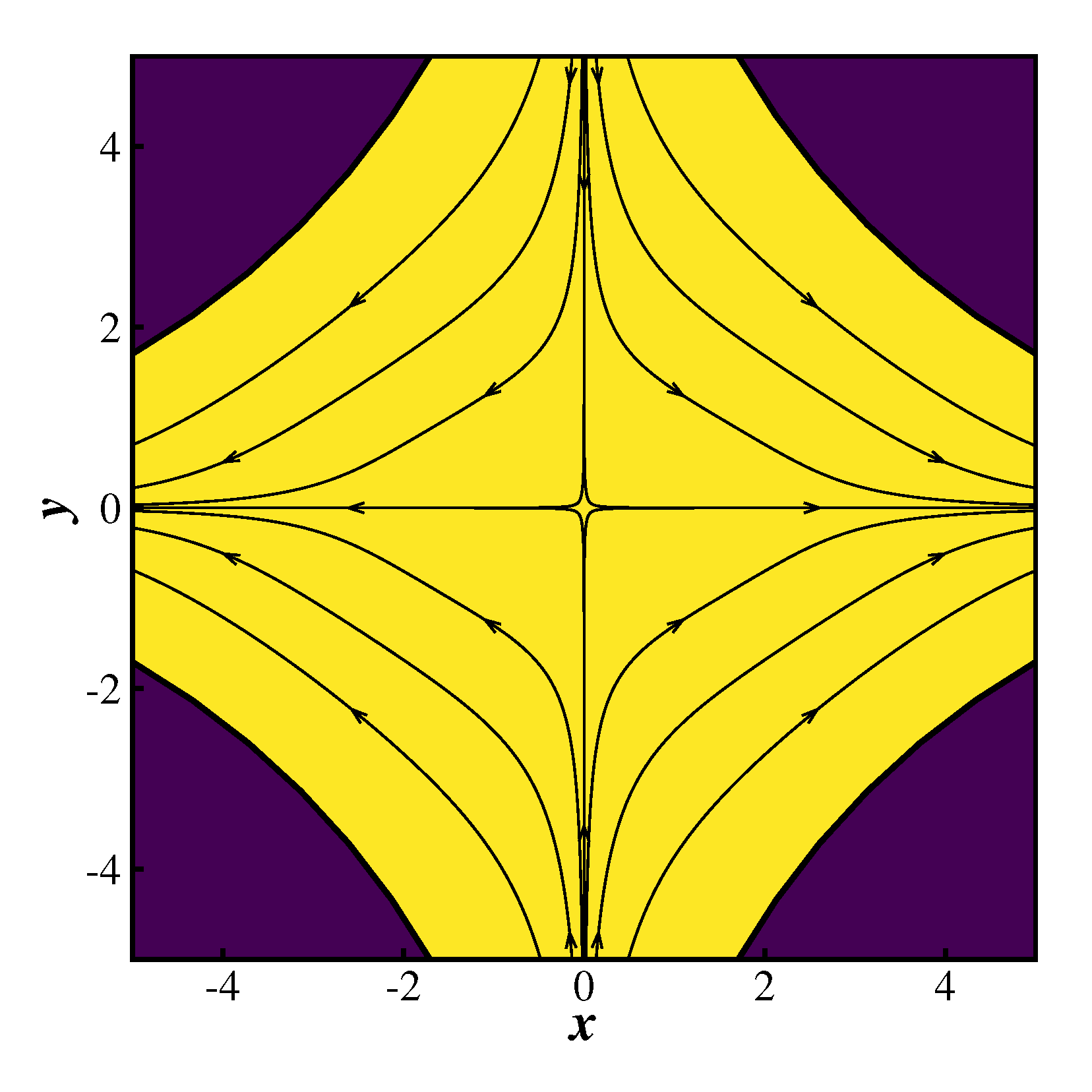}
    \caption{}
    \label{fig.4.2.4._c_h=1_n=0.7_Re=31_central_streamline_3200}
  \end{subfigure}  
  \caption{Central streamline in the four roll mill for $n=0.7$, $Re=31$, $h = 1$, $\delta \approx 2.56$, and $r = 9$ under different meshes including (\textit{a}) $800 \times 800$, (\textit{b}) $1600 \times 1600$, and (\textit{c}) $3200 \times 3200$.}
  \label{fig.4.2.4.n=0.7_Re=31_diff_mesh_central_streamline}
\end{figure*} 

\begin{figure*}
  \centering
  \begin{subfigure}{0.45\textwidth}
    \includegraphics[width=\linewidth]{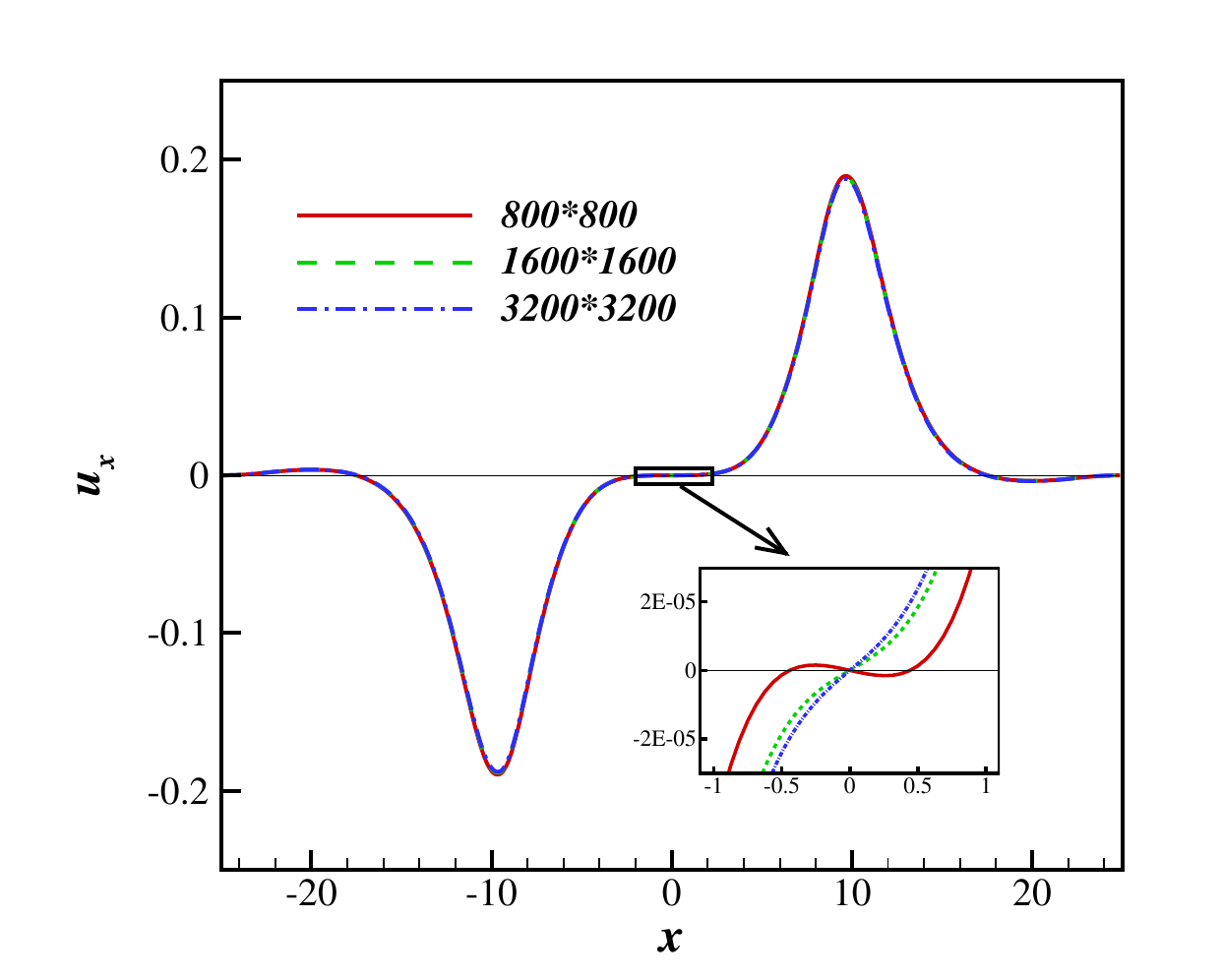}
    \caption{}
    \label{fig.4.2.5._b_n=0.7_Re=31_x-axis_compared_mesh}
  \end{subfigure}
  \hfill
  \begin{subfigure}{0.45\textwidth}
    \includegraphics[width=\linewidth]{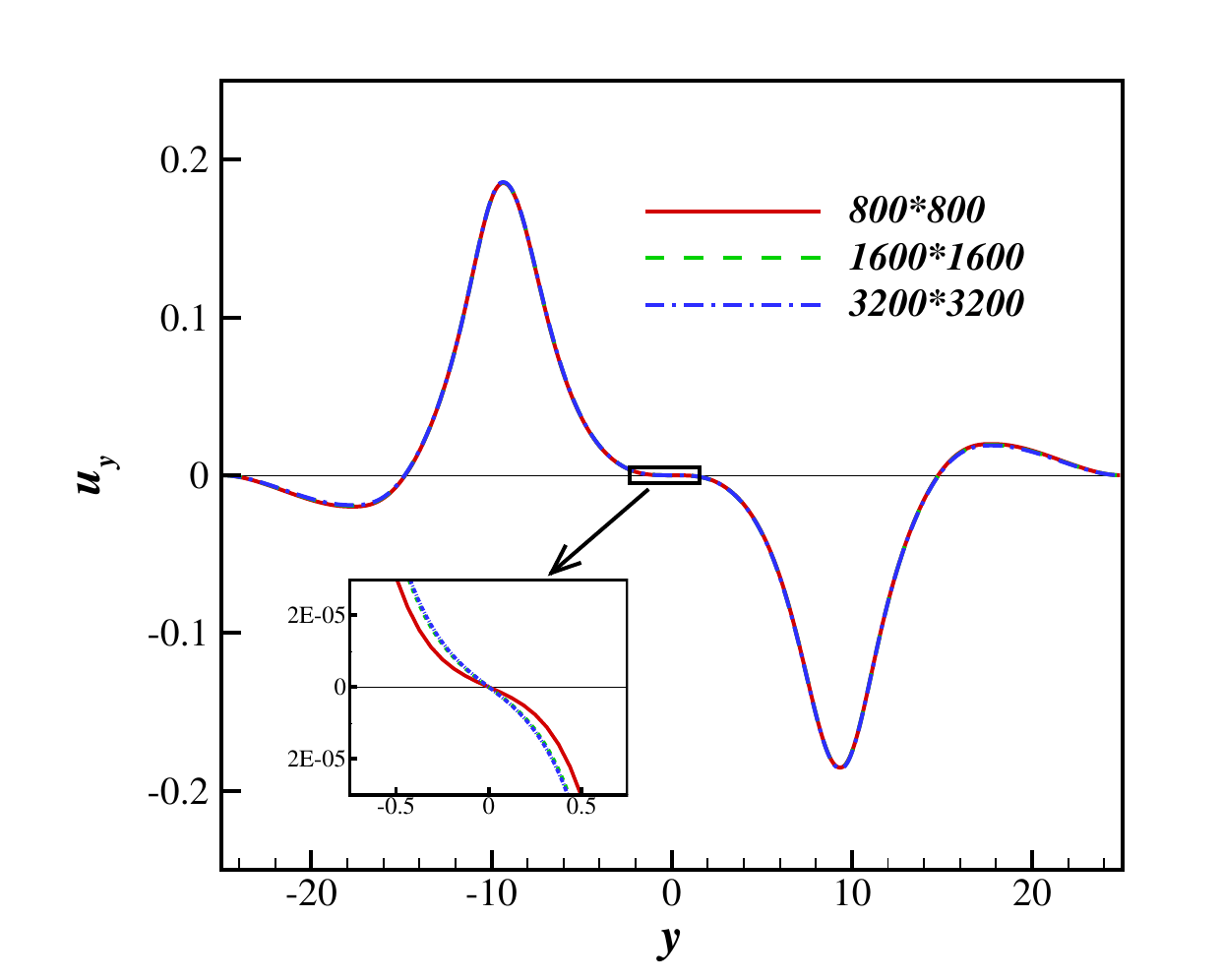}
    \caption{}
    \label{fig.4.2.5._b_n=0.7_Re=31_y-axis_compared_mesh}
  \end{subfigure}
  
  \caption{The velocity profiles (\textit{a}) $u_x$ in $x$-axis and (\textit{b}) $u_y$ in $y$-axis for $n=0.7$, $Re=31$, $h = 1$, $\delta \approx 2.56$, and $r = 9$ under different meshes.}
  \label{fig.4.2.5.n=0.7_Re=31_diff_mesh_velocity_profiles}
\end{figure*}

\begin{figure*}
  \centering
  \includegraphics[width=0.6\textwidth]{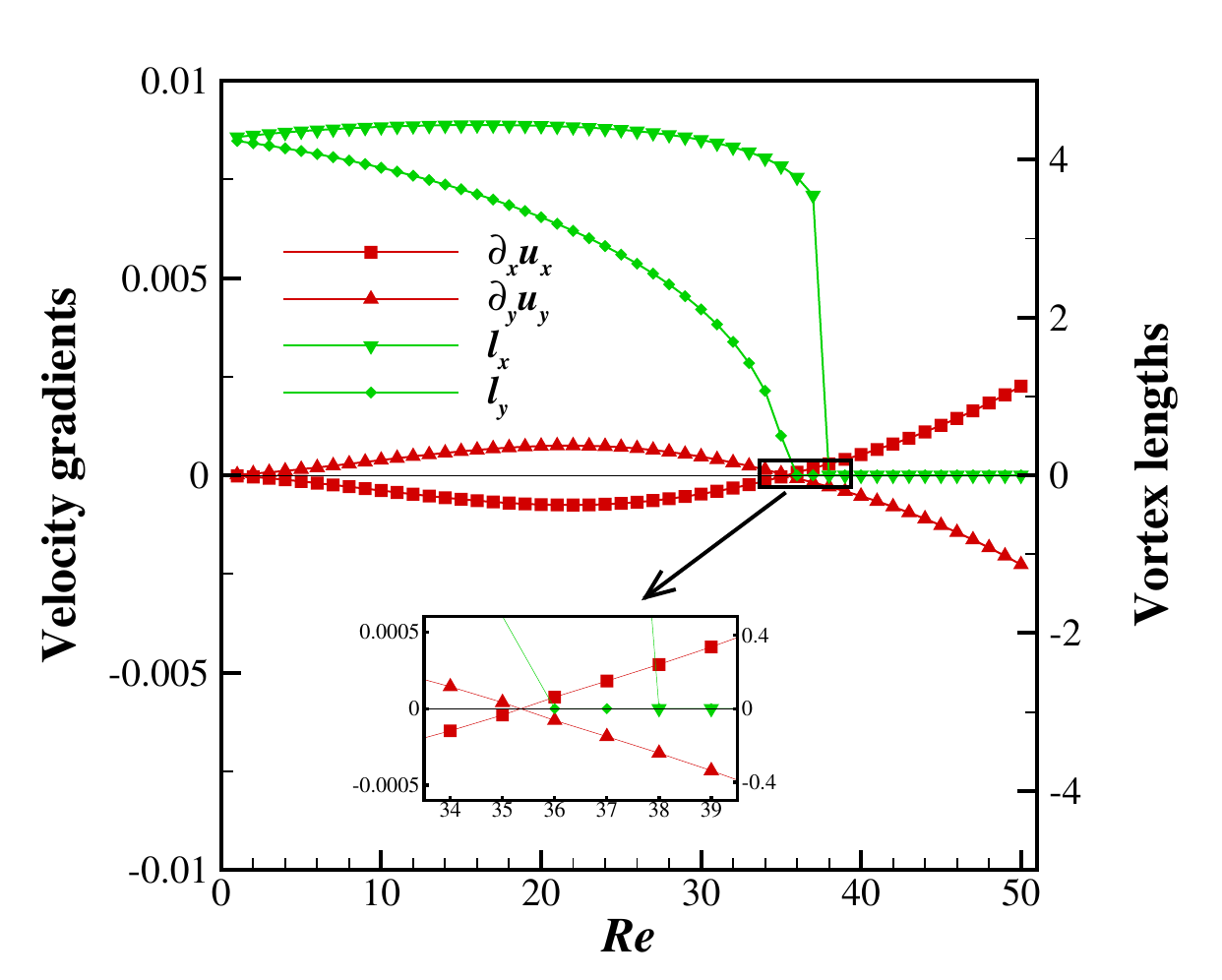}
  \hfill
  \caption{Variations of velocity gradients and vortex lengths with $Re$ for $n=1.3$.}
  \label{fig.4.2.6.h=1_n=1.3_velocity_gradients_vortex_lengths}
\end{figure*}

\begin{figure*}[htbp!]
  \centering
  \begin{subfigure}{0.4\textwidth}
    \includegraphics[width=\linewidth]{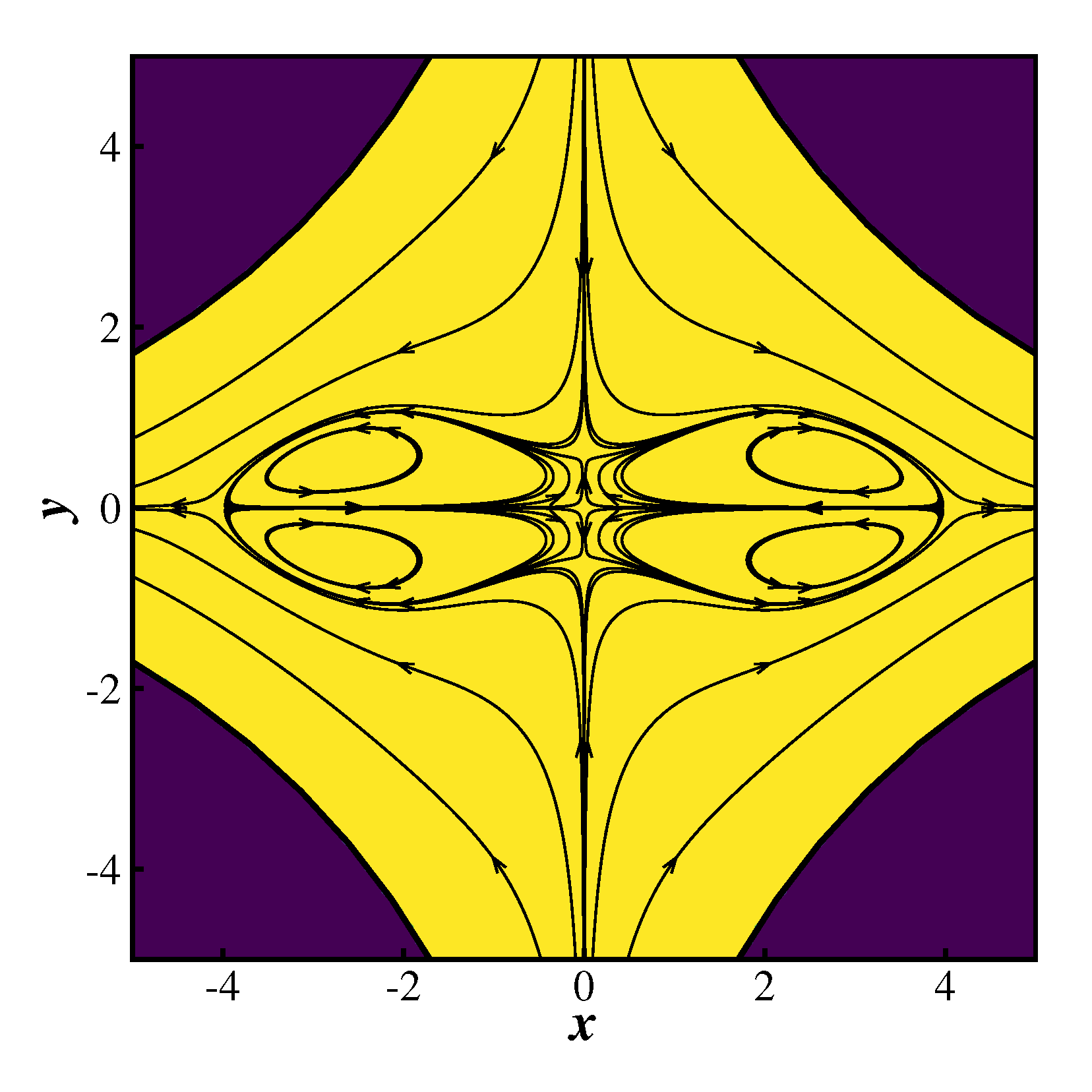}
    \caption{}
    \label{fig.4.2.7._b_h=1_n=1.3_Re=35_central_streamline}
  \end{subfigure}
  \hfill
  \begin{subfigure}{0.4\textwidth}
    \includegraphics[width=\linewidth]{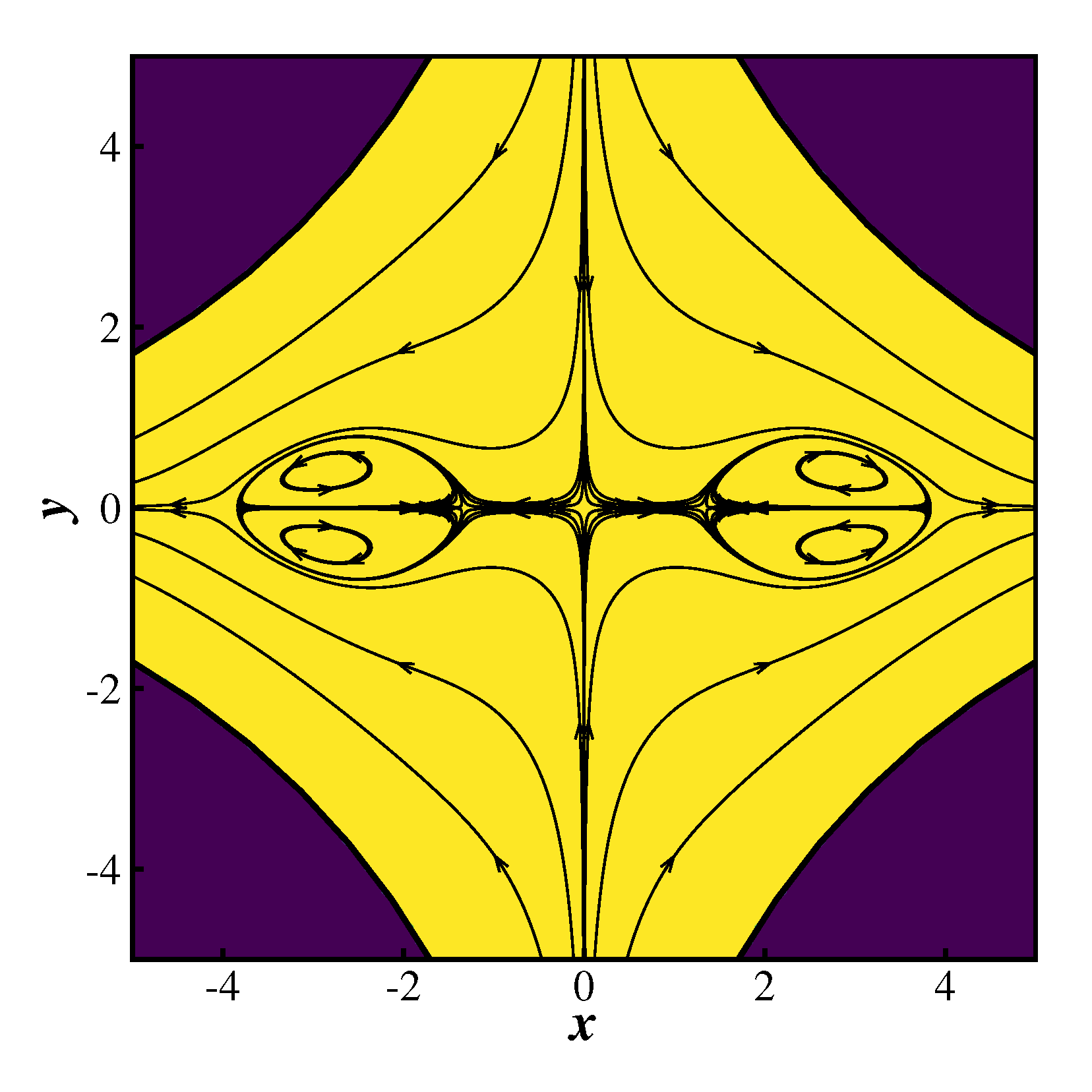}
    \caption{}
    \label{fig.4.2.7._b_h=1_n=1.3_Re=36_central_streamline}
  \end{subfigure}
  \hfill
  \begin{subfigure}{0.4\textwidth}
    \includegraphics[width=\linewidth]{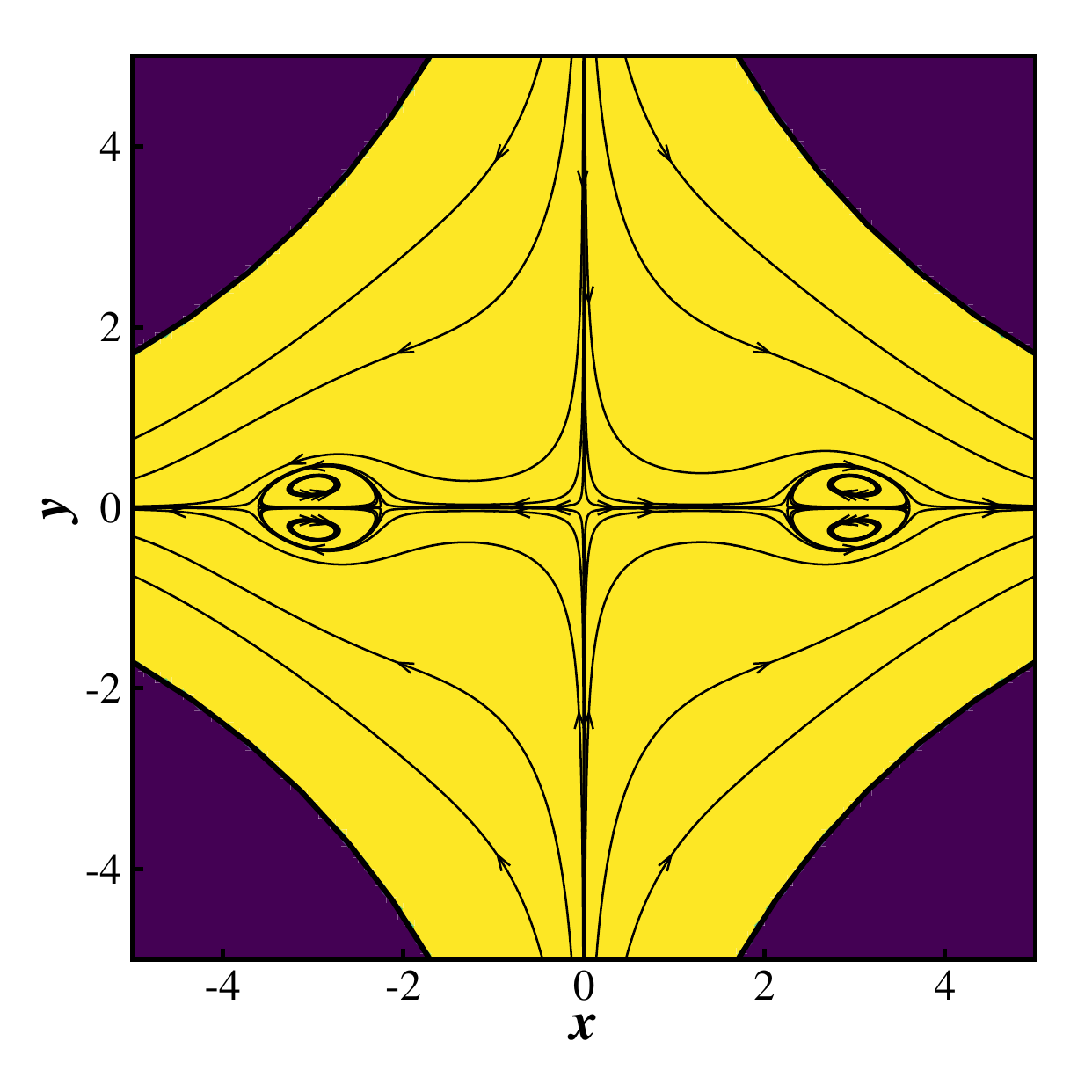}
    \caption{}
    \label{fig.4.2.7._c_h=1_n=1.3_Re=37_central_streamline}
  \end{subfigure}
  \hfill
  \begin{subfigure}{0.4\textwidth}
    \includegraphics[width=\linewidth]{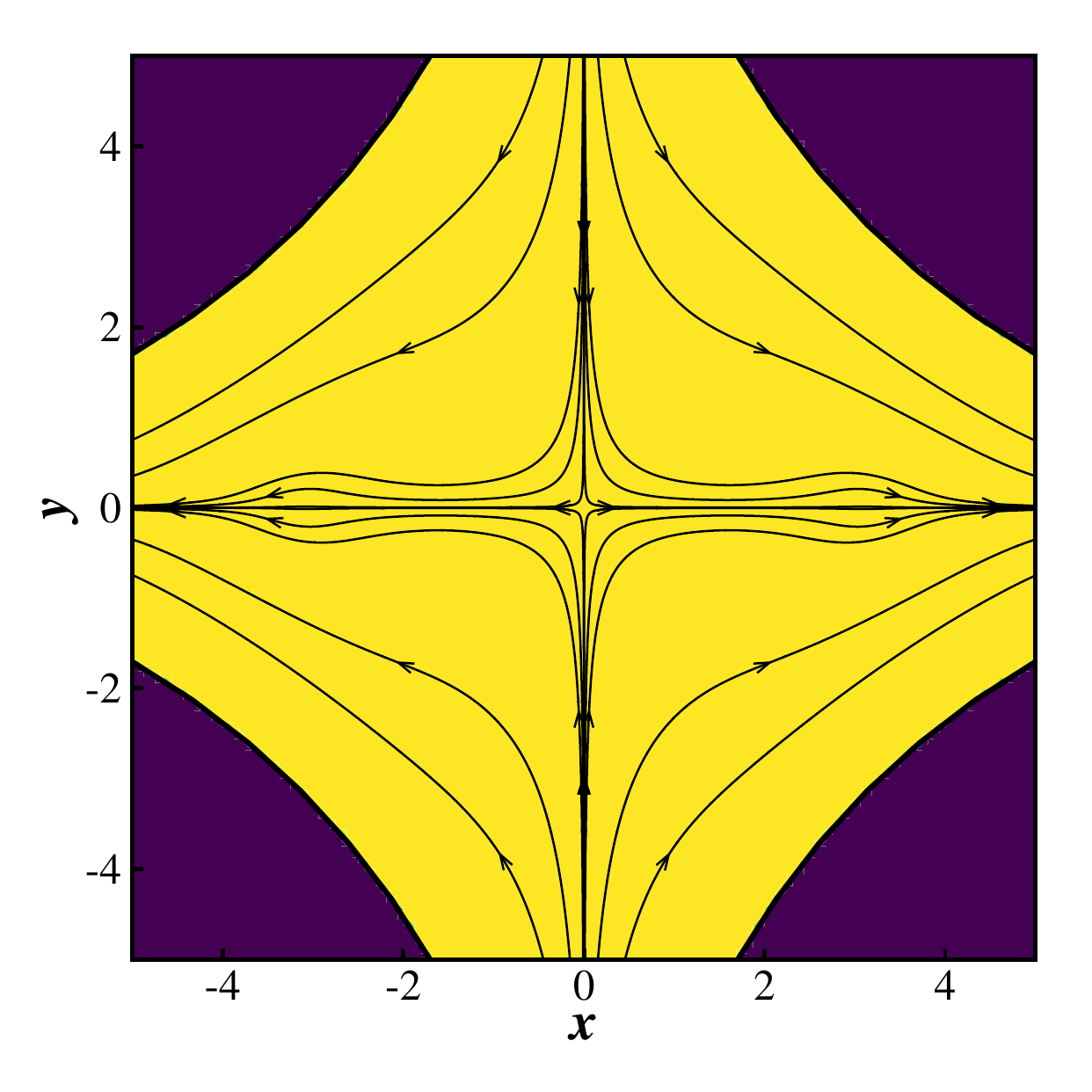}
    \caption{}
    \label{fig.4.2.7._d_h=1_n=1.3_Re=38_central_streamline}
  \end{subfigure}
  
  \caption{Central streamline in the four roll mill at (\textit{a}) $Re=35$, (\textit{b}) $Re=36$, (\textit{c}) $Re=37$, and (\textit{d}) $Re=38$ for $n=1.3$, $h = 1$, $\delta \approx 2.56$, and $r = 9$ under the mesh of $800 \times 800$.}
  \label{fig.4.2.7.n=0.7_Re=35-38_central_streamline}
\end{figure*} 

\begin{figure*}[htbp!]
  \centering
  \begin{subfigure}{0.45\textwidth}
    \includegraphics[width=\linewidth]{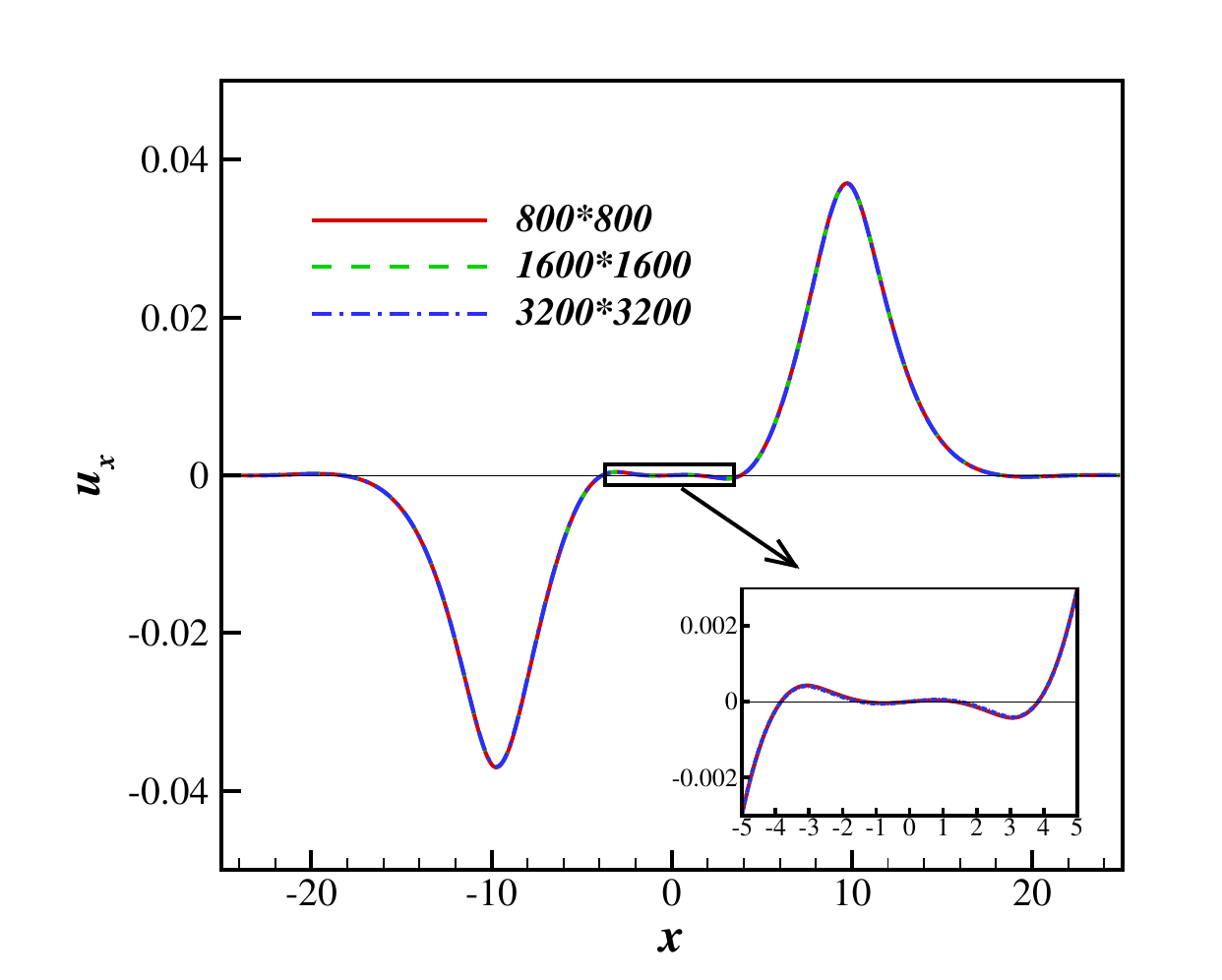}
    \caption{}
    \label{fig.4.2.8._b_n=1.3_Re=36_x-axis_compared_mesh}
  \end{subfigure}
  \hfill
  \begin{subfigure}{0.45\textwidth}
    \includegraphics[width=\linewidth]{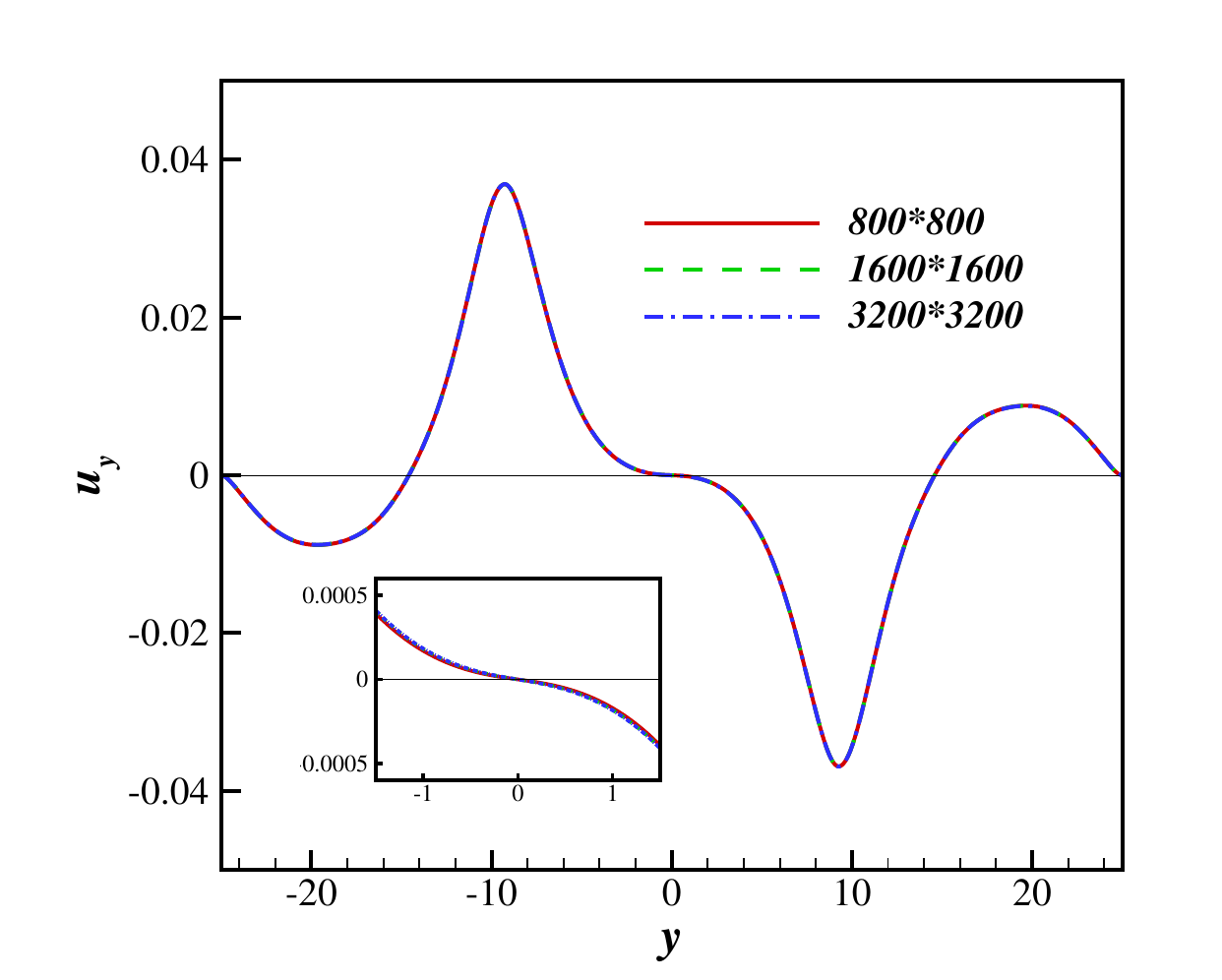}
    \caption{}
    \label{fig.4.2.8._b_n=1.3_Re=36_y-axis_compared_mesh}
  \end{subfigure}
  \caption{The velocity profiles (\textit{a}) $u_x$ in $x$-axis and (\textit{b}) $u_y$ in $y$-axis for $n=1.3$, $Re=36$, $h = 1$, $\delta \approx 2.56$, and $r = 9$ under different meshes.}
  \label{fig.4.2.8.n=1.3_Re=36_diff_mesh_velocity_profiles}
\end{figure*} 

\begin{figure*}[htbp!]
  \centering
  \includegraphics[width=1\textwidth]{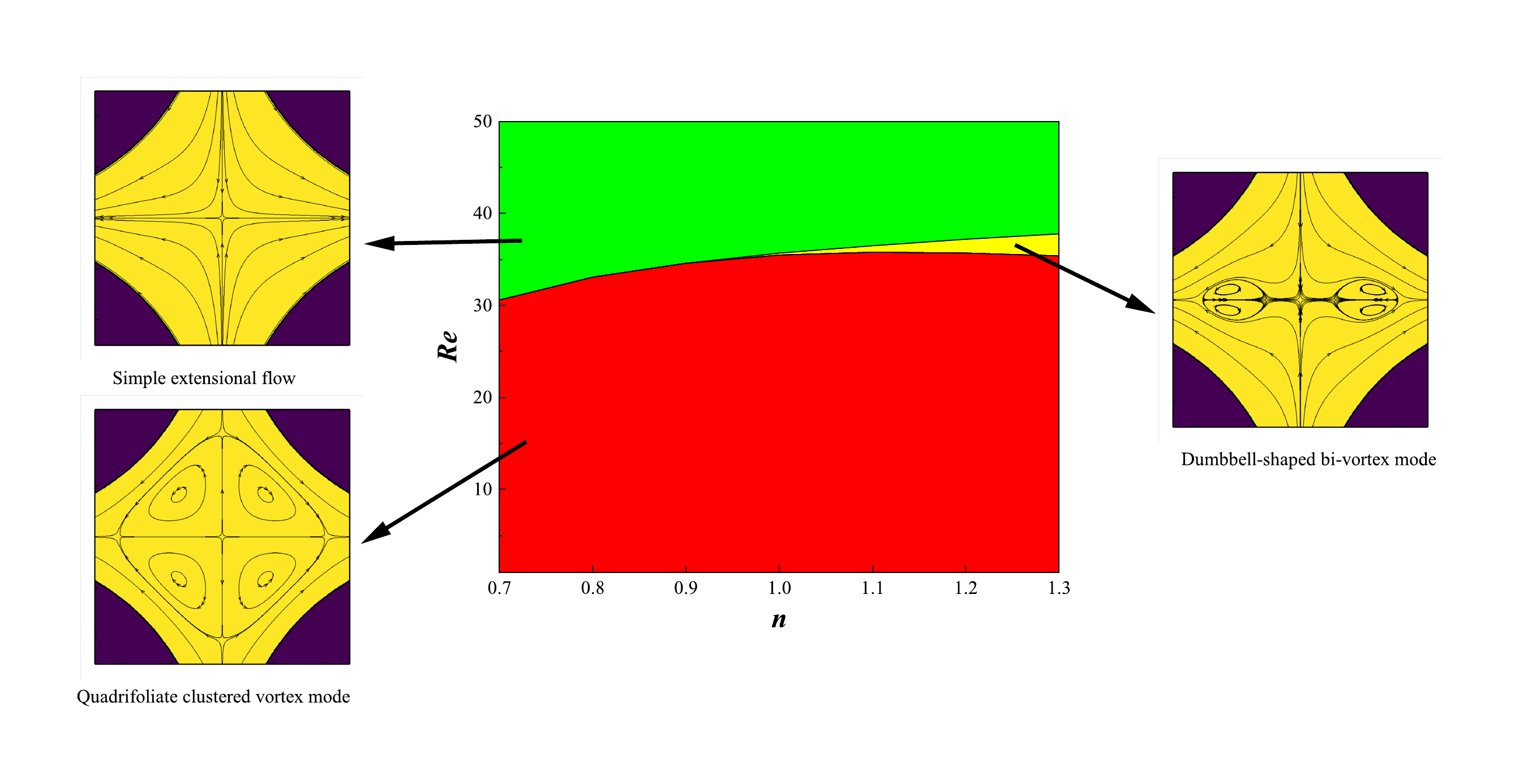}
  \caption{A phase diagram that illustrates the different flow modes in simulation of the four-roll mill as a function of power-law index $n$ and Reynolds number $Re$, with parameters $h = 1$, $\delta \approx 2.56$, and $r = 9$.}
  \label{fig.4.2.9.n=0.7-1.3_Rec_phase_diagram}
\end{figure*} 



\begin{figure*}
  \centering
    \centering
    \includegraphics[width=\textwidth]{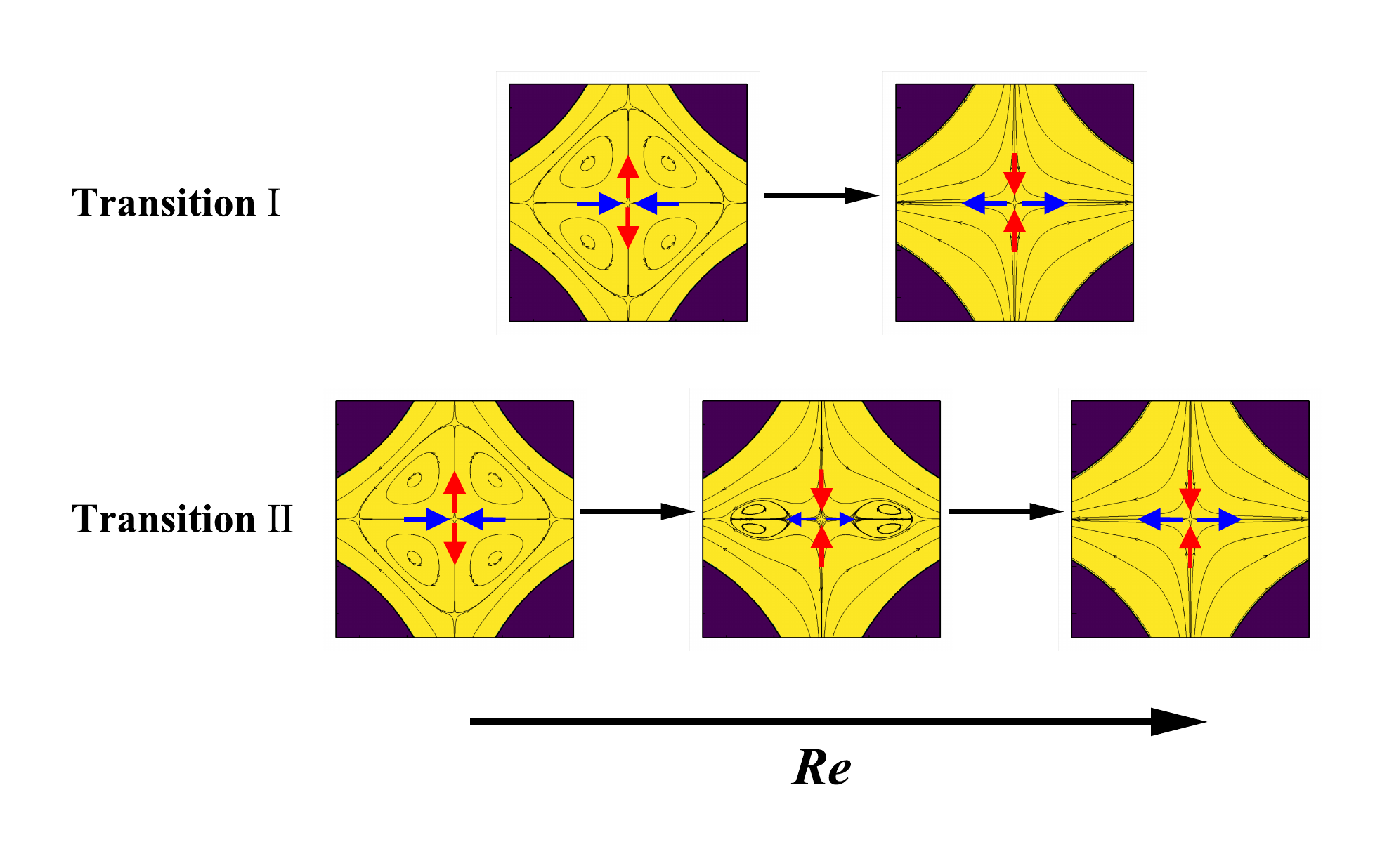}
  \hfill
    \captionsetup{justification=justified, singlelinecheck=false}
    \caption{Schematic representation of two distinct transition pathways in the four-roll mill. (a) Transition \uppercase\expandafter{\romannumeral1}: direct transformation from quadrifoliate vortex mode to simple extensional flow, typically observed at relatively small power-law indices. (b) Transition \uppercase\expandafter{\romannumeral2}: evolution from quadrifoliate vortex mode through an intermediate dumbbell-shaped quad-vortex mode to simple extensional flow, characteristic of relatively large power-law indices.}
    \label{fig.4.2.10.flow_transition12_examples}
\end{figure*}

\begin{figure*}[htbp!]
  \centering
  \begin{subfigure}{0.45\textwidth}
    \includegraphics[width=\linewidth]{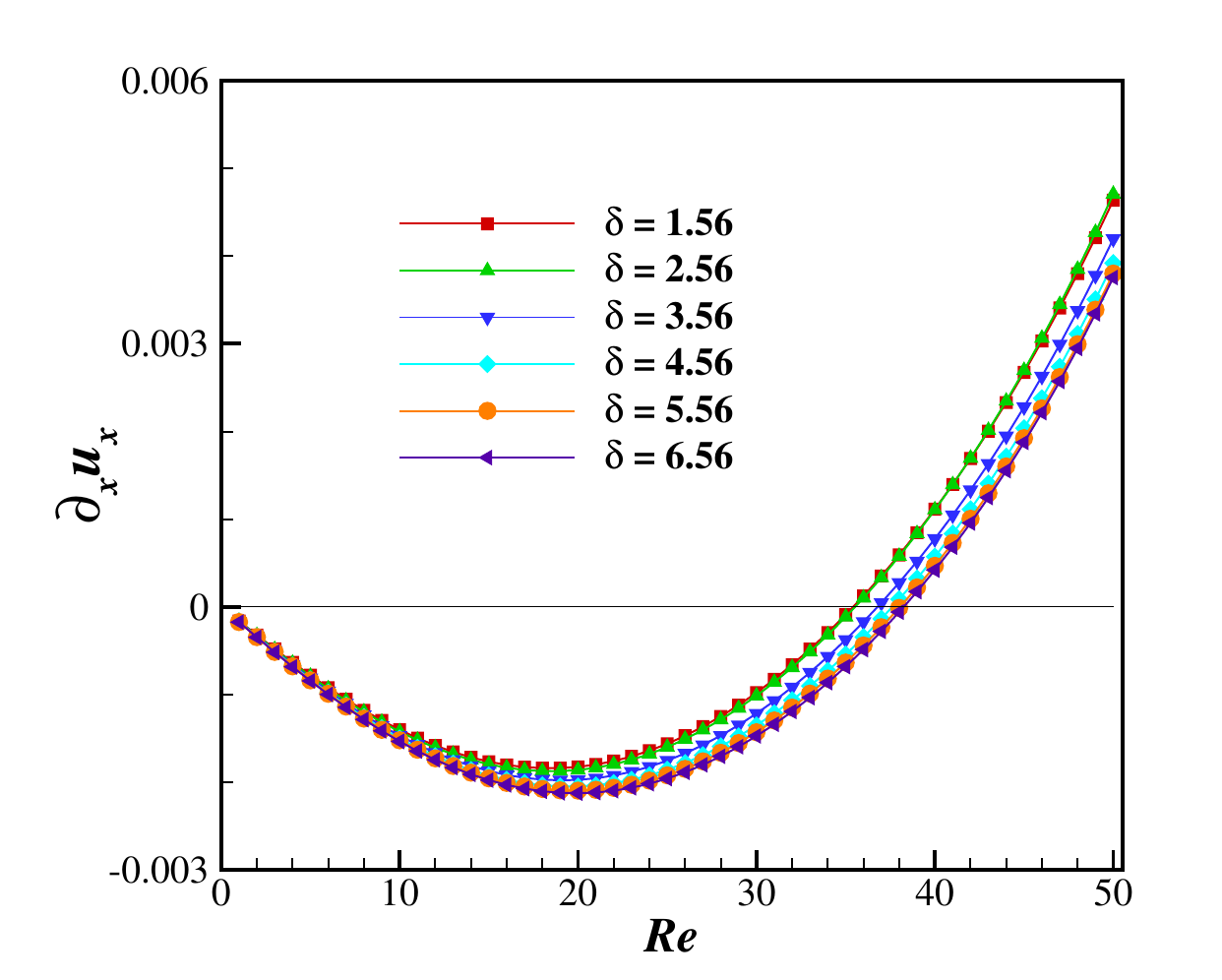}
    \caption{}
    \label{fig.4.3.2.1._b_delta=1.56-6.56_duxdx}
  \end{subfigure}
  \hfill
  \begin{subfigure}{0.45\textwidth}
    \includegraphics[width=\linewidth]{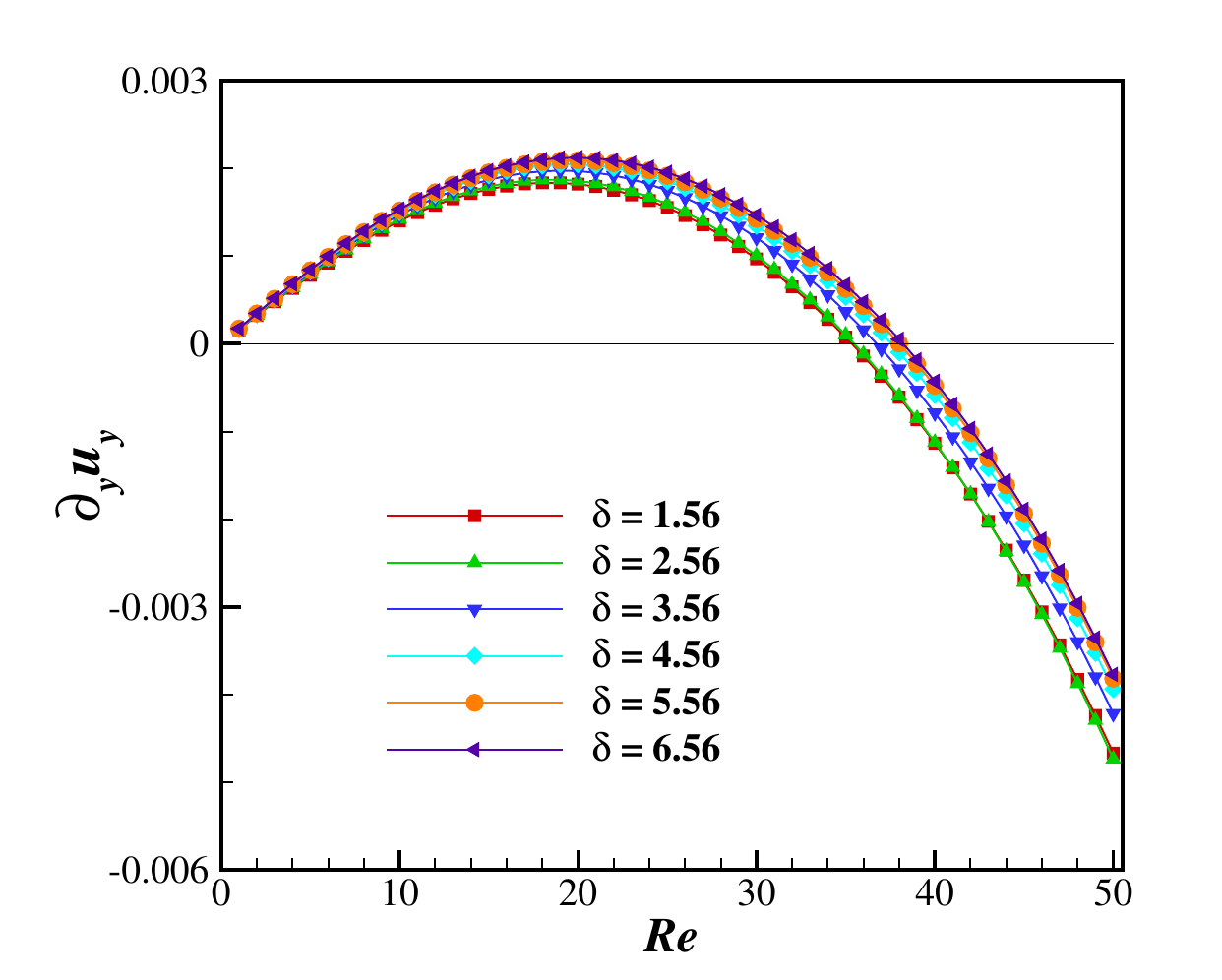}
    \caption{}
    \label{fig.4.3.2.1._b_delta=1.56-6.56_duydy}
  \end{subfigure}
  \hfill
  \begin{subfigure}{0.45\textwidth}
    \includegraphics[width=\linewidth]{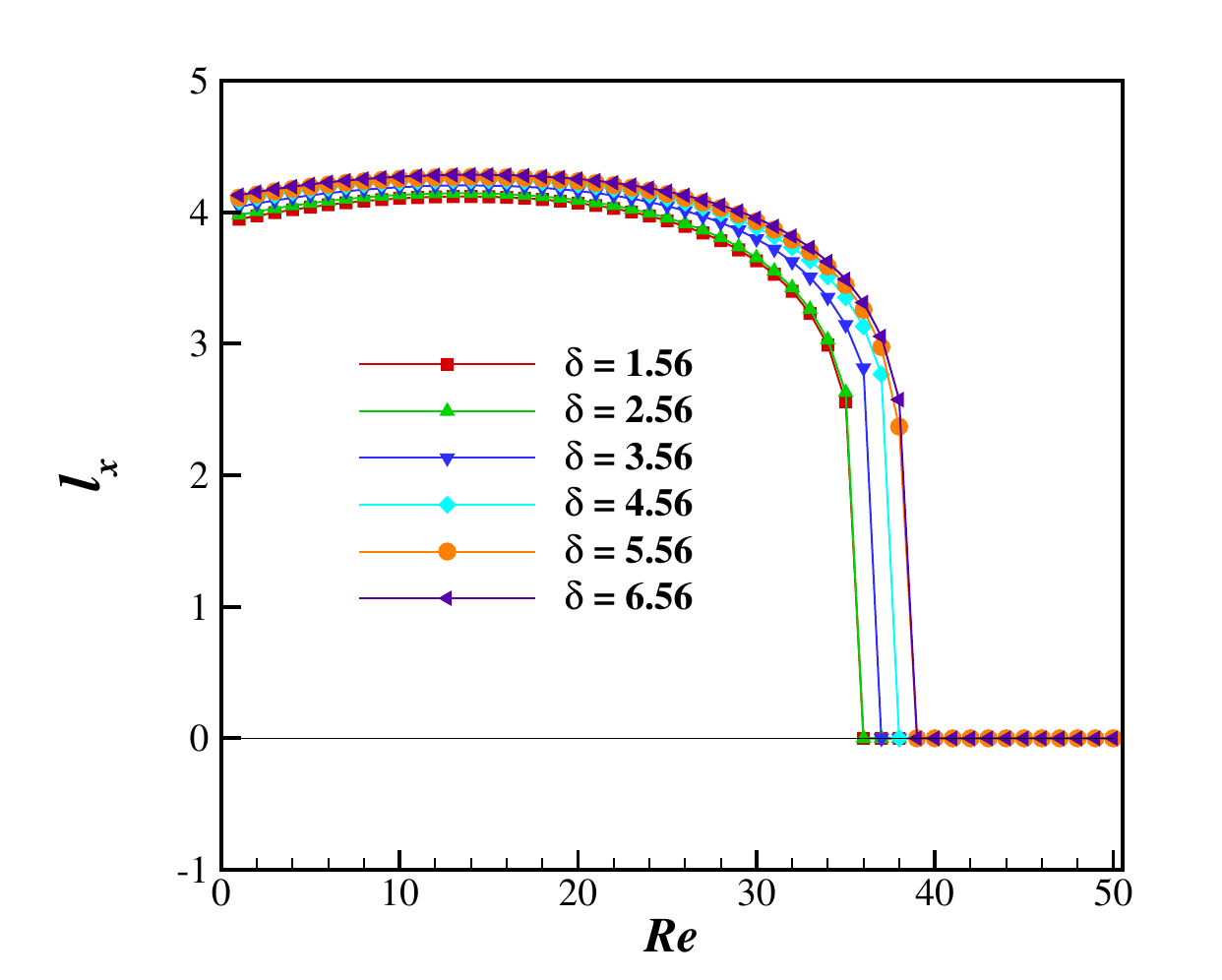}
    \caption{}
    \label{fig.4.3.2.1._c_delta=1.56-6.56_lx}
  \end{subfigure}
  \hfill
  \begin{subfigure}{0.45\textwidth}
    \includegraphics[width=\linewidth]{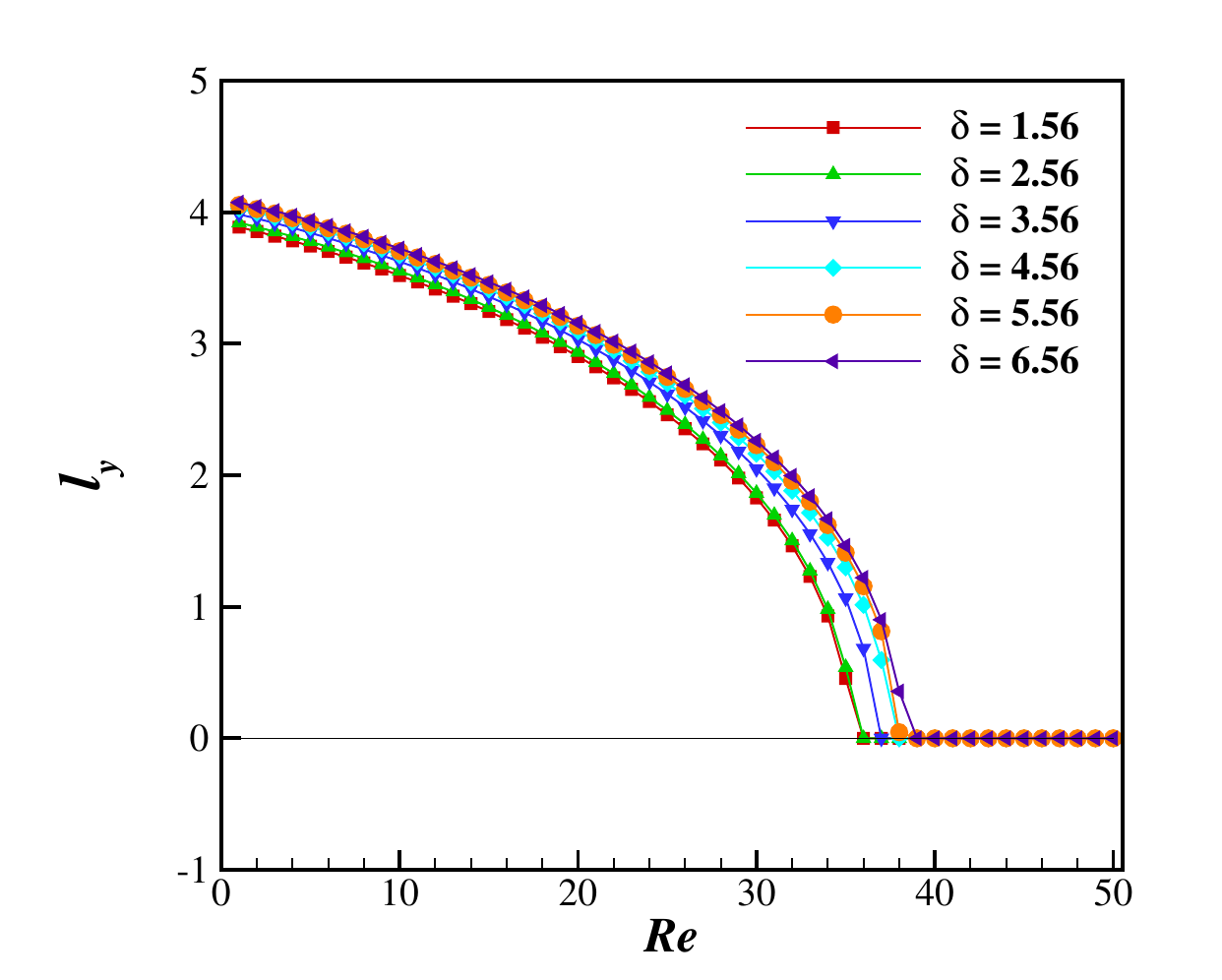}
    \caption{}
    \label{fig.4.3.2.1._d_delta=1.56-6.56_ly}
  \end{subfigure}
  \caption{Variation of velocity gradients (\textit{a}) $\partial{u_x}/\partial{x}$, (\textit{b}) $\partial{u_y}/\partial{y}$ at point $O$ and vortex lengths (\textit{c}) $l_x$, and (\textit{d}) $l_y$ with $Re$ for $\delta$ values of 1.56, 2.56, 3.56, 4.56, 5.56, and 6.56, with parameters $h=1$ and $r=9$.}
  \label{fig.4.3.2.1.Re=1-50_delta=1.56-6.56_duxdx_duydy_lx_ly}
\end{figure*} 

\begin{figure*}
  \includegraphics[width=0.6\textwidth]{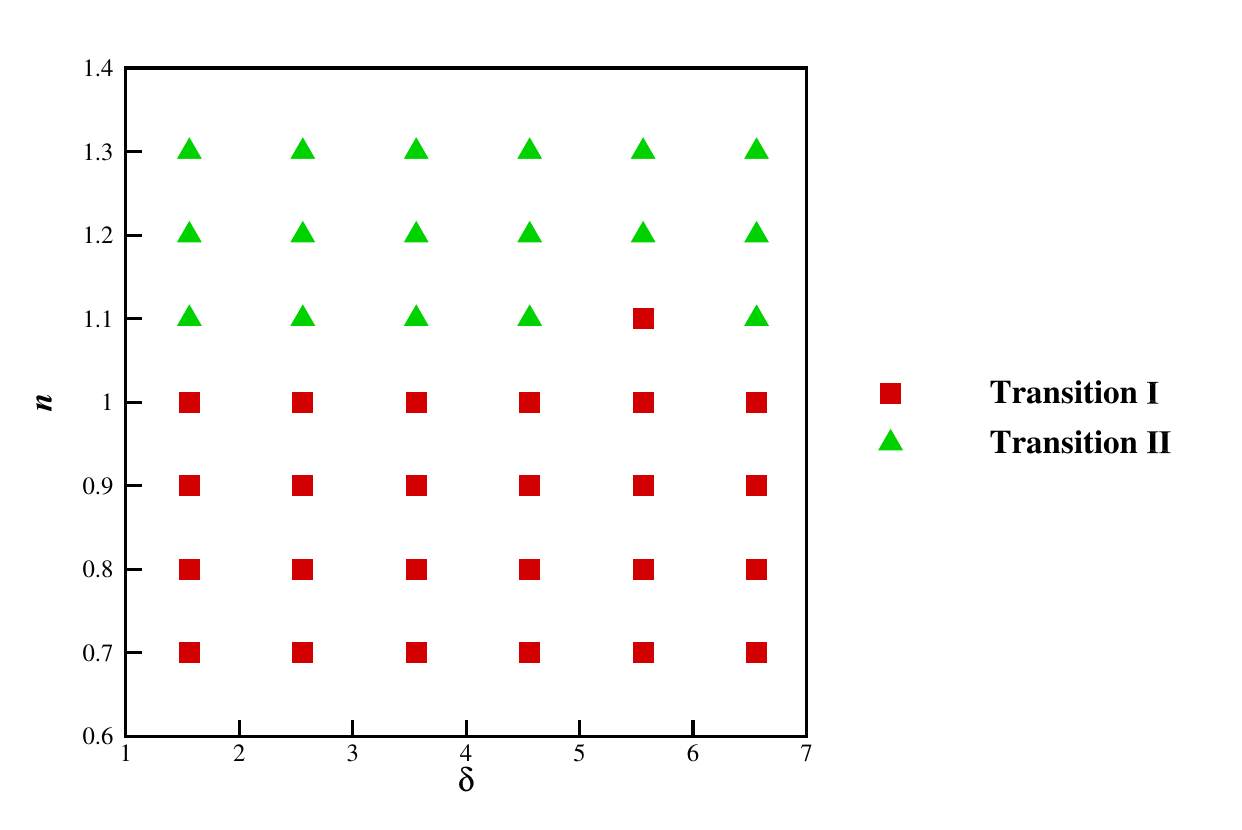}
  \caption{A phase diagram that illustrates the different mode transitions at $h=1$ and $r=9$ for $\delta$ values of 1.56, 2.56, 3.56, 4.56, 5.56, and 6.56 and $n$ ranging from 0.7 to 1.3.}
  \label{fig.4.3.2.2.delta=1.56-6.56_n=0.7-1.3_phase_diagram}
\end{figure*}

\subsection{Grid convergence study}\label{section3.4}

A grid convergence analysis was performed to ensure the spatial resolution independence of the numerical results for most flow conditions. Six different grid resolutions were examined: $N^2 = 500^2$, $600^2$, $700^2$, $800^2$, $900^2$, and $1000^2$. The geometric configuration of the four-roll mill was specified with $r = 9.5$, $L_c = 50$, and $h = 1$, yielding $I \approx 4.64$ and $\delta \approx 1.36$. The flow conditions were characterized by $Re = 10$ and $\mu_0 = 0.002$. Two quantities were monitored as convergence metrics: the $x$-direction velocity $u_x$ at the midpoint between rollers 1 and 4, and the shear rate magnitude $|\dot{\gamma}|$ at the central point $O$ (see Fig.~\ref{four_roll_mill}). The results are presented in Table~\ref{Grid_tab}. As the variations in both metrics become negligible for grid resolutions of $800^2$ and above, this resolution was adopted for subsequent simulations unless otherwise specified. It should be noted, however, that finer grid resolution may be required for cases with small power-law indices, where non-physical spindle-shaped streamline mode might emerge in the central region without adequate spatial resolution (this phenomenon will be discussed in detail in Section~\ref{section4.2}).

\begin{figure*}
  \centering
  \begin{subfigure}{0.49\textwidth}
    \includegraphics[width=\linewidth]{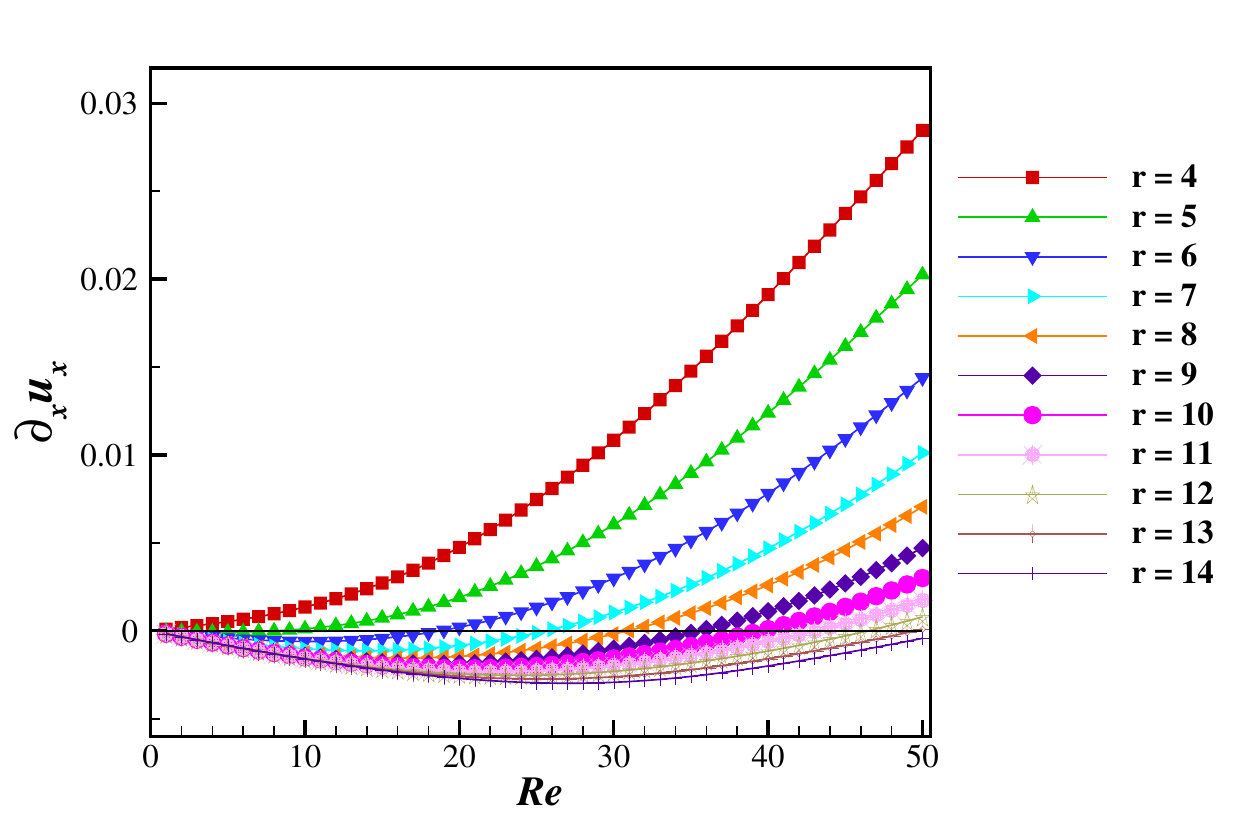}
    \caption{}
    \label{fig.4.3.3.1._b_r=4-14_duxdx}
  \end{subfigure}
  \hfill
  \begin{subfigure}{0.49\textwidth}
    \includegraphics[width=\linewidth]{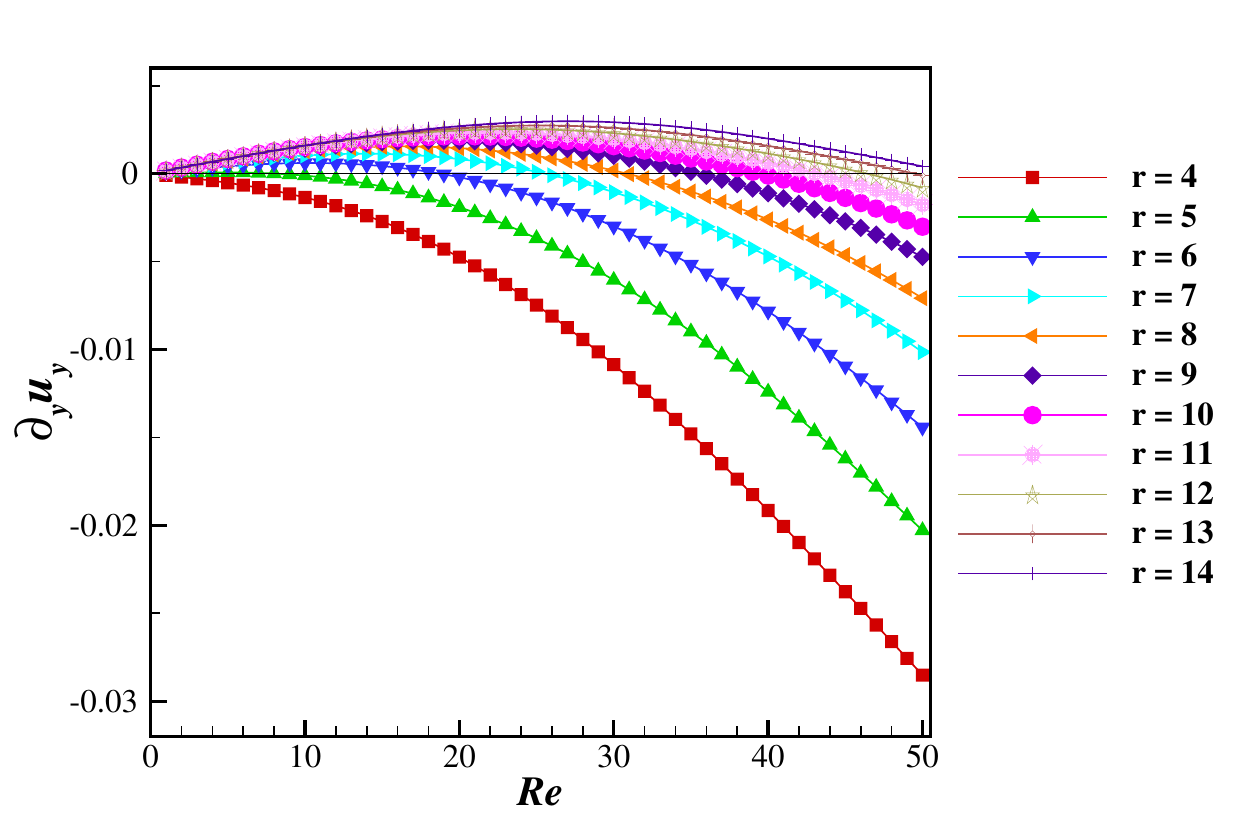}
    \caption{}
    \label{fig.4.3.3.1._b_r=4-14_duydy}
  \end{subfigure}
  \hfill
  \begin{subfigure}{0.49\textwidth}
    \includegraphics[width=\linewidth]{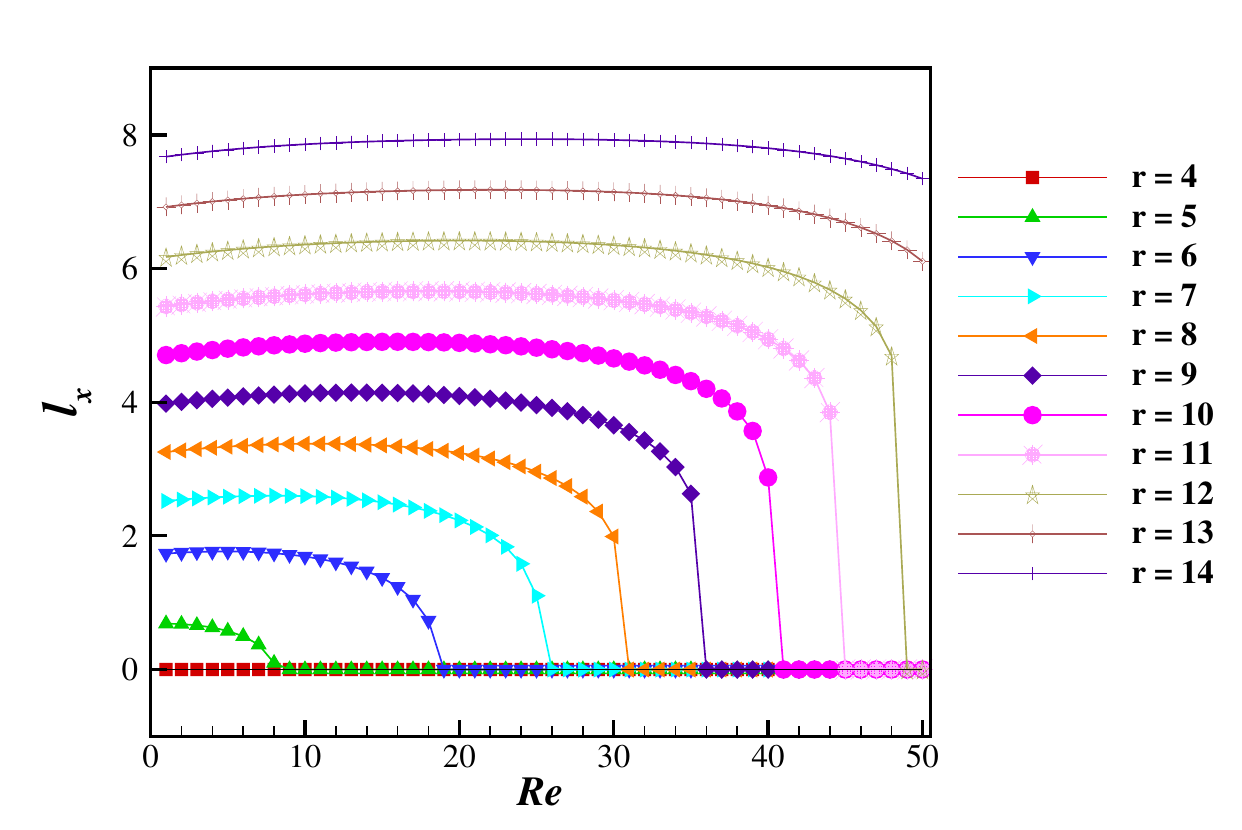}
    \caption{}
    \label{fig.4.3.3.1._c_r=4-14_lx}
  \end{subfigure}
  \hfill
  \begin{subfigure}{0.49\textwidth}
    \includegraphics[width=\linewidth]{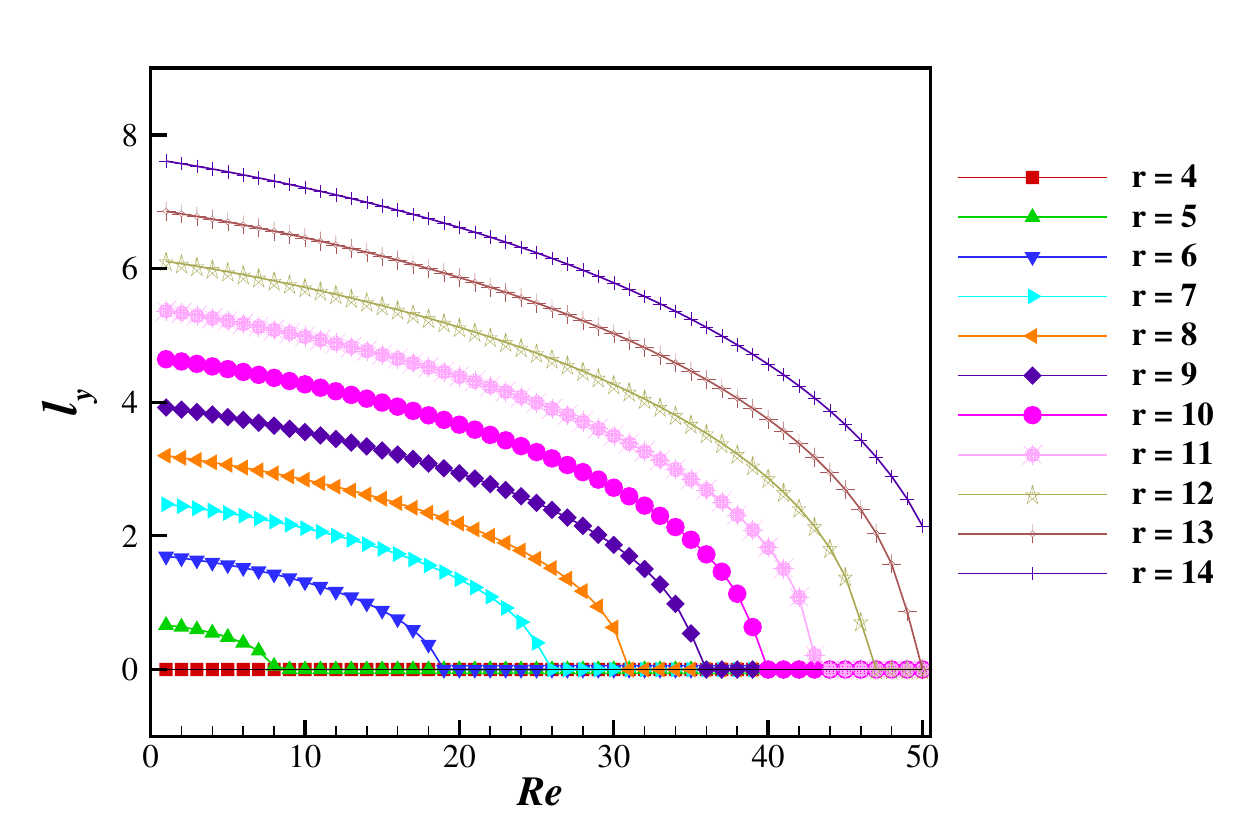}
    \caption{}
    \label{fig.4.3.3.1._d_r=4-14_ly}
  \end{subfigure}
  \caption{Variations of velocity gradients (\textit{a}) $\partial{u_x}/\partial{x}$, (\textit{b}) $\partial{u_y}/\partial{y}$ at point $O$ and vortex lengths (\textit{c}) $l_x$ and (\textit{d}) $l_y$ with $Re$ for $r$ values from 4 to 14, with parameters $h=1$ and $\delta \approx 2.56$.}
  \label{fig.4.3.3.1.Re=1-50_r=4-14_velocity_gradient_vortex_length}
\end{figure*} 

\begin{figure*}
  \centering
  \includegraphics[width=0.8\textwidth]{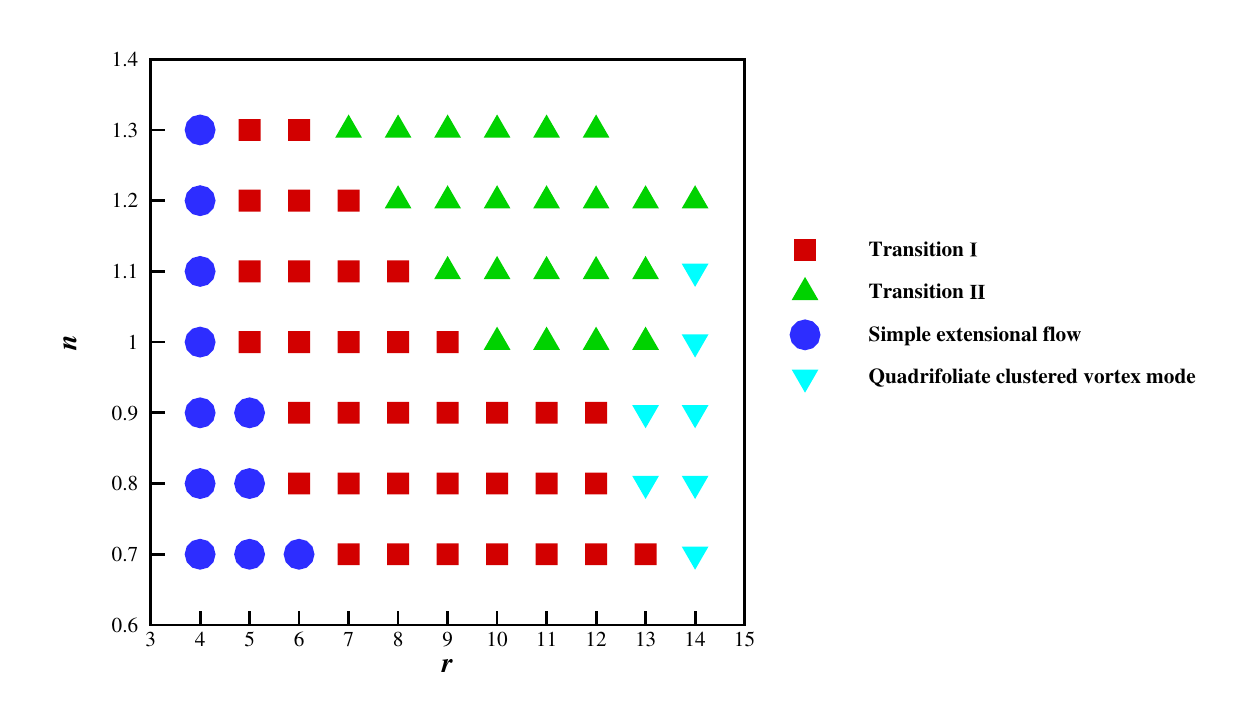}
  \caption{A phase diagram that illustrates the different mode transitions at $h=1$ and $\delta \approx 2.56$ for $r$ ranging from 4 to 14 and $n$ from 0.7 to 1.3.}
  \label{fig.4.3.3.2.r=4-14__n=0.7-1.3_phase_diagram}
\end{figure*}

\section{Results and discussion}\label{section4}
	
This section presents a systematic investigation of flow characteristics in the four-roll mill. In Section~\ref{section4.1}, we first analyze the flow modes of Newtonian fluids and examine how varying Reynolds numbers influence the emergence of different flow modes. Section~\ref{section4.2} extends this analysis to power-law fluids, focusing on their distinct mode transition behaviors. The effects of geometric parameters, specifically the roller-container gap $\delta$ and roller radius $r$, on flow characteristics are investigated in Section~\ref{section4.3}.

\subsection{Flow modes of Newtonian fluid under different \texorpdfstring{$Re$}{Re}}\label{section4.1}

The geometric parameters are initially set as $r = 9$, $L_c = 50$, and $h = 1$, which determine the corresponding values of $I \approx 4.44$ and $\delta \approx 2.56$. The viscosity coefficient $\mu_0$ is fixed at 0.002 unless otherwise specified. A series of numerical simulations are performed for a Newtonian fluid ($n=1.0$) over Reynolds numbers ranging from $1$ to $50$.

Fig.~\ref{fig.4.1.1.h=1_n=1_Re=50_streamline} illustrates the flow structure in the four-roll mill for Newtonian fluid at $Re=50$ through streamline plots of (\textit{a}) the entire domain and (\textit{b}) the central region. Eight vortices are observed in Fig.~\ref{fig.4.1.1._b_h=1_n=1_Re=50_entire_streamline}, which remain isolated from the central flow as they do not interact with the inlet and outlet of the central region. Fig.~\ref{fig.4.1.1._b_h=1_n=1_Re=50_central_streamline} demonstrates that the central region exhibits simple extensional flow, consistent with previous experimental \citep{taylor_formation_1934,bentleyComputercontrolledFourrollMill1986} and numerical investigations \citep{FENG1997187} of four-roll mills. The velocity profiles $u_x$ and $u_y$ along the $y$-axis and $x$-axis at $Re=50$ are presented in Fig.~\ref{fig.4.1.2.uxuy_along(a)y-axis(b)x-axis_Re=50}. Both velocity components vanish at point $O$, confirming its stagnation point nature, and exhibit central symmetry about this point. Moreover, the vanishing of $u_x$ along the $y$-axis and $u_y$ along the $x$-axis indicates the absence of fluid exchange between quadrants.

At $Re=10$, the streamline patterns in the central region exhibit a distinct flow mode, as shown in Fig.~\ref{fig.4.1.3.Streamline_Re=10}. Four symmetrical vortices form around point $O$, with adjacent vortices rotating in directions opposite to their neighboring rollers. Although these vortices dominate the central region, detailed examination reveals that the flow near point $O$ maintains extensional characteristics, albeit with reversed directions compared to the $Re=50$ case: the vertical direction transitions from compression to extension, while the horizontal direction shifts from extension to compression. This previously unreported flow configuration, which we term the quadrifoliate vortex mode due to its four-leaf pattern, represents a novel finding in four-roll mill dynamics.

The velocity profiles at $Re=10$, presented in Fig.~\ref{fig.4.1.4.uxuy_along(a)y-axis(b)x-axis_Re=10}, share certain characteristics with the simple extensional flow: both $u_x$ along the $y$-axis and $u_y$ along the $x$-axis vanish, indicating no cross-axial fluid transport. As shown in Fig.~\ref{fig.4.1.4._b_n=1_Re=10_ux_uy_y_axis}, $u_y$ reaches its maximum at the inlet ($y=-9.5$) due to the local flow constriction. The velocity $u_y$ exhibits a characteristic evolution from inlet to point $O$: it initially decreases to zero, reaches a minimum value, and finally returns to zero. The position where $u_y$ first vanishes defines the vortex generation point, with the vortex length $l_y$ measured as the distance from this point to point $O$. Similar velocity characteristics and vortex length definition ($l_x$) apply along the $x$-axis, as demonstrated in Fig.~\ref{fig.4.1.4._b_n=1_Re=10_ux_uy_x_axis}.

Comparing the cases of $ Re=10 $ and $ Re=50 $, it is evident that a reduction in the roller speed leads to the emergence of vortices in the central region of the flow field under the same structure and fluid. Additionally, the original vertical compression transforms into extension, and the horizontal extension becomes compression at point $O$. The states of compression and extension at point $O$ can be quantified using $\partial_x u_x$ and $\partial_y u_y$. These gradients, computed via the second-order central differencing, indicate the magnitude and direction of compression or extension at point $O$. Specifically, if $\partial_y u_y > 0$, it indicates extension in the vertical direction at point $O$; if $\partial_y u_y < 0$, it indicates compression. The same interpretation applies to $\partial_x u_x$ for the horizontal direction.

Fig.~\ref{fig.4.1.5.h=1_n=1_Re=1-50_shear_rate_and_velo_gradient} depicts the evolution of shear rate and velocity gradients at point $O$ with varying $Re$. The velocity gradients $\partial_x u_y$ and $\partial_y u_x$ consistently remain zero, confirming the absence of cross-axial fluid transport. Consequently, the shear rate is determined solely by $\partial_x u_x$ and $\partial_y u_y$. A critical transition is observed between $Re=35$ and $Re=36$, where $\partial_x u_x$ and $\partial_y u_y$ undergo sign reversal. This marks the transition from the quadrifoliate vortex mode ($\partial_x u_x > 0$ and $\partial_y u_y < 0$) to simple extensional flow ($\partial_x u_x < 0$ and $\partial_y u_y > 0$). For $Re \geq 36$, the flow maintains simple extensional characteristics. This transition point defines the critical Reynolds number $Re_c$, which equals 36 under the current configuration.

The evolution of vortex dimensions with Reynolds number is characterized through the lengths $l_x$ and $l_y$. Given the symmetry of the four vortices, we focus on the size variation of a single vortex. Fig.~\ref{fig.4.1.6.h=1_n=1_Re=1-50_vortex_length} illustrates the variation of $l_x$ and $l_y$ over $Re$ ranging from 1 to 50. Both lengths exhibit monotonic decrease with increasing $Re$, eventually vanishing at $Re = 36$. This point marks the disappearance of the quadrifoliate vortex mode and its transition to simple extensional flow. Notably, throughout the quadrifoliate vortex regime, $l_x$ consistently exceeds $l_y$. These observations establish that the critical Reynolds number $Re_c$ can be determined through two independent criteria: the reversal of velocity gradients ($\partial_x u_x$ and $\partial_y u_y$) and the vanishing of vortex lengths ($l_x$ and $l_y$).

\subsection{Effect of the power-law fluid}\label{section4.2}

This section investigates how power-law fluids influence the flow modes in the central region, maintaining the geometric configuration described in Section~\ref{section4.1}. The analysis spans Reynolds numbers from 1 to 50 and power-law indices from 0.7 to 1.3, encompassing both shear-thinning and shear-thickening fluid behaviors.

Fig.~\ref{fig.4.2.1.velocity_gradient_n=0.7-1.3} depicts the evolution of velocity gradients $\partial_x u_x$ and $\partial_y u_y$ across the specified ranges of $n$ and $Re$. For each power-law index, $\partial_x u_x$ exhibits a non-monotonic trend with increasing $Re$, first decreasing then increasing, while $\partial_y u_y$ shows the opposite behavior. As illustrated in Fig.~\ref{fig.4.2.2.vortex_length_n=0.7-1.3}, throughout the range $0.7 \leq n \leq 1.3$, both vortex lengths $l_x$ and $l_y$ monotonically decrease to zero with increasing $Re$ and remain zero thereafter. These characteristics mirror those observed for Newtonian fluids in Section~\ref{section4.1}, indicating a consistent transition mechanism from the quadrifoliate vortex mode to simple extensional flow.

However, numerical complexities arise near the critical Reynolds number, where discrepancies emerge between transition points identified by velocity gradients and vortex lengths. For instance, at $n=0.7$ (Fig.~\ref{fig.4.2.3.h=1_n=0.7_velocity_gradients_vortex_lengths}), the critical $Re$ values determined from $\partial_x u_x$, $\partial_y u_y$, $l_x$, and $l_y$ are $32$, $31$, $32$, and $31$, respectively. At $Re=31$, the flow field exhibits an anomalous behavior where both $\partial_x u_x$ and $\partial_y u_y$ are negative, suggesting physically impossible flow convergence toward point $O$ in both directions (Fig.~\ref{fig.4.2.4._b_h=1_n=0.7_Re=31_central_streamline_800}). This non-physical behavior is attributed to insufficient spatial resolution and can be eliminated by grid refinement. As demonstrated in Fig.~\ref{fig.4.2.4.n=0.7_Re=31_diff_mesh_central_streamline}, increasing the grid resolution to $1600 \times 1600$ and $3200 \times 3200$ resolves this anomaly. Fig.~\ref{fig.4.2.5.n=0.7_Re=31_diff_mesh_velocity_profiles} compares velocity profiles under different grid resolutions, revealing that while the overall velocity distributions remain consistent, the local $\partial_x u_x$ at point $O$ transitions from negative to positive with increasing resolution, while $\partial_y u_y$ remains relatively unchanged. This grid-dependent behavior occurs for $0.7 \leq n \leq 0.9$ and can be effectively addressed through mesh refinement.

A distinct transition phenomenon emerges for power-law indices between 1.0 and 1.3. Taking $n=1.3$ as a representative case, the critical Reynolds numbers determined from $\partial_x u_x$, $\partial_y u_y$, $l_x$, and $l_y$ are 36, 36, 38, and 36, respectively. Fig.~\ref{fig.4.2.6.h=1_n=1.3_velocity_gradients_vortex_lengths} illustrates the evolution of velocity gradients and vortex lengths with increasing $Re$ for $n=1.3$. At $Re=35$ and 36, while the velocity gradients ($\partial_x u_x < 0$ and $\partial_y u_y > 0$) exhibit characteristics of simple extensional flow, an asymmetry in vortex extinction is observed: $l_y$ vanishes while $l_x$ remains finite.
The flow field evolution for $Re$ ranging from 35 to 38 at $n=1.3$ is presented in Fig.~\ref{fig.4.2.7.n=0.7_Re=35-38_central_streamline}. The flow transitions from a well-defined quadrifoliate vortex mode at $Re=35$ to simple extensional flow at $Re=38$. However, at intermediate Reynolds numbers ($Re=36$ and 37), a previously unreported flow configuration emerges, distinctly different from the anomalous behavior observed in Fig.~\ref{fig.4.2.4._b_h=1_n=0.7_Re=31_central_streamline_800}. This intermediate state is characterized by four vortices symmetrically attached to the $x$-axis and displaced laterally from point $O$. As $Re$ increases from 36 to 37, these vortices progressively shrink before completely disappearing at $Re=38$, giving way to simple extensional flow.
Grid independence analysis confirms the physical nature of this phenomenon. Fig.~\ref{fig.4.2.8.n=1.3_Re=36_diff_mesh_velocity_profiles} demonstrates that velocity profiles along both axes remain consistent across different grid resolutions, with the stagnation points associated with the four vortices maintaining their positions along the $x$-axis. This intermediate flow configuration, observed for power-law indices from 1.0 to 1.3, serves as a transitional state between the quadrifoliate vortex mode and simple extensional flow. Due to its characteristic shape, we designate this newly identified flow state as the dumbbell-shaped quad-vortex mode.

Based on the above findings, we have identified two previously unreported flow modes in the four-roll mill: the quadrifoliate vortex mode and the dumbbell-shaped quad-vortex mode. To systematically characterize these flow modes and their transitions, we present a comprehensive phase diagram in Fig.~\ref{fig.4.2.9.n=0.7-1.3_Rec_phase_diagram}, mapping the flow regimes as functions of power-law index $n$ and Reynolds number $Re$. The diagram is constructed with refined resolution ($\Delta Re = 0.1$) near critical transitions and coarser resolution ($\Delta Re = 1$) elsewhere. The phase diagram reveals two distinct transition pathways depending on the power-law index:
For shear-thinning fluids ($0.7 \leq n < 1.0$), increasing $Re$ triggers a direct transition from the quadrifoliate vortex mode to simple extensional flow, with the transition Reynolds number increasing monotonically with $n$. In contrast, shear-thickening fluids ($1.0 \leq n \leq 1.3$) exhibit a more complex behavior, where the dumbbell-shaped quad-vortex mode emerges as an intermediate state. The critical $Re$ for the initial transition (quadrifoliate to dumbbell-shaped) shows a non-monotonic dependence on $n$, while the subsequent transition (dumbbell-shaped to simple extensional) occurs at progressively higher $Re$ as $n$ increases.
Furthermore, the dumbbell-shaped quad-vortex mode becomes increasingly prevalent at higher power-law indices, with its stability range expanding with increasing $n$. These distinct transition scenarios are schematically illustrated in Fig.~\ref{fig.4.2.10.flow_transition12_examples}.

\subsection{Effect of geometric configuration in four-roll mill}\label{section4.3}

Having characterized the flow modes and their transitions for a specific geometry, we now investigate how the geometric parameters of the four-roll mill influence these phenomena. Our analysis focuses on two key geometric parameters: the roller radius $r$ and the roller-container gap $\delta$. The investigation spans Reynolds numbers from 1 to 50 and power-law indices from 0.7 to 1.3, enabling a systematic comparison of flow modes in the central region across different geometric configurations.

\subsubsection{The gap \texorpdfstring{$\delta$}{delta} between the rollers and the container}\label{section4.3.2}

To examine how the outer region beyond the four rollers influences the central flow modes, we maintained fixed values of $h = 1$ and $r = 9$ while systematically varying $\delta$. The container diameter $L_c$ was set to 48, 50, 52, 54, 56, and 58, yielding corresponding $\delta$ values of 1.56, 2.56, 3.56, 4.56, 5.56, and 6.56.

Fig.~\ref{fig.4.3.2.1.Re=1-50_delta=1.56-6.56_duxdx_duydy_lx_ly} presents the Reynolds number dependence of (a) $\partial u_x/\partial x$, (b) $\partial u_y/\partial y$ at point $O$, and vortex lengths (c) $l_x$ and (d) $l_y$ for a Newtonian fluid across different $\delta$ values. These parameters exhibit evolution patterns consistent with our earlier observations: $\partial u_x/\partial x$ shows an initial decrease followed by an increase, while $\partial u_y/\partial y$ displays the opposite trend. Both $l_x$ and $l_y$ monotonically decrease to zero and remain there. For each $\delta$, the critical Reynolds numbers $Re_c$ derived from different criteria coincide, indicating that only the extensional flow and quadrifoliate vortex mode exist within this $Re$ range. The effect of varying $\delta$ manifests primarily in the magnitude of these parameters: at fixed $Re$, increasing $\delta$ leads to decreased $\partial{u_x}/\partial{x}$, increased $\partial{u_y}/\partial{y}$, and slightly enlarged vortex dimensions in the quadrifoliate mode.

Fig.~\ref{fig.4.3.2.2.delta=1.56-6.56_n=0.7-1.3_phase_diagram} illustrates the two transition types across different $\delta$ values and power-law indices. For each $\delta$, both transition mechanisms are observed across the range $0.7 \leq n \leq 1.3$, with Transition \uppercase\expandafter{\romannumeral1} predominating near $n = 0.7$ and Transition \uppercase\expandafter{\romannumeral2} near $n = 1.3$. Notably, the transition characteristics remain largely invariant across different $\delta$ values at fixed $n$.

\subsubsection{The roller radius \texorpdfstring{$r$}{r}}\label{section4.3.3}

To examine the influence of roller radius on flow characteristics, we investigated configurations with fixed $\delta = 2.56$ and $h = 1$ while varying $r$ from 4 to 14.

Fig.~\ref{fig.4.3.3.1.Re=1-50_r=4-14_velocity_gradient_vortex_length} presents the evolution of velocity gradients ($\partial u_x/\partial x$, $\partial u_y/\partial y$) and vortex lengths ($l_x$, $l_y$) for a Newtonian fluid across Reynolds numbers from 1 to 50. For roller radii between 5 and 12, these parameters exhibit familiar trends: $\partial u_x/\partial x$ shows an initial decrease followed by an increase, while $\partial u_y/\partial y$ displays the opposite behavior. Both vortex lengths monotonically decrease to zero. However, distinct behavior emerges at larger roller radii. At $r=13$, while $l_y$ vanishes and both velocity gradients reverse sign at $Re=50$, $l_x$ remains finite, suggesting $Re_c > 50$. This behavior mirrors Transition \uppercase\expandafter{\romannumeral2} described in Fig.~\ref{fig.4.2.10.flow_transition12_examples}. For $r=14$, both vortex lengths remain non-zero, indicating persistent quadrifoliate vortex mode throughout the examined Reynolds number range. At fixed $Re$, increasing $r$ enhances $\partial u_y/\partial y$ and both vortex lengths while reducing $\partial u_x/\partial x$. Notably, when present, the vortex dimensions scale nearly linearly with roller radius at constant $Re$.

Fig.~\ref{fig.4.3.3.2.r=4-14__n=0.7-1.3_phase_diagram} maps the flow behavior across different values of $r$ and $n$. As $Re$ increases from 1 to 50, we observe four distinct flow scenarios: two transition processes and two stable modes (simple extensional flow and quadrifoliate vortex mode without transitions). The transition processes correspond to those shown in Fig.~\ref{fig.4.2.10.flow_transition12_examples}. Cases for $n=1.3$ and $r=13,14$ are excluded due to numerical convergence issues. Notably, at $r = 4$, only simple extensional flow is observed across all power-law indices, consistent with previous findings in the literature. Clear patterns emerge in the parameter space: when $r < 7$ and $n < 1.0$, the flow tends to maintain simple extensional characteristics; conversely, near $r = 14$, the flow preferentially remains in the dumbbell-shaped quad-vortex mode. Transition \uppercase\expandafter{\romannumeral2} predominantly occurs in regions where $n > 1.0$ and $r > 7$, while Transition \uppercase\expandafter{\romannumeral1} appears in the intermediate regions between Transition \uppercase\expandafter{\romannumeral2} and the stable modes.

\section{Conclusion}\label{section5}
We have conducted systematic numerical investigations of flow characteristics in the central region of a four-roll mill containing Newtonian or power-law fluids. The simulations employed a two-relaxation-time regularized lattice Boltzmann model, representing the first successful extension of this methodology to power-law fluid dynamics with curved boundaries. After validating our numerical framework against canonical test cases, we discovered and characterized two previously unreported flow modes and their supercritical bifurcation transitions. Our parameter space investigation spans Reynolds numbers ($1 \leq Re \leq 50$), power-law indices ($0.7 \leq n \leq 1.3$), and geometric configurations. The key findings are:
\begin{enumerate}[label=(\textit{\alph*}), leftmargin=2em, labelwidth=2em, itemindent=0em, labelsep=0.5em]
  \item The central region of the four-roll mill exhibits a novel quadrifoliate vortex mode at low Reynolds numbers, characterized by four symmetrical vortices with alternating rotation directions. As Re increases, these vortices progressively shrink and eventually vanish, transitioning to simple extensional flow. This transition is accompanied by a reversal in the local deformation pattern at the stagnation point, where the flow switches from horizontal compression/vertical extension to horizontal extension/vertical compression.
  \item The transition from quadrifoliate vortex mode to simple extensional flow follows two distinct pathways depending on the power-law index. At relatively small power-law indices, a direct transition occurs (Transition \uppercase\expandafter{\romannumeral1}). In contrast, at relatively large power-law indices, the transition proceeds through an intermediate dumbbell-shaped quad-vortex mode (Transition \uppercase\expandafter{\romannumeral2}), where vortices detach from the stagnation point while maintaining their presence along the horizontal axis.
  \item These transition mechanisms can be quantitatively characterized through multiple bifurcation criteria. In Transition I, all characteristic parameters ($\partial u_x/\partial x$, $\partial u_y/\partial y$, $l_x$, and $l_y$) yield identical critical Reynolds numbers. Transition II shows a distinctive signature where $l_x$ persists beyond the critical point indicated by other parameters, reflecting the unique spatial structure of the dumbbell-shaped mode.
  \item The geometric parameters exhibit markedly different effects on flow structures. While the roller-container gap $\delta$ has minimal influence on flow transitions and vortex dimensions, the roller radius $r$ emerges as a crucial parameter. Larger $r$ values not only lead to proportionally larger vortices but can also fundamentally alter the transition pathways, with sufficiently large radii stabilizing the quadrifoliate vortex mode even at high Reynolds numbers.
  
  These findings significantly advance our understanding of flow physics in four-roll mills and provide practical guidance for achieving desired flow patterns in applications. The identification of supercritical bifurcations and distinct transition pathways offers new opportunities for flow control and optimization in extensional devices. Future work should explore the stability characteristics of these newly discovered flow modes and investigate their potential applications in material processing and rheological measurements.
\end{enumerate}

\section*{Acknowledgements.}
This research was supported in part by the High Performance Computing Platform of Xiangtan University. The authors gratefully acknowledge this computational resource that enabled the detailed numerical investigations presented in this work.

\section*{Funding.}
This work is financially supported by the National Natural Science Foundation of China (Grant Nos. 12101527, 12271464 and 11971414), the Science and Technology Innovation Program of Hunan Province (Program No. 2021RC2096), Project of Scientific Research Fund of Hunan Provincial Science and Technology Department (Grant No. 21B0159) and the Natural Science Foundation for Distinguished Young Scholars of Hunan Province (Grant No. 2023JJ10038).

\section*{Declaration of interests.} The authors report no conflict of interest.

\bibliography{aipsamp}

\providecommand{\noopsort}[1]{}\providecommand{\singleletter}[1]{#1}%
\begin{thebibliography}{43}%
\makeatletter
\providecommand \@ifxundefined [1]{%
 \@ifx{#1\undefined}
}%
\providecommand \@ifnum [1]{%
 \ifnum #1\expandafter \@firstoftwo
 \else \expandafter \@secondoftwo
 \fi
}%
\providecommand \@ifx [1]{%
 \ifx #1\expandafter \@firstoftwo
 \else \expandafter \@secondoftwo
 \fi
}%
\providecommand \natexlab [1]{#1}%
\providecommand \enquote  [1]{``#1''}%
\providecommand \bibnamefont  [1]{#1}%
\providecommand \bibfnamefont [1]{#1}%
\providecommand \citenamefont [1]{#1}%
\providecommand \href@noop [0]{\@secondoftwo}%
\providecommand \href [0]{\begingroup \@sanitize@url \@href}%
\providecommand \@href[1]{\@@startlink{#1}\@@href}%
\providecommand \@@href[1]{\endgroup#1\@@endlink}%
\providecommand \@sanitize@url [0]{\catcode `\\12\catcode `\$12\catcode
  `\&12\catcode `\#12\catcode `\^12\catcode `\_12\catcode `\%12\relax}%
\providecommand \@@startlink[1]{}%
\providecommand \@@endlink[0]{}%
\providecommand \url  [0]{\begingroup\@sanitize@url \@url }%
\providecommand \@url [1]{\endgroup\@href {#1}{\urlprefix }}%
\providecommand \urlprefix  [0]{URL }%
\providecommand \Eprint [0]{\href }%
\providecommand \doibase [0]{http://dx.doi.org/}%
\providecommand \selectlanguage [0]{\@gobble}%
\providecommand \bibinfo  [0]{\@secondoftwo}%
\providecommand \bibfield  [0]{\@secondoftwo}%
\providecommand \translation [1]{[#1]}%
\providecommand \BibitemOpen [0]{}%
\providecommand \bibitemStop [0]{}%
\providecommand \bibitemNoStop [0]{.\EOS\space}%
\providecommand \EOS [0]{\spacefactor3000\relax}%
\providecommand \BibitemShut  [1]{\csname bibitem#1\endcsname}%
\let\auto@bib@innerbib\@empty
\bibitem [{\citenamefont {Akbaridoust}, \citenamefont {Philip},\ and\
  \citenamefont {Marusic}(2018)}]{akbaridoustAssessmentMiniatureFourroll2018}%
  \BibitemOpen
  \bibfield  {author} {\bibinfo {author} {\bibnamefont {Akbaridoust},
  \bibfnamefont {F.}}, \bibinfo {author} {\bibnamefont {Philip}, \bibfnamefont
  {J.}}, \ and\ \bibinfo {author} {\bibnamefont {Marusic}, \bibfnamefont
  {I.}},\ }\bibfield  {title} {\enquote {\bibinfo {title} {Assessment of a
  miniature four-roll mill and a cross-slot microchannel for high-strain-rate
  stagnation point flows},}\ }\href {\doibase 10.1088/1361-6501/aaacf3}
  {\bibfield  {journal} {\bibinfo  {journal} {Measurement Science and
  Technology}\ }\textbf {\bibinfo {volume} {29}},\ \bibinfo {pages} {045302}
  (\bibinfo {year} {2018})}\BibitemShut {NoStop}%
\bibitem [{\citenamefont {Andreotti}, \citenamefont {Douady},\ and\
  \citenamefont {Couder}(2001)}]{andreottiExperimentTwoAspects2001}%
  \BibitemOpen
  \bibfield  {author} {\bibinfo {author} {\bibnamefont {Andreotti},
  \bibfnamefont {B.}}, \bibinfo {author} {\bibnamefont {Douady}, \bibfnamefont
  {S.}}, \ and\ \bibinfo {author} {\bibnamefont {Couder}, \bibfnamefont {Y.}},\
  }\bibfield  {title} {\enquote {\bibinfo {title} {An experiment on two aspects
  of the interaction between strain and vorticity},}\ }\href {\doibase
  10.1017/S0022112001005353} {\bibfield  {journal} {\bibinfo  {journal}
  {Journal of Fluid Mechanics}\ }\textbf {\bibinfo {volume} {444}},\ \bibinfo
  {pages} {151--174} (\bibinfo {year} {2001})}\BibitemShut {NoStop}%
\bibitem [{\citenamefont {Arratia}\ \emph {et~al.}(2006)\citenamefont
  {Arratia}, \citenamefont {Thomas}, \citenamefont {Diorio},\ and\
  \citenamefont {Gollub}}]{PhysRevLett.96.144502}%
  \BibitemOpen
  \bibfield  {author} {\bibinfo {author} {\bibnamefont {Arratia}, \bibfnamefont
  {P.~E.}}, \bibinfo {author} {\bibnamefont {Thomas}, \bibfnamefont {C.~C.}},
  \bibinfo {author} {\bibnamefont {Diorio}, \bibfnamefont {J.}}, \ and\
  \bibinfo {author} {\bibnamefont {Gollub}, \bibfnamefont {J.~P.}},\ }\bibfield
   {title} {\enquote {\bibinfo {title} {Elastic instabilities of polymer
  solutions in cross-channel flow},}\ }\href {\doibase
  10.1103/PhysRevLett.96.144502} {\bibfield  {journal} {\bibinfo  {journal}
  {Phys. Rev. Lett.}\ }\textbf {\bibinfo {volume} {96}},\ \bibinfo {pages}
  {144502} (\bibinfo {year} {2006})}\BibitemShut {NoStop}%
\bibitem [{\citenamefont {Balasuriya}\ and\ \citenamefont
  {{Padberg-Gehle}}(2013)}]{balasuriyaControllingUnsteadyAnalogue2013}%
  \BibitemOpen
  \bibfield  {author} {\bibinfo {author} {\bibnamefont {Balasuriya},
  \bibfnamefont {S.}}\ and\ \bibinfo {author} {\bibnamefont {{Padberg-Gehle}},
  \bibfnamefont {K.}},\ }\bibfield  {title} {\enquote {\bibinfo {title}
  {Controlling the {{Unsteady Analogue}} of {{Saddle Stagnation Points}}},}\
  }\href {\doibase 10.1137/120886042} {\bibfield  {journal} {\bibinfo
  {journal} {SIAM Journal on Applied Mathematics}\ }\textbf {\bibinfo {volume}
  {73}},\ \bibinfo {pages} {1038--1057} (\bibinfo {year} {2013})}\BibitemShut
  {NoStop}%
\bibitem [{\citenamefont {Bentley}(1985)}]{Bentley1985DropDA}%
  \BibitemOpen
  \bibfield  {author} {\bibinfo {author} {\bibnamefont {Bentley}, \bibfnamefont
  {B.}},\ }\emph {\bibinfo {title} {Drop Deformation and Burst in
  Two-Dimensional Flows}},\ \href@noop {} {Ph.D. thesis},\ \bibinfo  {school}
  {California Institute of Technology} (\bibinfo {year} {1985})\BibitemShut
  {NoStop}%
\bibitem [{\citenamefont {Bentley}\ and\ \citenamefont
  {Leal}(1986)}]{bentleyComputercontrolledFourrollMill1986}%
  \BibitemOpen
  \bibfield  {author} {\bibinfo {author} {\bibnamefont {Bentley}, \bibfnamefont
  {B.~J.}}\ and\ \bibinfo {author} {\bibnamefont {Leal}, \bibfnamefont
  {L.~G.}},\ }\bibfield  {title} {\enquote {\bibinfo {title} {A
  computer-controlled four-roll mill for investigations of particle and drop
  dynamics in two-dimensional linear shear flows},}\ }\href {\doibase
  10.1017/S002211208600280X} {\bibfield  {journal} {\bibinfo  {journal}
  {Journal of Fluid Mechanics}\ }\textbf {\bibinfo {volume} {167}},\ \bibinfo
  {pages} {219} (\bibinfo {year} {1986})}\BibitemShut {NoStop}%
\bibitem [{\citenamefont {Bryngelson}\ and\ \citenamefont
  {Freund}(2019)}]{BRYNGELSON2019171}%
  \BibitemOpen
  \bibfield  {author} {\bibinfo {author} {\bibnamefont {Bryngelson},
  \bibfnamefont {S.~H.}}\ and\ \bibinfo {author} {\bibnamefont {Freund},
  \bibfnamefont {J.~B.}},\ }\bibfield  {title} {\enquote {\bibinfo {title}
  {Non-modal floquet stability of capsules in large-amplitude oscillatory
  extensional flow},}\ }\href {\doibase
  https://doi.org/10.1016/j.euromechflu.2019.04.012} {\bibfield  {journal}
  {\bibinfo  {journal} {European Journal of Mechanics - B/Fluids}\ }\textbf
  {\bibinfo {volume} {77}},\ \bibinfo {pages} {171--176} (\bibinfo {year}
  {2019})}\BibitemShut {NoStop}%
\bibitem [{\citenamefont {Crowley}\ \emph {et~al.}(1976)\citenamefont
  {Crowley}, \citenamefont {Frank}, \citenamefont {Mackley},\ and\
  \citenamefont {Stephenson}}]{crowley_localized_1976}%
  \BibitemOpen
  \bibfield  {author} {\bibinfo {author} {\bibnamefont {Crowley}, \bibfnamefont
  {D.~G.}}, \bibinfo {author} {\bibnamefont {Frank}, \bibfnamefont {F.~C.}},
  \bibinfo {author} {\bibnamefont {Mackley}, \bibfnamefont {M.~R.}}, \ and\
  \bibinfo {author} {\bibnamefont {Stephenson}, \bibfnamefont {R.~G.}},\
  }\bibfield  {title} {\enquote {\bibinfo {title} {Localized flow birefringence
  of polyethylene oxide solutions in a four roll mill},}\ }\href {\doibase
  10.1002/pol.1976.180140613} {\bibfield  {journal} {\bibinfo  {journal}
  {Journal of Polymer Science: Polymer Physics Edition}\ }\textbf {\bibinfo
  {volume} {14}},\ \bibinfo {pages} {1111--1119} (\bibinfo {year}
  {1976})}\BibitemShut {NoStop}%
\bibitem [{\citenamefont {D'Avino}\ and\ \citenamefont
  {Maffettone}(2017)}]{davinoNumericalSimulationsDispersive2017}%
  \BibitemOpen
  \bibfield  {author} {\bibinfo {author} {\bibnamefont {D'Avino}, \bibfnamefont
  {G.}}\ and\ \bibinfo {author} {\bibnamefont {Maffettone}, \bibfnamefont
  {P.~L.}},\ }\bibfield  {title} {\enquote {\bibinfo {title} {Numerical
  simulations of dispersive mixing of viscoelastic suspensions in a four-roll
  mill},}\ }\href {\doibase 10.1007/s00397-017-1028-x} {\bibfield  {journal}
  {\bibinfo  {journal} {Rheologica Acta}\ }\textbf {\bibinfo {volume} {56}},\
  \bibinfo {pages} {695--706} (\bibinfo {year} {2017})}\BibitemShut {NoStop}%
\bibitem [{\citenamefont {Dunlap}\ and\ \citenamefont
  {Leal}(1987)}]{dunlap_dilute_1987}%
  \BibitemOpen
  \bibfield  {author} {\bibinfo {author} {\bibnamefont {Dunlap}, \bibfnamefont
  {P.}}\ and\ \bibinfo {author} {\bibnamefont {Leal}, \bibfnamefont {L.}},\
  }\bibfield  {title} {\enquote {\bibinfo {title} {Dilute polystyrene solutions
  in extensional flows: {Birefringence} and flow modification},}\ }\href
  {\doibase 10.1016/0377-0257(87)80009-5} {\bibfield  {journal} {\bibinfo
  {journal} {Journal of Non-Newtonian Fluid Mechanics}\ }\textbf {\bibinfo
  {volume} {23}},\ \bibinfo {pages} {5--48} (\bibinfo {year}
  {1987})}\BibitemShut {NoStop}%
\bibitem [{\citenamefont {Feng}\ and\ \citenamefont
  {Leal}(1997)}]{FENG1997187}%
  \BibitemOpen
  \bibfield  {author} {\bibinfo {author} {\bibnamefont {Feng}, \bibfnamefont
  {J.}}\ and\ \bibinfo {author} {\bibnamefont {Leal}, \bibfnamefont {L.}},\
  }\bibfield  {title} {\enquote {\bibinfo {title} {Numerical simulations of the
  flow of dilute polymer solutions in a four-roll mill},}\ }\href {\doibase
  https://doi.org/10.1016/S0377-0257(97)00027-X} {\bibfield  {journal}
  {\bibinfo  {journal} {Journal of Non-Newtonian Fluid Mechanics}\ }\textbf
  {\bibinfo {volume} {72}},\ \bibinfo {pages} {187--218} (\bibinfo {year}
  {1997})}\BibitemShut {NoStop}%
\bibitem [{\citenamefont {Fuller}\ and\ \citenamefont
  {Leal}(1980)}]{fuller_flow_1980}%
  \BibitemOpen
  \bibfield  {author} {\bibinfo {author} {\bibnamefont {Fuller}, \bibfnamefont
  {G.~G.}}\ and\ \bibinfo {author} {\bibnamefont {Leal}, \bibfnamefont
  {L.~G.}},\ }\bibfield  {title} {\enquote {\bibinfo {title} {Flow
  birefringence of dilute polymer solutions in two-dimensional flows},}\ }\href
  {\doibase 10.1007/BF01517512} {\bibfield  {journal} {\bibinfo  {journal}
  {Rheologica Acta}\ }\textbf {\bibinfo {volume} {19}},\ \bibinfo {pages}
  {580--600} (\bibinfo {year} {1980})}\BibitemShut {NoStop}%
\bibitem [{\citenamefont {Fuller}\ and\ \citenamefont
  {Leal}(1981)}]{fuller_flow_1981}%
  \BibitemOpen
  \bibfield  {author} {\bibinfo {author} {\bibnamefont {Fuller}, \bibfnamefont
  {G.~G.}}\ and\ \bibinfo {author} {\bibnamefont {Leal}, \bibfnamefont
  {L.~G.}},\ }\bibfield  {title} {\enquote {\bibinfo {title} {Flow
  birefringence of concentrated polymer solutions in two‐dimensional
  flows},}\ }\href {\doibase 10.1002/pol.1981.180190402} {\bibfield  {journal}
  {\bibinfo  {journal} {Journal of Polymer Science: Polymer Physics Edition}\
  }\textbf {\bibinfo {volume} {19}},\ \bibinfo {pages} {557--587} (\bibinfo
  {year} {1981})}\BibitemShut {NoStop}%
\bibitem [{\citenamefont {Ginzburg}, \citenamefont {Verhaeghe},\ and\
  \citenamefont {d'Humi{\`e}res}(2008)}]{Ginzburg2008TworelaxationtimeLB}%
  \BibitemOpen
  \bibfield  {author} {\bibinfo {author} {\bibnamefont {Ginzburg},
  \bibfnamefont {I.}}, \bibinfo {author} {\bibnamefont {Verhaeghe},
  \bibfnamefont {F.}}, \ and\ \bibinfo {author} {\bibnamefont {d'Humi{\`e}res},
  \bibfnamefont {D.}},\ }\bibfield  {title} {\enquote {\bibinfo {title}
  {Two-relaxation-time lattice boltzmann scheme: About parametrization,
  velocity, pressure and mixed boundary conditions},}\ }\href
  {https://api.semanticscholar.org/CorpusID:123717273} {\bibfield  {journal}
  {\bibinfo  {journal} {Communications in Computational Physics}\ }\textbf
  {\bibinfo {volume} {3}},\ \bibinfo {pages} {427--478} (\bibinfo {year}
  {2008})}\BibitemShut {NoStop}%
\bibitem [{\citenamefont {Ha}\ and\ \citenamefont
  {Leal}(2001)}]{haExperimentalStudyDrop2001}%
  \BibitemOpen
  \bibfield  {author} {\bibinfo {author} {\bibnamefont {Ha}, \bibfnamefont
  {J.-W.}}\ and\ \bibinfo {author} {\bibnamefont {Leal}, \bibfnamefont
  {L.~G.}},\ }\bibfield  {title} {\enquote {\bibinfo {title} {An experimental
  study of drop deformation and breakup in extensional flow at high capillary
  number},}\ }\href {\doibase 10.1063/1.1358306} {\bibfield  {journal}
  {\bibinfo  {journal} {Physics of Fluids}\ }\textbf {\bibinfo {volume} {13}},\
  \bibinfo {pages} {1568--1576} (\bibinfo {year} {2001})}\BibitemShut {NoStop}%
\bibitem [{\citenamefont {Haward}, \citenamefont {McKinley},\ and\
  \citenamefont {Shen}(2016)}]{hawardElasticInstabilitiesPlanar2016a}%
  \BibitemOpen
  \bibfield  {author} {\bibinfo {author} {\bibnamefont {Haward}, \bibfnamefont
  {S.~J.}}, \bibinfo {author} {\bibnamefont {McKinley}, \bibfnamefont {G.~H.}},
  \ and\ \bibinfo {author} {\bibnamefont {Shen}, \bibfnamefont {A.~Q.}},\
  }\bibfield  {title} {\enquote {\bibinfo {title} {Elastic instabilities in
  planar elongational flow of monodisperse polymer solutions},}\ }\href
  {\doibase 10.1038/srep33029} {\bibfield  {journal} {\bibinfo  {journal}
  {Scientific Reports}\ }\textbf {\bibinfo {volume} {6}},\ \bibinfo {pages}
  {33029} (\bibinfo {year} {2016})}\BibitemShut {NoStop}%
\bibitem [{\citenamefont {Haward}\ \emph
  {et~al.}(2012{\natexlab{a}})\citenamefont {Haward}, \citenamefont {Ober},
  \citenamefont {Oliveira}, \citenamefont {Alves},\ and\ \citenamefont
  {McKinley}}]{Haward2012ExtensionalRA}%
  \BibitemOpen
  \bibfield  {author} {\bibinfo {author} {\bibnamefont {Haward}, \bibfnamefont
  {S.~J.}}, \bibinfo {author} {\bibnamefont {Ober}, \bibfnamefont {T.~J.}},
  \bibinfo {author} {\bibnamefont {Oliveira}, \bibfnamefont {M.~S.~A.}},
  \bibinfo {author} {\bibnamefont {Alves}, \bibfnamefont {M.~A.}}, \ and\
  \bibinfo {author} {\bibnamefont {McKinley}, \bibfnamefont {G.~H.}},\
  }\bibfield  {title} {\enquote {\bibinfo {title} {Extensional rheology and
  elastic instabilities of a wormlike micellar solution in a microfluidic
  cross-slot device},}\ }\href
  {https://api.semanticscholar.org/CorpusID:18081127} {\bibfield  {journal}
  {\bibinfo  {journal} {Soft Matter}\ }\textbf {\bibinfo {volume} {8}},\
  \bibinfo {pages} {536--555} (\bibinfo {year}
  {2012}{\natexlab{a}})}\BibitemShut {NoStop}%
\bibitem [{\citenamefont {Haward}\ \emph
  {et~al.}(2012{\natexlab{b}})\citenamefont {Haward}, \citenamefont {Oliveira},
  \citenamefont {Alves},\ and\ \citenamefont
  {McKinley}}]{Haward2012OptimizedCF}%
  \BibitemOpen
  \bibfield  {author} {\bibinfo {author} {\bibnamefont {Haward}, \bibfnamefont
  {S.~J.}}, \bibinfo {author} {\bibnamefont {Oliveira}, \bibfnamefont
  {M.~S.~A.}}, \bibinfo {author} {\bibnamefont {Alves}, \bibfnamefont {M.~A.}},
  \ and\ \bibinfo {author} {\bibnamefont {McKinley}, \bibfnamefont {G.~H.}},\
  }\bibfield  {title} {\enquote {\bibinfo {title} {Optimized cross-slot flow
  geometry for microfluidic extensional rheometry.}}\ }\href
  {https://api.semanticscholar.org/CorpusID:2096605} {\bibfield  {journal}
  {\bibinfo  {journal} {Physical review letters}\ }\textbf {\bibinfo {volume}
  {109 12}},\ \bibinfo {pages} {128301} (\bibinfo {year}
  {2012}{\natexlab{b}})}\BibitemShut {NoStop}%
\bibitem [{\citenamefont {Higdon}(1993)}]{higdonKinematicsFourrollMill1993}%
  \BibitemOpen
  \bibfield  {author} {\bibinfo {author} {\bibnamefont {Higdon}, \bibfnamefont
  {J.~J.~L.}},\ }\bibfield  {title} {\enquote {\bibinfo {title} {The kinematics
  of the four-roll mill},}\ }\href {\doibase 10.1063/1.858782} {\bibfield
  {journal} {\bibinfo  {journal} {Physics of Fluids A: Fluid Dynamics}\
  }\textbf {\bibinfo {volume} {5}},\ \bibinfo {pages} {274--276} (\bibinfo
  {year} {1993})}\BibitemShut {NoStop}%
\bibitem [{\citenamefont {Hsiao}\ \emph {et~al.}(2016)\citenamefont {Hsiao},
  \citenamefont {Sasmal}, \citenamefont {Ravi~Prakash},\ and\ \citenamefont
  {Schroeder}}]{10.1122/1.4972236}%
  \BibitemOpen
  \bibfield  {author} {\bibinfo {author} {\bibnamefont {Hsiao}, \bibfnamefont
  {K.-W.}}, \bibinfo {author} {\bibnamefont {Sasmal}, \bibfnamefont {C.}},
  \bibinfo {author} {\bibnamefont {Ravi~Prakash}, \bibfnamefont {J.}}, \ and\
  \bibinfo {author} {\bibnamefont {Schroeder}, \bibfnamefont {C.~M.}},\
  }\bibfield  {title} {\enquote {\bibinfo {title} {{Direct observation of DNA
  dynamics in semidilute solutions in extensional flow}},}\ }\href {\doibase
  10.1122/1.4972236} {\bibfield  {journal} {\bibinfo  {journal} {Journal of
  Rheology}\ }\textbf {\bibinfo {volume} {61}},\ \bibinfo {pages} {151--167}
  (\bibinfo {year} {2016})}\BibitemShut {NoStop}%
\bibitem [{\citenamefont {Hymel}, \citenamefont {Lan},\ and\ \citenamefont
  {Khismatullin}(2020)}]{HYMEL2020493}%
  \BibitemOpen
  \bibfield  {author} {\bibinfo {author} {\bibnamefont {Hymel}, \bibfnamefont
  {S.~J.}}, \bibinfo {author} {\bibnamefont {Lan}, \bibfnamefont {H.}}, \ and\
  \bibinfo {author} {\bibnamefont {Khismatullin}, \bibfnamefont {D.~B.}},\
  }\bibfield  {title} {\enquote {\bibinfo {title} {Elongation index as a
  sensitive measure of cell deformation in high-throughput microfluidic
  systems},}\ }\href {\doibase https://doi.org/10.1016/j.bpj.2020.06.027}
  {\bibfield  {journal} {\bibinfo  {journal} {Biophysical Journal}\ }\textbf
  {\bibinfo {volume} {119}},\ \bibinfo {pages} {493--501} (\bibinfo {year}
  {2020})}\BibitemShut {NoStop}%
\bibitem [{\citenamefont {Kumar}\ \emph {et~al.}(2019)\citenamefont {Kumar},
  \citenamefont {Walkama}, \citenamefont {Guasto},\ and\ \citenamefont
  {Ardekani}}]{PhysRevE.100.063107}%
  \BibitemOpen
  \bibfield  {author} {\bibinfo {author} {\bibnamefont {Kumar}, \bibfnamefont
  {M.}}, \bibinfo {author} {\bibnamefont {Walkama}, \bibfnamefont {D.~M.}},
  \bibinfo {author} {\bibnamefont {Guasto}, \bibfnamefont {J.~S.}}, \ and\
  \bibinfo {author} {\bibnamefont {Ardekani}, \bibfnamefont {A.~M.}},\
  }\bibfield  {title} {\enquote {\bibinfo {title} {Flow-induced buckling
  dynamics of sperm flagella},}\ }\href {\doibase 10.1103/PhysRevE.100.063107}
  {\bibfield  {journal} {\bibinfo  {journal} {Phys. Rev. E}\ }\textbf {\bibinfo
  {volume} {100}},\ \bibinfo {pages} {063107} (\bibinfo {year}
  {2019})}\BibitemShut {NoStop}%
\bibitem [{\citenamefont {Lagnado}\ and\ \citenamefont
  {Leal}(1990)}]{lagnadoVisualizationThreedimensionalFlow1990}%
  \BibitemOpen
  \bibfield  {author} {\bibinfo {author} {\bibnamefont {Lagnado}, \bibfnamefont
  {R.~R.}}\ and\ \bibinfo {author} {\bibnamefont {Leal}, \bibfnamefont
  {L.~G.}},\ }\bibfield  {title} {\enquote {\bibinfo {title} {Visualization of
  three-dimensional flow in a four-roll mill},}\ }\href {\doibase
  10.1007/BF00575332} {\bibfield  {journal} {\bibinfo  {journal} {Experiments
  in Fluids}\ }\textbf {\bibinfo {volume} {9}},\ \bibinfo {pages} {25--32}
  (\bibinfo {year} {1990})}\BibitemShut {NoStop}%
\bibitem [{\citenamefont {Lee}\ and\ \citenamefont
  {Muller}(1999)}]{leeFlowLightScattering1999}%
  \BibitemOpen
  \bibfield  {author} {\bibinfo {author} {\bibnamefont {Lee}, \bibfnamefont
  {E.~C.}}\ and\ \bibinfo {author} {\bibnamefont {Muller}, \bibfnamefont
  {S.~J.}},\ }\bibfield  {title} {\enquote {\bibinfo {title} {Flow {{Light
  Scattering Studies}} of {{Polymer Coil Conformation}} in {{Solutions}} in
  {{Extensional Flow}}},}\ }\href {\doibase 10.1021/ma981277a} {\bibfield
  {journal} {\bibinfo  {journal} {Macromolecules}\ }\textbf {\bibinfo {volume}
  {32}},\ \bibinfo {pages} {3295--3305} (\bibinfo {year} {1999})}\BibitemShut
  {NoStop}%
\bibitem [{\citenamefont {Milliken}\ and\ \citenamefont
  {Leal}(1991)}]{millikenDeformationBreakupViscoelastic1991}%
  \BibitemOpen
  \bibfield  {author} {\bibinfo {author} {\bibnamefont {Milliken},
  \bibfnamefont {W.}}\ and\ \bibinfo {author} {\bibnamefont {Leal},
  \bibfnamefont {L.}},\ }\bibfield  {title} {\enquote {\bibinfo {title}
  {Deformation and breakup of viscoelastic drops in planar extensional
  flows},}\ }\href {\doibase 10.1016/0377-0257(91)87018-S} {\bibfield
  {journal} {\bibinfo  {journal} {Journal of Non-Newtonian Fluid Mechanics}\
  }\textbf {\bibinfo {volume} {40}},\ \bibinfo {pages} {355--379} (\bibinfo
  {year} {1991})}\BibitemShut {NoStop}%
\bibitem [{\citenamefont {Narayan}\ \emph {et~al.}(2020)\citenamefont
  {Narayan}, \citenamefont {Moravec}, \citenamefont {Dallas},\ and\
  \citenamefont {Dutcher}}]{PhysRevFluids.5.113603}%
  \BibitemOpen
  \bibfield  {author} {\bibinfo {author} {\bibnamefont {Narayan}, \bibfnamefont
  {S.}}, \bibinfo {author} {\bibnamefont {Moravec}, \bibfnamefont {D.~B.}},
  \bibinfo {author} {\bibnamefont {Dallas}, \bibfnamefont {A.~J.}}, \ and\
  \bibinfo {author} {\bibnamefont {Dutcher}, \bibfnamefont {C.~S.}},\
  }\bibfield  {title} {\enquote {\bibinfo {title} {Droplet shape relaxation in
  a four-channel microfluidic hydrodynamic trap},}\ }\href {\doibase
  10.1103/PhysRevFluids.5.113603} {\bibfield  {journal} {\bibinfo  {journal}
  {Phys. Rev. Fluids}\ }\textbf {\bibinfo {volume} {5}},\ \bibinfo {pages}
  {113603} (\bibinfo {year} {2020})}\BibitemShut {NoStop}%
\bibitem [{\citenamefont {Pontrelli}, \citenamefont {Ubertini},\ and\
  \citenamefont {Succi}(2009)}]{pontrelli_unstructured_2009}%
  \BibitemOpen
  \bibfield  {author} {\bibinfo {author} {\bibnamefont {Pontrelli},
  \bibfnamefont {G.}}, \bibinfo {author} {\bibnamefont {Ubertini},
  \bibfnamefont {S.}}, \ and\ \bibinfo {author} {\bibnamefont {Succi},
  \bibfnamefont {S.}},\ }\bibfield  {title} {\enquote {\bibinfo {title} {The
  unstructured lattice {Boltzmann} method for non-{Newtonian} flows},}\ }\href
  {\doibase 10.1088/1742-5468/2009/06/P06005} {\bibfield  {journal} {\bibinfo
  {journal} {Journal of Statistical Mechanics: Theory and Experiment}\ }\textbf
  {\bibinfo {volume} {2009}},\ \bibinfo {pages} {P06005} (\bibinfo {year}
  {2009})}\BibitemShut {NoStop}%
\bibitem [{\citenamefont {Poole}, \citenamefont {Alves},\ and\ \citenamefont
  {Oliveira}(2007)}]{PhysRevLett.99.164503}%
  \BibitemOpen
  \bibfield  {author} {\bibinfo {author} {\bibnamefont {Poole}, \bibfnamefont
  {R.~J.}}, \bibinfo {author} {\bibnamefont {Alves}, \bibfnamefont {M.~A.}}, \
  and\ \bibinfo {author} {\bibnamefont {Oliveira}, \bibfnamefont {P.~J.}},\
  }\bibfield  {title} {\enquote {\bibinfo {title} {Purely elastic flow
  asymmetries},}\ }\href {\doibase 10.1103/PhysRevLett.99.164503} {\bibfield
  {journal} {\bibinfo  {journal} {Phys. Rev. Lett.}\ }\textbf {\bibinfo
  {volume} {99}},\ \bibinfo {pages} {164503} (\bibinfo {year}
  {2007})}\BibitemShut {NoStop}%
\bibitem [{\citenamefont {Reis}(2020)}]{reis_lattice_2020}%
  \BibitemOpen
  \bibfield  {author} {\bibinfo {author} {\bibnamefont {Reis}, \bibfnamefont
  {T.}},\ }\bibfield  {title} {\enquote {\bibinfo {title} {On the {Lattice}
  {Boltzmann} {Deviatoric} {Stress}: {Analysis}, {Boundary} {Conditions}, and
  {Optimal} {Relaxation} {Times}},}\ }\href {\doibase 10.1137/19M1244846}
  {\bibfield  {journal} {\bibinfo  {journal} {SIAM Journal on Scientific
  Computing}\ }\textbf {\bibinfo {volume} {42}},\ \bibinfo {pages} {B397--B424}
  (\bibinfo {year} {2020})}\BibitemShut {NoStop}%
\bibitem [{\citenamefont {Sasmal}\ \emph {et~al.}(2017)\citenamefont {Sasmal},
  \citenamefont {Hsiao}, \citenamefont {Schroeder},\ and\ \citenamefont
  {Ravi~Prakash}}]{10.1122/1.4972237}%
  \BibitemOpen
  \bibfield  {author} {\bibinfo {author} {\bibnamefont {Sasmal}, \bibfnamefont
  {C.}}, \bibinfo {author} {\bibnamefont {Hsiao}, \bibfnamefont {K.-W.}},
  \bibinfo {author} {\bibnamefont {Schroeder}, \bibfnamefont {C.~M.}}, \ and\
  \bibinfo {author} {\bibnamefont {Ravi~Prakash}, \bibfnamefont {J.}},\
  }\bibfield  {title} {\enquote {\bibinfo {title} {{Parameter-free prediction
  of DNA dynamics in planar extensional flow of semidilute solutions}},}\
  }\href {\doibase 10.1122/1.4972237} {\bibfield  {journal} {\bibinfo
  {journal} {Journal of Rheology}\ }\textbf {\bibinfo {volume} {61}},\ \bibinfo
  {pages} {169--186} (\bibinfo {year} {2017})}\BibitemShut {NoStop}%
\bibitem [{\citenamefont {Schroeder}\ \emph
  {et~al.}(2003{\natexlab{a}})\citenamefont {Schroeder}, \citenamefont
  {Babcock}, \citenamefont {Shaqfeh},\ and\ \citenamefont
  {Chu}}]{10.1126/science.1086070}%
  \BibitemOpen
  \bibfield  {author} {\bibinfo {author} {\bibnamefont {Schroeder},
  \bibfnamefont {C.~M.}}, \bibinfo {author} {\bibnamefont {Babcock},
  \bibfnamefont {H.~P.}}, \bibinfo {author} {\bibnamefont {Shaqfeh},
  \bibfnamefont {E.~S.~G.}}, \ and\ \bibinfo {author} {\bibnamefont {Chu},
  \bibfnamefont {S.}},\ }\bibfield  {title} {\enquote {\bibinfo {title}
  {Observation of polymer conformation hysteresis in extensional flow},}\
  }\href {\doibase 10.1126/science.1086070} {\bibfield  {journal} {\bibinfo
  {journal} {Science}\ }\textbf {\bibinfo {volume} {301}},\ \bibinfo {pages}
  {1515--1519} (\bibinfo {year} {2003}{\natexlab{a}})}\BibitemShut {NoStop}%
\bibitem [{\citenamefont {Schroeder}\ \emph
  {et~al.}(2003{\natexlab{b}})\citenamefont {Schroeder}, \citenamefont
  {Babcock}, \citenamefont {Shaqfeh},\ and\ \citenamefont
  {Chu}}]{schroederObservationPolymerConformation2003}%
  \BibitemOpen
  \bibfield  {author} {\bibinfo {author} {\bibnamefont {Schroeder},
  \bibfnamefont {C.~M.}}, \bibinfo {author} {\bibnamefont {Babcock},
  \bibfnamefont {H.~P.}}, \bibinfo {author} {\bibnamefont {Shaqfeh},
  \bibfnamefont {E.~S.~G.}}, \ and\ \bibinfo {author} {\bibnamefont {Chu},
  \bibfnamefont {S.}},\ }\bibfield  {title} {\enquote {\bibinfo {title}
  {Observation of {{Polymer Conformation Hysteresis}} in {{Extensional
  Flow}}},}\ }\href {\doibase 10.1126/science.1086070} {\bibfield  {journal}
  {\bibinfo  {journal} {Science}\ }\textbf {\bibinfo {volume} {301}},\ \bibinfo
  {pages} {1515--1519} (\bibinfo {year} {2003}{\natexlab{b}})}\BibitemShut
  {NoStop}%
\bibitem [{\citenamefont {Tao}\ \emph {et~al.}(2018)\citenamefont {Tao},
  \citenamefont {He}, \citenamefont {Chen}, \citenamefont {Yang},\ and\
  \citenamefont {Huang}}]{tao_one-point_2018}%
  \BibitemOpen
  \bibfield  {author} {\bibinfo {author} {\bibnamefont {Tao}, \bibfnamefont
  {S.}}, \bibinfo {author} {\bibnamefont {He}, \bibfnamefont {Q.}}, \bibinfo
  {author} {\bibnamefont {Chen}, \bibfnamefont {B.}}, \bibinfo {author}
  {\bibnamefont {Yang}, \bibfnamefont {X.}}, \ and\ \bibinfo {author}
  {\bibnamefont {Huang}, \bibfnamefont {S.}},\ }\bibfield  {title} {\enquote
  {\bibinfo {title} {One-point second-order curved boundary condition for
  lattice {Boltzmann} simulation of suspended particles},}\ }\href {\doibase
  10.1016/j.camwa.2018.07.013} {\bibfield  {journal} {\bibinfo  {journal}
  {Computers \& Mathematics with Applications}\ }\textbf {\bibinfo {volume}
  {76}},\ \bibinfo {pages} {1593--1607} (\bibinfo {year} {2018})}\BibitemShut
  {NoStop}%
\bibitem [{\citenamefont {Taylor}(1934)}]{taylor_formation_1934}%
  \BibitemOpen
  \bibfield  {author} {\bibinfo {author} {\bibnamefont {Taylor}, \bibfnamefont
  {G.~I.}},\ }\bibfield  {title} {\enquote {\bibinfo {title} {The formation of
  emulsions in definable fields of flow},}\ }\href {\doibase
  https://doi.org/10.1098/rspa.1934.0169} {\bibfield  {journal} {\bibinfo
  {journal} {Proceedings of the Royal Society of London. Series A, Containing
  Papers of a Mathematical and Physical Character}\ ,\ \bibinfo {pages}
  {146:501--523}} (\bibinfo {year} {1934})}\BibitemShut {NoStop}%
\bibitem [{\citenamefont {Tretheway}\ and\ \citenamefont
  {Leal}(2001)}]{trethewayDeformationRelaxationNewtonian2001}%
  \BibitemOpen
  \bibfield  {author} {\bibinfo {author} {\bibnamefont {Tretheway},
  \bibfnamefont {D.~C.}}\ and\ \bibinfo {author} {\bibnamefont {Leal},
  \bibfnamefont {L.}},\ }\bibfield  {title} {\enquote {\bibinfo {title}
  {Deformation and relaxation of {{Newtonian}} drops in planar extensional
  flows of a {{Boger}} fluid},}\ }\href {\doibase
  10.1016/S0377-0257(01)00123-9} {\bibfield  {journal} {\bibinfo  {journal}
  {Journal of Non-Newtonian Fluid Mechanics}\ }\textbf {\bibinfo {volume}
  {99}},\ \bibinfo {pages} {81--108} (\bibinfo {year} {2001})}\BibitemShut
  {NoStop}%
\bibitem [{\citenamefont {Tu}\ \emph {et~al.}(2023)\citenamefont {Tu},
  \citenamefont {Nguyen}, \citenamefont {Foley}, \citenamefont {Jacobs},\ and\
  \citenamefont {Schroeder}}]{10.1122/8.0000600}%
  \BibitemOpen
  \bibfield  {author} {\bibinfo {author} {\bibnamefont {Tu}, \bibfnamefont
  {M.~Q.}}, \bibinfo {author} {\bibnamefont {Nguyen}, \bibfnamefont {H.~V.}},
  \bibinfo {author} {\bibnamefont {Foley}, \bibfnamefont {E.}}, \bibinfo
  {author} {\bibnamefont {Jacobs}, \bibfnamefont {M.~I.}}, \ and\ \bibinfo
  {author} {\bibnamefont {Schroeder}, \bibfnamefont {C.~M.}},\ }\bibfield
  {title} {\enquote {\bibinfo {title} {{3D manipulation and dynamics of soft
  materials in 3D flows}},}\ }\href {\doibase 10.1122/8.0000600} {\bibfield
  {journal} {\bibinfo  {journal} {Journal of Rheology}\ }\textbf {\bibinfo
  {volume} {67}},\ \bibinfo {pages} {877--} (\bibinfo {year}
  {2023})}\BibitemShut {NoStop}%
\bibitem [{\citenamefont {Vona}\ and\ \citenamefont
  {Lauga}(2021)}]{vonaStabilizingViscousExtensional2021}%
  \BibitemOpen
  \bibfield  {author} {\bibinfo {author} {\bibnamefont {Vona}, \bibfnamefont
  {M.}}\ and\ \bibinfo {author} {\bibnamefont {Lauga}, \bibfnamefont {E.}},\
  }\bibfield  {title} {\enquote {\bibinfo {title} {Stabilizing viscous
  extensional flows using reinforcement learning},}\ }\href {\doibase
  10.1103/PhysRevE.104.055108} {\bibfield  {journal} {\bibinfo  {journal}
  {Physical Review E}\ }\textbf {\bibinfo {volume} {104}},\ \bibinfo {pages}
  {055108} (\bibinfo {year} {2021})}\BibitemShut {NoStop}%
\bibitem [{\citenamefont {Wang}, \citenamefont {Han},\ and\ \citenamefont
  {Yu}(2012)}]{wang2012numerical}%
  \BibitemOpen
  \bibfield  {author} {\bibinfo {author} {\bibnamefont {Wang}, \bibfnamefont
  {J.}}, \bibinfo {author} {\bibnamefont {Han}, \bibfnamefont {J.}}, \ and\
  \bibinfo {author} {\bibnamefont {Yu}, \bibfnamefont {D.}},\ }\bibfield
  {title} {\enquote {\bibinfo {title} {Numerical studies of geometry effects of
  a two-dimensional microfluidic four-roll mill on droplet elongation and
  rotation},}\ }\href@noop {} {\bibfield  {journal} {\bibinfo  {journal}
  {Engineering analysis with boundary elements}\ }\textbf {\bibinfo {volume}
  {36}},\ \bibinfo {pages} {1453--1464} (\bibinfo {year} {2012})}\BibitemShut
  {NoStop}%
\bibitem [{\citenamefont {Xie}\ and\ \citenamefont
  {Leonetti}(2023)}]{CRMECA_2023__351_S2_163_0}%
  \BibitemOpen
  \bibfield  {author} {\bibinfo {author} {\bibnamefont {Xie}, \bibfnamefont
  {K.}}\ and\ \bibinfo {author} {\bibnamefont {Leonetti}, \bibfnamefont {M.}},\
  }\bibfield  {title} {\enquote {\bibinfo {title} {Mechanical characterization
  of core-shell microcapsules},}\ }\href {\doibase 10.5802/crmeca.148}
  {\bibfield  {journal} {\bibinfo  {journal} {Comptes Rendus. M\'ecanique}\
  }\textbf {\bibinfo {volume} {351}},\ \bibinfo {pages} {163--182} (\bibinfo
  {year} {2023})}\BibitemShut {NoStop}%
\bibitem [{\citenamefont {Yu}\ \emph {et~al.}(2023)\citenamefont {Yu},
  \citenamefont {Qin}, \citenamefont {Chen}, \citenamefont {Shu},\ and\
  \citenamefont {Yuan}}]{yu_two-relaxation-time_2024}%
  \BibitemOpen
  \bibfield  {author} {\bibinfo {author} {\bibnamefont {Yu}, \bibfnamefont
  {Y.}}, \bibinfo {author} {\bibnamefont {Qin}, \bibfnamefont {Z.}}, \bibinfo
  {author} {\bibnamefont {Chen}, \bibfnamefont {S.}}, \bibinfo {author}
  {\bibnamefont {Shu}, \bibfnamefont {S.}}, \ and\ \bibinfo {author}
  {\bibnamefont {Yuan}, \bibfnamefont {H.-Z.}},\ }\bibfield  {title} {\enquote
  {\bibinfo {title} {Two-relaxation-time regularized lattice boltzmann model
  for navier-stokes equations},}\ }\href@noop {} {\bibfield  {journal}
  {\bibinfo  {journal} {arXiv preprint arXiv:2312.10318}\ } (\bibinfo {year}
  {2023})}\BibitemShut {NoStop}%
\bibitem [{\citenamefont {Yu}\ \emph {et~al.}(2025)\citenamefont {Yu},
  \citenamefont {Qin}, \citenamefont {Yuan},\ and\ \citenamefont
  {Shu}}]{YU2025129135}%
  \BibitemOpen
  \bibfield  {author} {\bibinfo {author} {\bibnamefont {Yu}, \bibfnamefont
  {Y.}}, \bibinfo {author} {\bibnamefont {Qin}, \bibfnamefont {Z.}}, \bibinfo
  {author} {\bibnamefont {Yuan}, \bibfnamefont {H.}}, \ and\ \bibinfo {author}
  {\bibnamefont {Shu}, \bibfnamefont {S.}},\ }\bibfield  {title} {\enquote
  {\bibinfo {title} {Two-relaxation-time regularized lattice boltzmann model
  for convection-diffusion equation with spatially dependent coefficients},}\
  }\href {\doibase https://doi.org/10.1016/j.amc.2024.129135} {\bibfield
  {journal} {\bibinfo  {journal} {Applied Mathematics and Computation}\
  }\textbf {\bibinfo {volume} {488}},\ \bibinfo {pages} {129135} (\bibinfo
  {year} {2025})}\BibitemShut {NoStop}%
\bibitem [{\citenamefont {Zhao}\ and\ \citenamefont
  {Yong}(2017)}]{zhao_single-node_2017}%
  \BibitemOpen
  \bibfield  {author} {\bibinfo {author} {\bibnamefont {Zhao}, \bibfnamefont
  {W.}}\ and\ \bibinfo {author} {\bibnamefont {Yong}, \bibfnamefont {W.-A.}},\
  }\bibfield  {title} {\enquote {\bibinfo {title} {Single-node second-order
  boundary schemes for the lattice {Boltzmann} method},}\ }\href {\doibase
  10.1016/j.jcp.2016.10.049} {\bibfield  {journal} {\bibinfo  {journal}
  {Journal of Computational Physics}\ }\textbf {\bibinfo {volume} {329}},\
  \bibinfo {pages} {1--15} (\bibinfo {year} {2017})}\BibitemShut {NoStop}%
\bibitem [{\citenamefont {Zhou}\ and\ \citenamefont
  {Schroeder}(2016)}]{PhysRevFluids.1.053301}%
  \BibitemOpen
  \bibfield  {author} {\bibinfo {author} {\bibnamefont {Zhou}, \bibfnamefont
  {Y.}}\ and\ \bibinfo {author} {\bibnamefont {Schroeder}, \bibfnamefont
  {C.~M.}},\ }\bibfield  {title} {\enquote {\bibinfo {title} {Single polymer
  dynamics under large amplitude oscillatory extension},}\ }\href {\doibase
  10.1103/PhysRevFluids.1.053301} {\bibfield  {journal} {\bibinfo  {journal}
  {Phys. Rev. Fluids}\ }\textbf {\bibinfo {volume} {1}},\ \bibinfo {pages}
  {053301} (\bibinfo {year} {2016})}\BibitemShut {NoStop}%
\end{thebibliography}%

\end{document}